\documentclass[12pt]{article}
\usepackage{jheppub}

\pdfoutput=1

\usepackage{amsmath,bbm,array,amsfonts,graphicx,wrapfig,lscape,float,mathtools,multirow,longtable}
\usepackage[dvipsnames]{xcolor}
\usepackage{stackrel}
\usepackage[all]{xy}

\newcommand{\be}{\begin{equation}}
\newcommand{\ee}{\end{equation}}
\newcommand{\beq}{\begin{equation}}
\newcommand{\beql}[1]{\begin{equation}\label{#1}}
\newcommand{\eeq}{\end{equation}}
\newcommand{\ba}{\begin{array}}
\newcommand{\ea}{\end{array}}
\newcommand{\bea}{\begin{eqnarray}}
\newcommand{\beal}[1]{\begin{eqnarray}\label{#1}}
\newcommand{\eea}{\end{eqnarray}}
\newcommand{\ben}{\begin{enumerate}}
\newcommand{\een}{\end{enumerate}}
\newcommand{\bean}{\begin{eqnarray*}}
\newcommand{\eean}{\end{eqnarray*}}
\newcommand{\eref}[1]{(\ref{#1})}
\newcommand{\sref}[1]{\S\ref{#1}}

\newcommand{\fref}[1]{Figure \ref{#1}}
\newcommand{\btab}[1]{\begin{tabular}{#1}}
\newcommand{\etab}{\end{tabular}}

\def \Black {\pdfsetcolor{0 0 0 1}}

\newcommand{\comment}[1]{}

\newcommand{\qed}{\nobreak \ifvmode \relax \else
      \ifdim\lastskip<1.5em \hskip-\lastskip
      \hskip1.5em plus0em minus0.5em \fi \nobreak
      \vrule height0.75em width0.5em depth0.25em\fi}

\newcommand{\old}[1]{}

\definecolor{darkspringgreen}{rgb}{0.09, 0.45, 0.27}
\definecolor{forestgreen}{rgb}{0.13, 0.55, 0.13}
\definecolor{blue2}{rgb}{0.2, 0.59, 0.98}

\usepackage{array}
\newcolumntype{C}[1]{>{\centering\let\newline\\\arraybackslash\hspace{0pt}}m{#1}}

\title{Higher Cluster Categories and QFT Dualities} 

\author[a,b]{Sebasti\'an Franco,} 
\author[c]{Gregg Musiker}

\affiliation[a]{
Physics Department, The City College of the CUNY \\
160 Convent Avenue, New York, NY 10031, USA}

\affiliation[b]{The Graduate School and University Center, The City University of New York  \\
365 Fifth Avenue, New York NY 10016, USA}

\affiliation[c]{
School of Mathematics, University of Minnesota, Minneapolis, MN 55455, USA
}

\emailAdd{sfranco@ccny.cuny.edu}
\emailAdd{musiker@math.umn.edu}

\preprint{
\begin{flushright}
CCNY-HEP-17-05
\end{flushright}
}

\abstract{We present a unified mathematical framework that elegantly describes minimally SUSY gauge theories in even dimension, ranging from $6d$ to $0d$, and their dualities. This approach combines recent developments on graded quiver with potentials, higher Ginzburg algebras and higher cluster categories (also known as $m$-cluster categories). Quiver mutations studied in the context of mathematics precisely correspond to the order $(m+1)$ dualities of the gauge theories. Our work suggests that these equivalences of quiver gauge theories sit inside an infinite family of such generalized dualities, whose physical interpretation is yet to be understood.
}

\begin{document}

\maketitle

\section{Introduction}

Recently, it was realized that minimally supersymmetric gauge theories in $6-2m$ dimensions exhibit order $(m+1)$ dualities, generalizing the well known case of Seiberg duality for $4d$ $\mathcal{N}=1$ theories \cite{Seiberg:1994pq}.\footnote{By ``order $(m+1)$ duality" we mean a generalization of duality relating $(m+1)$ different theories. Furthermore, in this case, $(m+1)$ consecutive applications of an elementary duality transformation amounts to the identity.} The first hint in this direction was the discovery that $2d$ $\mathcal{N}=(0,2)$ gauge theories enjoy an order 3 duality named {\it triality} \cite{Gadde:2013lxa}. This was soon followed by the proposal of {\it quadrality}, an order 4 duality, for $0d$ $\mathcal{N}=1$ gauge theories \cite{Franco:2016tcm}.

There has also been significant progress in the brane engineering of $2d$ $\mathcal{N}=(0,2)$ and $0d$ $\mathcal{N}=1$ theories. These constructions include D-brane probes of toric Calabi-Yau (CY) singularities \cite{Franco:2015tna}, T-dual brane configurations generalizing brane tilings \cite{Franco:2015tya,Franco:2016nwv,Franco:2016fxm, Franco:2017cjj} and D-branes in the mirror geometries \cite{Franco:2016qxh,Franco:2016tcm}.\footnote{See \cite{Tatar:2015sga,Benini:2015bwz, Schafer-Nameki:2016cfr, Apruzzi:2016iac} for alternative constructions of $2d$ $\mathcal{N}=(0,2)$ theories.} These brane configurations have been useful for both understanding and postulating some of these dualities.

In parallel, there have been interesting mathematical developments concerning graded quivers with potentials \cite{2008arXiv0809.0691B, 2015arXiv150402617O}, higher Ginzburg algebras \cite{2010arXiv1008.0599V,2015arXiv150402617O} and higher cluster categories \cite{2008arXiv0809.0691B}. While these topics are closely related to each other, its presentation in the literature has not been fully integrated. In this paper, we will show that they can be combined into a unified mathematical framework that elegantly describes minimally SUSY gauge theories in even dimension, ranging from $6d$ to $0d$. Moreover, quiver mutations studied in the mathematical context precisely correspond to the order $(m+1)$ dualities of the gauge theories. Higher Ginzburg algebras thus provide an algebraic unification of gauge theories in different dimensions and their dualities, which is similar to the geometric unification attained in \cite{Cachazo:2001sg,Franco:2016qxh,Franco:2016tcm} using mirror symmetry. Interestingly, this realization suggests that these equivalences of quiver gauge theories sit inside an {\it infinite family} of such generalized dualities, whose physical interpretation is yet to be uncovered. Our presentation will try to make the mathematical concepts accessible to the physics audience and vice versa. 

This paper is organized as follows. \sref{section_graded_quivers} introduces graded quivers with potentials. These are quivers containing different types of arrows, whose number is controlled by an integer $m\geq 1$. \sref{section_Ginzburg_algebras} discusses the higher Ginzburg algebras associated to such quivers. \sref{section_mutations_quivers} contains some of the key ideas of this paper, introducing mutations of graded quivers and their potentials. The order of such mutations is established in \sref{section_order_mutation}. In \sref{section_gen_anom}, the field theoretic concept of anomalies is generalized to graded quivers with arbitrary $m$. \sref{section_map_to_physics} outlines the connection between graded quivers with $m\leq 3$ and physics, in terms of  gauge theories in various dimensions and D-brane probing CY singularities. \sref{section_gauge_theories} discusses at length the connection between graded quivers with $m=0,1,2,3$ and minimally supersymmetric gauge theories in $d=6,4,2,0$. \sref{section_mutations_QFT_dualities} explains how the mutations of graded quivers unify the order $(m+1)$ dualities of the corresponding gauge theories. In \sref{section_mirror_symmetry} we discuss the class of graded quivers coming from toric CY's and explain how they are described using mirror symmetry. \sref{section_dimensional_reduction} generalizes the physical notion of dimensional reduction to arbitrary $m$. We conclude and present directions for future research in \sref{section_conclusion}. We also include four appendices, discussing the mathematics of potentials, the mutation of differentials, cluster categories and silting.

\section{Graded Quivers with Potentials}

\label{section_graded_quivers}

In this section we introduce {\it graded quivers} with potentials. Our treatment combines the ideas developed in \cite{2008arXiv0809.0691B, 2015arXiv150402617O}.  Buan and Thomas \cite{2008arXiv0809.0691B} defined graded quivers, called {\it colored quivers} therein, motivated by their generalization of \emph{cluster categories} to \emph{higher cluster categories} (or $m$-cluster categories).  Further mathematical details of (higher) cluster categories and (higher) \emph{tilting theory} are included in Appendix \ref{sec:CC}.  Oppermann \cite{2015arXiv150402617O} was motivated by a variant of tilting theory known as \emph{silting}, see Appendix \ref{sec:silting}, and \emph{higher Ginzburg algebras}.  Because of its closer connection to Calabi-Yau manifolds and physics, we utilize this second perspective for the majority of this paper.

\subsection{Graded Quivers}

\label{gradquivs}

A {\it quiver} $Q = (Q_0,Q_1,s,t)$ consists of a set of nodes, $Q_0,$\footnote{The nodes are often indexed as $\{1,2,\dots, n\}$.} a set of arrows $Q_1$, and two functions, $s$ and $t$ that denote the start and target of every arrow.  In particular, $\varphi: v \to w \in Q_1$ has $s(\varphi) = v$ and $t(\varphi)= w$.  We say that a quiver is {\it finite} if it consists of a finite number of nodes and arrows.  A {\it path} is a concatenation of arrows $\varphi_1 \varphi_2\cdots \varphi_k$ such that  $s(\varphi_{i+1}) = t(\varphi_i)$.  We say that $k$ is the length of such a path.  A path is known as a {\it cycle} if in addition it satisfies the identity $s(\varphi_1) = t(\varphi_k)$.

Given an algebraically closed field $k$, e.g. $\mathbb{C}$, we let $kQ$ be the {\it path algebra} of $Q$.  The path algebra is defined as the algebra whose elements are paths plus idempotents $e_i$ for $i \in Q_0$.  By convention, we consider $e_i$ to be a path of length $0$ and set $s(e_i) = t(e_i) = i$. We define multiplication in the path algebra by concatenation, i.e. $p\cdot q = pq$ if $s(q) = t(p)$ and $p\cdot q = 0$ otherwise (where $p$ or $q =e_i$ possible).  In particular, $e_i\cdot e_j = \delta_{ij} e_i$, i.e. the $\{e_i\}$'s indeed a form an orthogonal collection of elements that are unchanged by taking their power.

We now fix $m$ to be a nonnegative integer and use this parameter to turn $Q$ into a graded quiver $\overline{Q}$. In particular $\overline{Q_0}=Q_0$ is the same set of nodes, but $\overline{Q_1}$ is now the set of graded arrows. A graded arrow $\varphi: i \to j \in Q_1$ has a start $s(\varphi) = i \in Q_0$, a target $t(\varphi) = j \in Q_0$, and a degree $|\varphi|$ which we will assume is an integer from the set $\{0,1,2,\dots, m\}$.\footnote{In other references, such as \cite{2008arXiv0809.0691B}, the degree of an arrow is instead referred to as the color.  Graded quivers are consequently also called colored quivers.}  For every graded arrow $\varphi \in Q_1$, we also adjoin its opposite $\varphi_{op} : j \to i$ with its start and target reversed, and degree given as $|\varphi_{op}| = m - |\varphi|$. Since the integer $m$ determines the possible degrees or colors, different values of $m$ give rise to qualitatively different classes of graded quivers.

Lastly, for every node $i \in Q_0$, we adjoin a loop $\ell_i$ based at node $i$, i.e. with $s(\ell_i)=t(\ell_i)=i$ and degree $|\ell_i|= -1$. These special loops are the only arrows of $\overline{Q}$ with degree greater than $m$. Since such loops are present at every node, we will leave them implicit whenever we draw a quiver diagram.

We let $\overline{Q}$ denote the resulting graded quiver after adjoining the opposite arrows and loops.\footnote{ \label{fn:opop} In \cite{2015arXiv150402617O}, Oppermann also includes the identity 
$(\varphi_{op})_{op} = (-1)^{|\varphi| (m - |\varphi|) + 1}~ \varphi$ as part of the definition.  We gloss over this technicality for the time being.}

We will soon incorporate \emph{potentials}, which are linear combinations of cycles.  These will allow us to consider quivers that contain loops (i.e. adjoints) and 2-cycles.

\paragraph{Double arrows.}

\begin{figure}[ht]
	\centering
	\includegraphics[width=4cm]{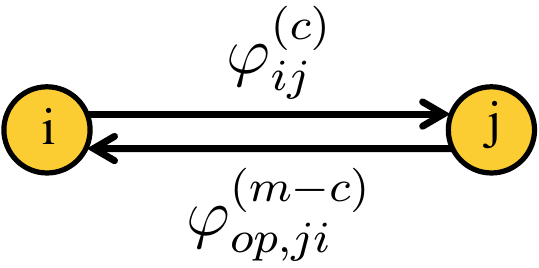}
\caption{Double arrow in a graded quiver.}
	\label{double_arrow}
\end{figure}

It is convenient to combine every $\varphi^{(c)}_{ij}$ with its corresponding $\varphi^{(m-c)}_{op,ji}$ to form a {\it double arrow}, as illustrated in \fref{double_arrow}.\footnote{This also applies to adjoint arrows, for which $i=j$.} In this extended notation, the subscripts are the nodes connected by an arrow and the superscript indicates its degree. We will refer to such a pair as a $(c,m-c)$ arrow. The number of different types of double arrows, i.e the number of $(c,m-c)$ pairs with $0\leq c \leq m/2$, is
\beq
n_f=\lfloor (m+1)/2 \rfloor .
\eeq

Double arrows of $(0,m)$ type exists for any $m$. Motivated by physics, we refer to such arrows as {\it chiral fields}. Furthermore, for even $m$ there is always a type of $(m/2,m/2)$ arrows, in which both components have the same degree.

The signed adjacency matrix is {\it skew-symmetric}, which in the context of graded quivers means that
\beq
q_{ij}^{(c)}=q_{ji}^{(m-c)} .
\eeq

\paragraph{Remark.} 
We will not impose the {\it monochromaticity} condition of \cite{2008arXiv0809.0691B}, i.e. we will not require that if $q_{ij}^{(c)}\neq 0$, then $q_{ij}^{(c')}=0$ for $c\neq c'$.

\paragraph{Cyclic order.}

\begin{figure}[ht]
	\centering
	\includegraphics[width=9cm]{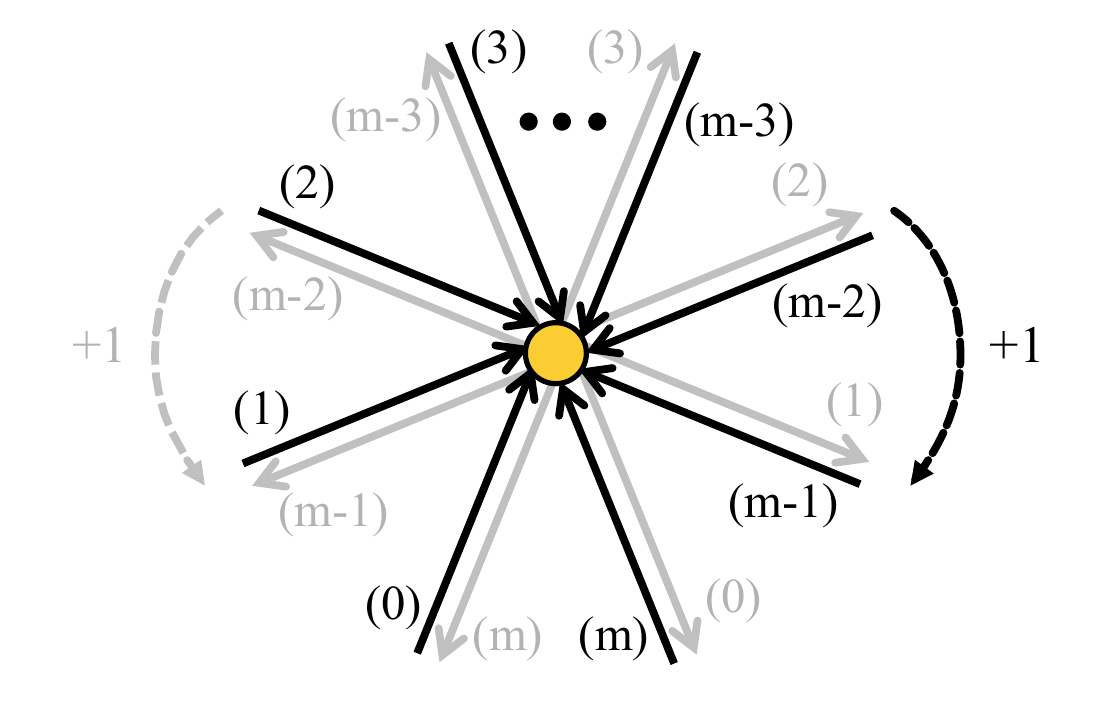}
\caption{Cyclic ordering of arrows connected to a node. The degree of incoming arrows increases clockwise.}
	\label{ordering_arrows}
\end{figure}

Let us focus on a node in the quiver. Taking into consideration both the degrees of arrows and their incidence orientation with respect to the node, there are $(m+1)$ different possibilities. There is a natural cyclic order for arrows around the node, in which the degree of incoming arrows increases clockwise, as shown in \fref{ordering_arrows}. There might be multiple or no arrows of each type. 

This order will become handy when discussing mutations and, as explained in \sref{section_mirror_symmetry}, also arises from mirror symmetry.

 \paragraph{Double arrows as quantum fields and their orientation.}

\label{sec_doub-arr_orientation}

Perhaps it is not surprising that a framework that allows multiple types of arrows can be useful for describing gauge theories in different dimensions. Such theories can contain different type of superfields which, as we discuss below, are captured by the different types, i.e. degrees, of arrows in the quiver.

Given a double arrow, the distinction between $\varphi^{(c)}_{ij}$ and $\varphi^{(m-c)}_{op,ji}$ is arbitrary. As we will explain in \sref{section_gauge_theories}, in cases with known physical interpretations as quantum field theories, every double arrow corresponds to a matter superfield. More generally, a double arrow should be regarded as a single entity. With this in mind, it is natural to associate an {\it orientation} to double arrows. Without loss of generality, let us assume that $0 \leq c\leq m/2$. We will adopt the following convention: 
\beq
(\varphi^{(c)}_{ij},\varphi^{(m-c)}_{op,ji}) \ \ \to \ \ 
\left\{\begin{array}{rc}
\mbox{even c:} & \Phi_{ij}^{(c)} \\[.2cm]
\mbox{odd c:} & \Phi_{ji}^{(c)}
\end{array}\right.
\eeq
which is defined such that it coincides with the orientation of fields in quantum field theories. In other words, $\varphi^{(c)}_{ij}$ and $\varphi^{(m-c)}_{op,ji}$ should be identified with physical fields or their conjugates as follows:
\beq
\begin{array}{c|c|c}
&  \ \ \ \varphi^{(c)}_{ij} \ \ \ & \ \ \ \varphi^{(m-c)}_{op,ji} \ \ \ \\  \hline 
\mbox{even c} & \Phi_{ij}^{(c)} & \bar{\Phi}_{ij}^{(c)} \\ \hline
\mbox{odd c} & \bar{\Phi}_{ji}^{(c)}& \Phi_{ji}^{(c)}
\end{array}
\label{orientation_fields}
\eeq 
For brevity, in what follows we will often use the terms double arrow and field interchangeably.  

In the special case of even $m$, there is an ambiguity in identifying the field associated to an $(m/2,m/2)$ double arrow. We can either pick the corresponding field to be $\Phi_{ij}^{(m/2)}= \varphi^{(c)}_{ij}$ or $\Phi_{ji}^{(m/2)}= \varphi^{(c)}_ {op,ji} $. This is possible because such fields are unoriented but, more importantly, is a manifestation of a $\Phi_{ij}^{(m/2)} \leftrightarrow \overline{\Phi}_{ij}^{(m/2)}$ symmetry of such theories.\footnote{Notice that we are not saying that $\Phi_{ij}^{(m/2)}$ and $\overline{\Phi}_{ij}^{(m/2)}$ are equal. This symmetry is a generalization to all even $m$ of the well-known Fermi-conjugate Fermi symmetry of $2d$ $\mathcal{N}=(0,2)$ gauge theories. This symmetry acts on each unoriented field independently.} This issue will be revisited in coming sections.

The orientation of fields becomes more significant in physics, where it enters the determination of anomalies. The orientation in \eref{orientation_fields} is nicely consistent with the generalization of anomalies to arbitrary $m$ that we will introduce in \sref{section_gen_anom}.

Finally, in the cases with a gauge theory interpretation, we can identify the loops $\ell_i$ at every node with vector superfields.\footnote{More precisely, we mean gauge supermultiplets. In particular, we refer to the corresponding superfield in $0d$ as a gaugino superfield, since it has no vector component. For brevity, this distinction will be implicit throughout most of the paper.} Since these loops are in one-to-one correspondence with nodes in the quiver and we will leave them implicit we can simply say, as it is standard, that nodes correspond to vector multiplets. Notice that the $\ell_{i}$'s are special in that graded quivers do not include their conjugates, they would be $\ell_{op,i}$'s. This is in nice agreement with physics, where vector superfields satisfy a reality condition.

 \paragraph{Single arrow representation.}

For simplicity, throughout the paper we will often focus on a single arrow representative for every double arrow. A way of doing so is by simply picking any of the two arrows, let us call it $\varphi^{(c)}_{ij}$, keeping $\varphi^{(m-c)}_{op,ji}$ implicit. In what follows, the choice of representative in a double arrow will be guided by practical purposes. Another natural way of picking a single arrow representation is by using the physical orientation we introduced above. This will be the approach we will use when connecting to gauge theories in \sref{section_gauge_theories}.

\subsection{Ranks}

\label{sec:ranks}

We complete the definition of a graded quiver by assigning an integer $N_i$ to every node. This ingredient is typically absent in the math literature. In physics, each node corresponds to an $U(|N_i|)$ gauge group, so we refer to these integers as {\it ranks}.

It is natural to restrict ourselves to positive ranks. Negative ranks can even be generated when starting from quivers with non-negative ranks and applying a sequence of mutations. In physics, the presence of negative ranks is typically an indication of SUSY breaking. It would be interesting to determine whether SUSY breaking has a mathematical counterpart.

For non-trivial ranks, arrows connecting nodes $i$ and $j$ become $|N_i| \times |N_j|$ matrices.\footnote{This is also the case when $i=j$.} The matrix structure of arrows will be implicit in our presentation.

\subsection{Potentials}

\label{section_potentials}

In this section we extend the theory of graded quivers to include potentials. Our discussion is closely related to the one in \cite{2015arXiv150402617O}. 

The potential $W$ is a $\mathbb{C}$-linear combination of certain cycles in the path algebra $k\overline{Q}$, excluding the loops $\ell_i$.\footnote{More generally, open paths terminating on frozen nodes can also be terms in the potential. In what follows, this additional possibility will be implicit whenever we refer to cycles. In physics, the cycle condition corresponds to gauge invariance.} More precisely, $W \in k\overline{Q} / [k\overline{Q}, k\overline{Q}]$, where we quotient by the supercommutator 
$[u, v] = uv - (-1)^{|u| |v|} v u$. In other words, a cycle is an equivalence class of words made by closed paths of arrows up to sign as defined by 
\beq
\label{eq:sgneq}
(u_1u_2\cdots u_k)(v_1v_2\cdots v_\ell) \sim (-1)^{(|u_1|+|u_2|+\dots + |u_k|)(|v_1|+|v_2|+\dots + |v_\ell|)} (v_1v_2\cdots v_\ell)(u_1u_2\cdots u_k) .
\eeq
This is well-defined since we assume the path is a cycle, i.e. closed with $t(v_\ell)=s(u_1)$.

In addition, cycles must have degree\footnote{The degree of a product of arrows is the sum of the degrees of the constituent arrows.} $(m-1)$ in order to be allowed potential terms. The reasons for this will be explained momentarily in \sref{section_Ginzburg_algebras}. It is important to note that the potential $W$ does not necessarily contain all degree $(m-1)$ cycles. As an immediate important consequence of this restriction on the degree of the potential, there cannot be two arrows of the same type going in opposite directions along a cycle in the potential. I.e., when considered with the same orientation, we cannot simultaneously have degree $c$ and $(m-c)$ for any $c$. Physically, no potential term can simultaneously contain a type of superfield and its conjugate.  Additionally, arrows of degree $m$ cannot appear in potentials of such degrees.

As we explain in \sref{section_Ginzburg_algebras}, potentials give rise to relations in the path algebra. In physics, they encode non-gauge interactions.

\subsubsection{Further Constraints on the Potential: Kontsevich Bracket}

The {\it Kontsevich bracket}, sometimes also referred to as {\it necklace bracket}, between two functions $f$ and $g$ of the arrows in a quiver is defined as
\beq \label{eq:Kontsevich}
\{f,g \} = \sum_{\Phi \in \overline{Q}} \left({\partial f \over \partial \Phi} {\partial g \over \partial \bar{\Phi}} - {\partial f \over \partial \bar{\Phi}} {\partial g \over \partial \Phi} \right) .
\eeq
It is a generalization of a Poisson bracket defined by a quiver \cite{MR1906711, MR1839485, MR1247289}. When evaluating this bracket, it is necessary to take into account the commutation rules for arrows, which for a pair of them is given by $uv \sim (-1)^{|u||v|} vu$.

The potential is required to satisfy the condition that the Kontsevich bracket vanishes
\beq
\{W,W \} = 0 .
\label{KB_W}
\eeq
In \sref{section_Ginzburg_algebras} we will explain that this condition is necessary for the differential on the Ginzburg algebra to square to zero. The condition \eref{KB_W} leads to non-trivial constraints on the potential. In \sref{section_gauge_theories} we will explicitly consider the Kontsevich bracket for $m=1,2,3$.

\subsubsection{Mass Terms and Removable 2-Cycles} 

\label{sec:mass-term}

\begin{figure}[ht]
	\centering
	\includegraphics[width=6cm]{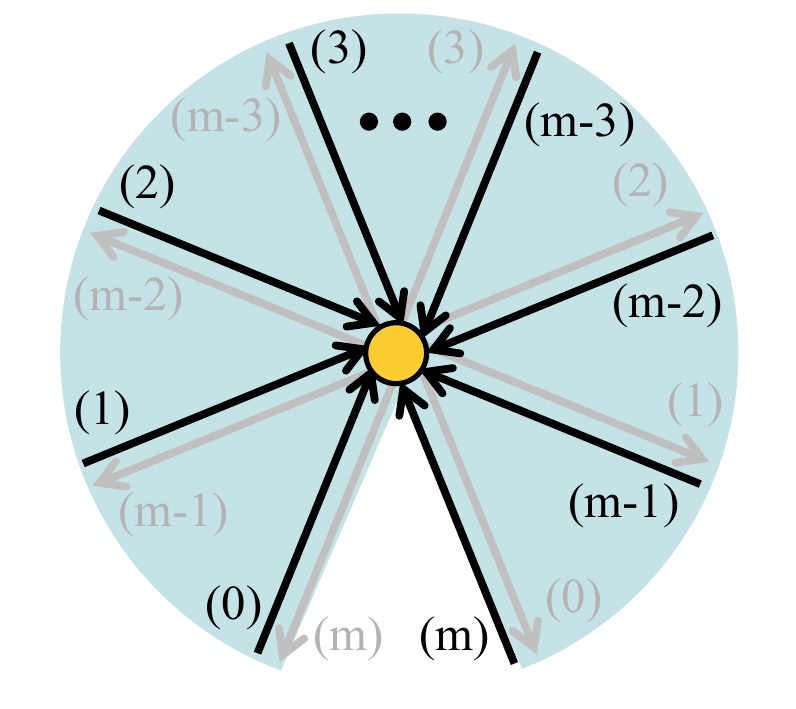}
\caption{The upper wedge.}
	\label{upper_wedge}
\end{figure}

Quadratic terms in the potential are of particular significance. In physics, they correspond to {\it mass terms}. Since the potential must have degree $(m-1)$, it is straightforward to classify all possible mass terms. For this, it is convenient to define the {\it upper wedge} of the plane. It corresponds to the wedge containing the $(0,m),(1,m-1),\ldots,(m,0)$ sequence of arrows, as shown in \fref{upper_wedge}. 

Every consecutive pair of fields in the upper wedge forming a closed path in the quiver can be combined into a quadratic term in the potential. We refer to such potential terms as {\it mass terms}. 

We define a {\it removable 2-cycle} as a length-2 closed path in the quiver that, in addition, appears in a mass term in the potential. In this case, it is possible to integrate out the corresponding arrows, as we explain in \sref{section_mutation_potential}.\footnote{It is important to emphasize that 2-cycles cannot be removed if the corresponding term is not present in the potential.} In physics, a removable 2-cycle corresponds to a massive pair of fields. Note that that chiral-chiral pairs can only form removable 2-cycles for $m=1$, for which $(0,m)$ and $(m,0)$ fields are consecutive on the upper wedge.

\section{Differential Graded Structures}

\label{section_Ginzburg_algebras}

We now introduce further structure that can be layered on top of a graded quiver with potential, combining together the treatments in \cite{2010arXiv1008.0599V,2015arXiv150402617O}.

\subsection{Differential Operators}

As our final ingredient, we introduce a differential operator $d: k\overline{Q} \to k\overline{Q}$ which lowers the degree of a given term by one and then extends linearly.  Furthermore, $d$ satisfies the graded Leibniz rule on products 
\beq
d(uv) = d(u)v + (-1)^{|u|}u~d(v).  
\eeq
We define the differential $d$ on graded arrows as follows:
\begin{eqnarray}
d(\alpha) & = & 0 \mathrm{~~~~~~~if~}\alpha\mathrm{~has~degree~zero.} \label{differential_1} \\
d(\alpha_{op}) & = & \partial_{\alpha}W \mathrm{~~if~}\alpha\mathrm{~has~degree}\in \{0,1,2,\dots,m-1\}. \label{differential_2} \\
d(\ell_{i}) & = & e_i(\sum_{\alpha \in Q_1} [\alpha, \alpha_{op}])e_i. \label{differential_3}
\end{eqnarray}
Here $e_i$ is the idempotent in the path algebra $k\overline{Q}$ at node $i$. The notation $\partial_\alpha W$ signifies the cyclic derivative which is defined as 
$\partial_\alpha (v_1\cdots v_{k-1}\alpha) = v_1 \cdots v_{k-1}$ and extended linearly. In the case that $v = v_1v_2\cdots v_k$ is not written ending with $\alpha$, we use signed cyclic equivalence \eref{eq:sgneq} to move $\alpha$ to the rightmost position.\footnote{When a term $v= v_1v_2\cdots v_k$ of the potential $W$ contains $\alpha$ at multiple places, $\partial_\alpha v$ is the sum of the remainders after each such removal. More precisely, $\partial_\alpha v$ is defined as 
$\sum_{v = p\alpha q} (-1)^{|p\alpha| |q|} ~qp$ in \cite{2015arXiv150402617O}.} As we will discuss later, the vanishings of all these differentials have important physical counterparts. 

For later use, we also note that the right-hand-side of \eref{differential_3} can be expressed as
\beq
e_i(\sum_{\alpha \in Q_1} [\alpha, \alpha_{op}])e_i=\sum_{\alpha:~ i ~\to ~?~ \in Q_1} \alpha \alpha_{op} - \sum_{\beta:~ ? ~\to ~i~ \in Q_1} \beta_{op} \beta\, 
\label{D-term}
\eeq
for fixed $i \in Q_0$. 

We now state two claims about differentials in our setting that will provide the backdrop for the correspondence between higher Ginzburg algebras and SUSY quantum field theories in various dimensions.  

\bigskip

\noindent \paragraph{Claim.} 
The fact that $W$ has degree $(m-1)$ implies that $d(\alpha_{op}) = \partial_\alpha W$ indeed has degree one less than that of $\alpha_{op}$ (whenever $\alpha$ has degree\footnote{In the special case when $|\alpha|=m$, then $|\alpha_{op}|=0$, and we use \eref{differential_1} to compute $d(\alpha_{op})$ instead.} $\in \{0,1,2,\dots, m-1\}$). 

\paragraph{Proof.} 

Suppose that $|\alpha| = c \in \{0,1,2,\dots, m-1\}$ so that $|\alpha_{op}| =(m-c) \in \{1,2,\dots, m\}$.
Then in $\partial_\alpha W$, only the terms containing $\alpha$ at least once survive.  Furthermore each term resulting from $\partial_\alpha W$ is a term $v$ of $W$ with exactly one copy of $\alpha$ removed, which we abbreviate as $v \setminus \alpha$.  Hence if $W$ is homogenous of degree $(m-1)$ then $|v| = m -1$ and $|v \setminus \alpha| = (m-1) - |\alpha| = m-1-c$.  We conclude that $d(\alpha_{op}) = \partial_\alpha W$ is homogeneous of degree $(m-c)-1$ as desired.

\bigskip

\noindent \paragraph{Claim.} If the potential $W$ vanishes under the Kontsevich backet, i.e. $\{W,W\}=0$, then the differential defined above indeed squares to zero, i.e. $d^2=0$.\footnote{This property is in fact crucial because otherwise the map $d: k\overline{Q} \to k\overline{Q}$ defined above would not be a \emph{differential operator}.}  

\paragraph{Proof.}  From \cite{2010arXiv1008.0599V}, it follows that for $f \in k\overline{Q}$, we have $df = \{W,f\}$.  Hence, as in (10.6) of \cite{2010arXiv1008.0599V}, we obtain 
\beq
d^2 f = \{W, \{W,f\}\} = \frac{1}{2}\{\{W,W\},f\} = 0.
\eeq
Furthermore, we have $d\ell_i = d\left(\sum_{\alpha:i \to ? \in Q_1} \alpha \alpha_{op} - \sum_{\beta: ? \to i \in Q_1} \beta_{op} \beta\right)=$
$$\sum_{\alpha:i \to ? \in Q_1} \left(d(\alpha) \alpha_{op} + (-1)^{|\alpha|} \alpha d(\alpha_{op})\right)
- \sum_{\beta: ? \to i \in Q_1}  \left(d(\beta_{op}) \beta + (-1)^{|\beta_{op}|} \beta_{op} d(\beta)\right)$$
\beq
+ \sum_{\alpha:i \to ? \in Q_1} \left((\partial_{\alpha_{op}}W) \alpha_{op} + (-1)^{|\alpha|} \alpha (\partial_{\alpha}W)\right)
- \sum_{\beta: ? \to i \in Q_1}  \left((\partial_{\beta}W) \beta + (-1)^{|\beta_{op}|} \beta_{op} (\partial_{\beta_{op}}W)\right). 
\eeq

Substituting in $\beta = \alpha_{op}$, we note that these two sums are actually over the same arrows and cancel out each other.\footnote{Technically, one uses footnote \ref{fn:opop} to simplify the instances of $(\alpha_{op})_{op}$ to $\pm \alpha$, but since the same sign change is applied twice, we still get zero.} Hence applying $d^2$ to the central element $\sum_{i\in Q_0} \ell_i$ yields zero.

\subsection{Higher Ginzburg Algebras}

Given the following three ingredients: 
\begin{itemize}
\item[1)] a graded quiver $\overline{Q}$ built from a quiver $Q$ with arrows $\alpha \in Q_1$, $|\alpha| \in \{0,1,2,\dots, m\}$, opposite arrows $\alpha_{op}$ for every such $\alpha$, and loops $\ell_i$ of degree $(m+1)$,
\item[2)] a potential $W \in k\overline{Q}/[k\overline{Q},k\overline{Q}]$ of degree $(m-1)$ satisfying $\{W,W\} = 0$, and
\item[3)] a differential operator $d: k\overline{Q}^{(j)} \to k\overline{Q}^{(j-1)}$ respecting the grading as above,\footnote{By convention, $k\overline{Q}^{(-1)} = 0$, so using \eref{differential_1} we have $d: k\overline{Q}^{(j)} \to k\overline{Q}^{(j-1)}$ even when $j=0$.} 
\end{itemize}
we let the {\it higher Ginzburg algebra} $\Gamma_{m+2}(Q,W)$ denote the dg (differential graded) algebra given as the direct sum 
\beq
\Gamma_{m+2}(Q,W) = \oplus_{j\geq 0} k\overline{Q}^{(j)} \, ,
\eeq
where $k\overline{Q}^{(j)}$ denotes the space of paths of degree $j$ in the graded path algebra $k\overline{Q}$, and quotienting by the ideal of arrows.  

The ordinary Ginzburg algebra from \cite{Ginzburg,KellerYang} corresponds to the $m=1$ case of the above.\footnote{Additionally, in \cite{KellerYang}, the authors focus on the \emph{completed} Ginzburg dg algebra $\widehat{\Gamma}(Q,W)$, i.e. the graded path algebra taking the limit of including paths of infinite length. This subtlety will not be needed in our work.}

\subsection{Jacobian Algebras and Vacuum Moduli Spaces}

\label{section_moduli_spaces}

We now consider the following result of Ladkani, which in turn is a generalization of \cite[Lemma 2.8]{KellerYang}.  In particular, the result implies that it is sufficient to consider a quotient algebra formed by quotienting only by the relations arising from cyclic derivatives with respect to arrows of degree $(m-1)$.

\noindent \paragraph{Claim \cite[Lemma 2.21]{Ladkani}.} Let $(Q,W)$ be a quiver with potential where $\overline{Q}$ is the associated graded quiver with degrees in $\{0,1,2,\dots, m\}$. Then the Jacobian algebra (with respect to the graded arrows of degree $(m-1)$) is the $0^{th}$ cohomology\footnote{In our language, this actually would be homology rather than cohomology, since we give arrows degrees which are positive rather than negative.} of the complete Ginzburg dg algebra $\widehat{\Gamma}_{m+2}(Q,W)$, i.e. 
\begin{equation} k\overline{Q}\bigg /\left(\{\partial_\alpha W: \alpha \in \overline{Q}_1^{(m-1)}\}\right) = H^0(\widehat{\Gamma}_{m+2}(Q,W)). \label{eq:Jacobian_algebra}\end{equation}

\paragraph{Proof.} The $0^{th}$ homology is defined as 
\beq
Ker ~d: \widehat{\Gamma}_{m+2}(Q,W)^{(0)} \to  \widehat{\Gamma}_{m+2}(Q,W)^{(-1)} \bigg / Im ~d: \widehat{\Gamma}_{m+2}(Q,W)^{(1)} \to \widehat{\Gamma}_{m+2}(Q,W)^{(0)} , 
\eeq
 where the superscripts indicate restricting to elements of the dg algebra $\widehat{\Gamma}_{m+2}(Q,W)$ of certain degrees.  Since we have no elements of degree $(-1)$, we get 
$\widehat{\Gamma}_{m+2}(Q,W)^{(-1)}= 0$ and hence the kernel is all of $\widehat{\Gamma}_{m+2}(Q,W)^{(0)}$, i.e. the component of the graded path algebra $k\overline{Q}^{(0)}$ on arrows of degree zero.  We get the immediate equality $k\overline{Q}^{(0)} = k\overline{Q}$, the ordinary path algebra. Furthermore, the image consists of $\{d(\alpha_{op})\}$ where $\alpha_{op}$ has degree one.  Hence the image consists of $\{\partial_\alpha W: \alpha \in \overline{Q}_1^{(m-1)}\}$, where the $(m-1)^{th}$ component yields arrows in $\overline{Q}_1$ of degree $(m-1)$.  In conclusion, we obtain the desired relations in the modified Jacobian algebra.

In the special case of $m=1$, the result in \cite{KellerYang} relates the ordinary Ginzburg algebra to the ordinary Jacobian algebra, i.e. the quotient algebra $kQ / \left(\{\partial_\alpha W: \alpha \in Q_1\}\right)$.

The mathematical importance of the Jacobian algebra with respect to arrows of degree $(m-1)$, i.e. of next to maximal degree, has a physical counterpart. As we will see in \sref{section_gauge_theories}, it is all we need for computing the (classical) moduli spaces of the corresponding quantum field theories. The underlying reason is that since the degree of the potential is $(m-1)$, $H^0(\widehat{\Gamma}_{m+2}(Q,W))$ consists exclusively of chiral fields, which are the only fields containing scalar components. When determining the moduli space, we also demand the vanishing of \eref{differential_3}. When it is expressed as in \eref{D-term}, it becomes clear that, when restricted to chiral fields, this condition corresponds to the vanishing of $D$-terms.

\subsubsection*{The Role of Higher Degree Arrows}

Since we focused solely on the Jacobian algebra, i.e. the $0^{th}$ homology, in the above, the reader might wonder what the roles of the higher degree arrows and components of the (higher) Ginzburg algebra are. These higher degree arrows and the differential operator are exactly defined so that all higher homologies of the (higher) Ginzburg algebra vanish.

In particular, in the $m=1$ case, notice that 
{\footnotesize 
\beq 
H_1(\widehat{\Gamma}_{m+2}(Q,W)) =  Ker ~d: \widehat{\Gamma}_{m+2}(Q,W)^{(1)} \to \widehat{\Gamma}_{m+2}(Q,W)^{(0)} / Im ~d: \widehat{\Gamma}_{m+2}(Q,W)^{(2)} \to \widehat{\Gamma}_{m+2}(Q,W)^{(1)}.
\eeq
}
We note that $\widehat{\Gamma}_{m+2}(Q,W)^{(2)}$ is generated by the loops $\ell_{i}$ and hence the image under $d$ is precisely generated by 
$e_i(\sum_{\alpha \in Q_1} [\alpha, \alpha_{op}])e_i$ as $i$ runs over possible nodes in $Q_0$.  Meanwhile the elements of $\widehat{\Gamma}_{m+2}(Q,W)^{(1)}$ are the expressions 
in $k\overline{Q}$ whose terms contain exactly one arrow $\alpha_{op} \in \overline{Q}_1^{(1)}$ and the rest in $\overline{Q}_1^{(0)}=Q_1$.  The kernel of $d$ acting on this set are precisely the elements which become zero in $k\overline{Q}^{(0)}=kQ$ when $\alpha_{op}$ is replaced with $\partial_\alpha W$.\footnote{This calculation uses $d(u_1u_2\dots u_k \alpha_{op}) = d(u_1u_2\dots u_k) \alpha_{op} + (-1)^0 u_1u_2 \dots u_k d(\alpha_{op}) = u_1u_2\dots u_k \partial_{\alpha} W$ if $|u_1|=|u_2|=\dots = |u_k| = |\alpha|=0$.  In fact for any such cyclic ordering, we have
$d(u_{j+1}u_{j+2}\cdots u_k \alpha_{op}u_1u_2\dots u_j) = u_{j+1}u_{j+2}\cdots u_k  \partial_{\alpha} W u_1u_2\dots u_j$.} We will show that the kernel is in fact also generated by the elements of the form $e_i(\sum_{\alpha \in Q_1} [\alpha, \alpha_{op}])e_i$, which has image 
\beq
\sum_{\alpha: ~i~ \to ~? ~ \in Q_1} \alpha (\partial_\alpha W) -  \sum_{\beta: ~?~ \to ~i ~ \in Q_1} (\partial_\beta W)\beta
\eeq 
under $d$.  This quantity indeed equals zero because the two summands express the sum over all terms in the potential $W$ incident to node $i$, in two different ways.  In the positive summand, we sum over all terms of $W$ that contain an arrow $\alpha$ starting at $i$ while in the negative summand we sum over all terms of $W$ that contain an arrow $\beta$ ending at $i$.  However, since terms in $W$ are cycles, these two summands cancel out each other.  For higher $m$, a related approach applies.

In Construction 2.6 of \cite{2015arXiv150402617O}, it is described in more detail how one can start with a basic finite dimensional algebra, for example one of the form $kQ / (R)$ where $kQ$ is a path algebra of a quiver $Q$ and $R$ is a minimal set of relations given by elements of the path algebra, and build a differential graded algebra.  In particular, let the arrows of the original quiver $Q$ be of degree $0$ and adjoin new arrows $\alpha_r$ of degree $1$ for each of the relations $r \in R$, such that the differential map sends $\alpha_r \in k\overline{Q}^{(1)}$ to $d(\alpha_r) = r \in k\overline{Q}^{(0)}$.  Consequently $H_0(k\overline{Q}^{(1)}) = k\overline{Q}^{(0)} / (R) = \Lambda$.  We then consider a generating set for $H_1(k\overline{Q}^{(1)})$ and adjoin arrows of degree $2$ for each element therein, defining the differential accordingly.  Iterating this process, considering relations of relations, we get a differential graded algebra such that all higher homologies vanish.

Because all higher homologies vanish, the sequence of homologies of the graded (higher) Ginzburg algebra agree identically with the sequence of homologies of the basic algebra $\Lambda$, treating this vacuously as a graded algebra (concentrated in degree zero).  We conclude that the higher Ginzburg algebra $\Gamma_{m+2}(Q,W)$ and $\Lambda$ are \emph{quasi-isomorphic}.  
As an application, when $\Lambda$ is the Jacobian algebra, it is quasi-isomorphic to $\Gamma_{m+2}(Q,W)$ while also agreeing to the $0^{th}$ homology of $\Gamma_{m+2}(Q,W)$.

However, because the higher Ginzburg algebra $\Gamma_{m+2}(Q,W)$ has the extra structure of a dg (differential graded) algebra, we observe that the higher Ginzburg algebra is \emph{$(m+2)$-Calabi-Yau}.  This means that $\Gamma_{m+2}(Q,W)$ is homology smooth and the shift functor $[m+2]$ is a Serre functor on the bounded derived category of finite length $\Gamma_{m+2}(Q,W)$-modules.  In particular, the composition of the suspension $[m+2]$ and duality yields a bimodule quasi-isomorphism \cite[Def. 3.2.3]{Ginzburg}, i.e. a transformation that induces certain symmetries between spaces of homomorphisms under duality.

\section{Order $(m+1)$ Mutations}

\label{section_mutations_quivers}

In this section we introduce mutations of graded quivers with potentials. Our treatment builds on the work in \cite{2008arXiv0809.0691B, 2015arXiv150402617O}. We explain how all the defining elements of a theory transform: the quiver in \sref{section_mutation_quiver}, the potential in \sref{section_mutation_potential} and the ranks in \sref{section_c-vectors_and_ranks}.  We postpone the explanation of how the differential transforms to Appendix \ref{section_mutation_diff}. 

We will restrict to mutations on nodes without adjoint fields, i.e. without loops. Preliminary studies of this case have appeared in mathematics literature \cite{2015arXiv150402617O,2011-BertaniOppermann}. However we consider the understanding of such cases to be incomplete. It would be very interesting to revisit this problem.

\subsection{Mutation of the Quiver}

\label{section_mutation_quiver}

Let us first explain how the graded quiver transforms under mutation.

\paragraph{1. Flavors.} 
Let us consider a mutation on node $j$. In physics, the arrows connected to the mutated node are usually referred to as {\it flavors}. The flavors are transformed as follows:

\begin{itemize}
\item[{\bf 1.a)}] {Replace every incoming arrow} \xymatrix{ i \ar@{->}[r]^{(c)} & j} with the arrow 
\xymatrix{ i \ar@{->}[r]^{(c-1)} & j}. 

\item[{\bf 1.b)}] Equivalently, replace every outgoing arrow \xymatrix{j \ar@{->}[r]^{(c)} & k} with an arrow \xymatrix{j \ar@{->}[r]^{(c+1)} & k}. 
\end{itemize}
\noindent Both of these values are taken {modulo $(m+1)$}.

\begin{figure}[H]
	\centering
	\includegraphics[width=14.5cm]{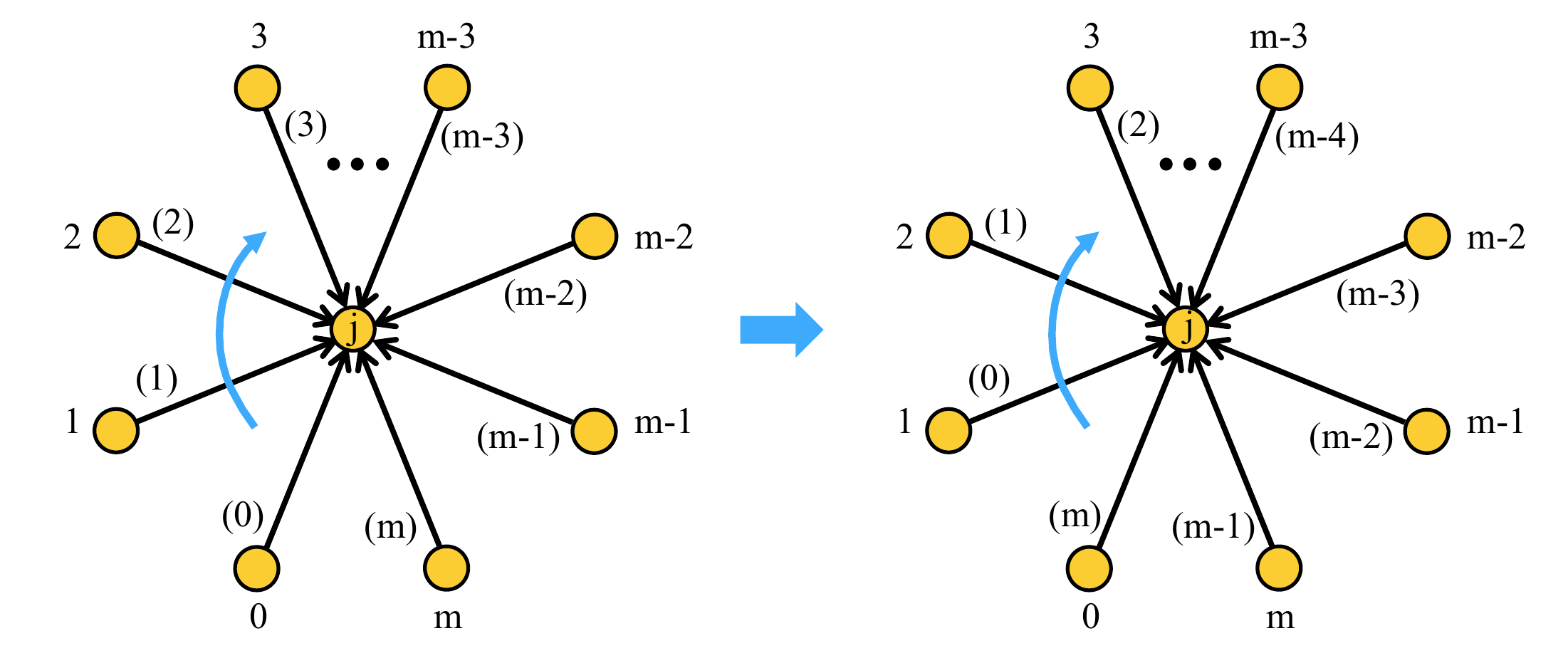}
\caption{The transformation of flavors upon a mutation on node $j$ can be implemented as a rotation of the degree of the arrows. The nodes remain fixed.}
	\label{mutation_flavors}
\end{figure}

This transformation has an elegant implementation in terms of the cyclic ordering of arrows introduced in \sref{gradquivs}. It simply becomes a rotation, as shown in \fref{mutation_flavors}, where we have numbered the spectator nodes to emphasize that they remain fixed under the mutation.

\paragraph{2. Composite Arrows.} 

The second step in the transformation of the quiver involves the addition of composite arrows as follows. For every $2$-path \xymatrix{ i \ar@{->}[r]^{(0)} & j \ar@{->}[r]^{(c)} & k} in $\overline{Q}$, where $c\neq m$, {add a new arrow} \xymatrix{ i  \ar@/^0.5pc/[rr]^{(c)} & j & k}.

In other words, we generate all possible composite arrows consisting of chiral fields, of degree zero, incoming into the mutated node and all other types of fields attached to it. In physics, such composite arrows are referred to as {\it mesons}. 

\begin{figure}[H]
	\centering
	\includegraphics[width=12cm]{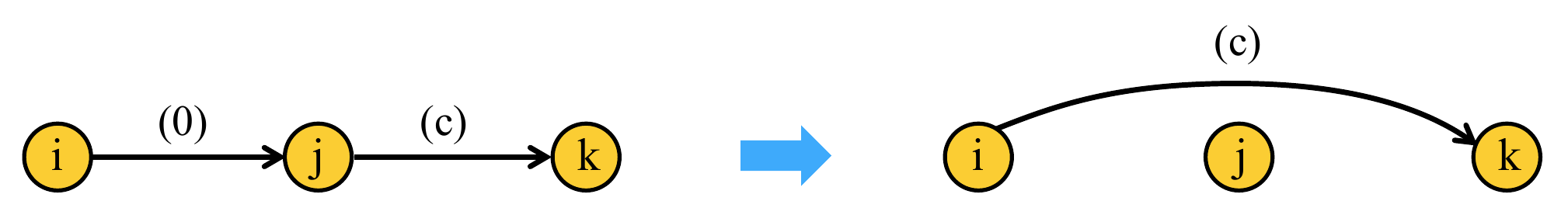}
\caption{Composite arrow, i.e. meson, generated by a mutation.}
	\label{mutation_mesons}
\end{figure}

\subsubsection*{Anticomposition} 

\begin{figure}[H]
	\centering
	\includegraphics[width=12cm]{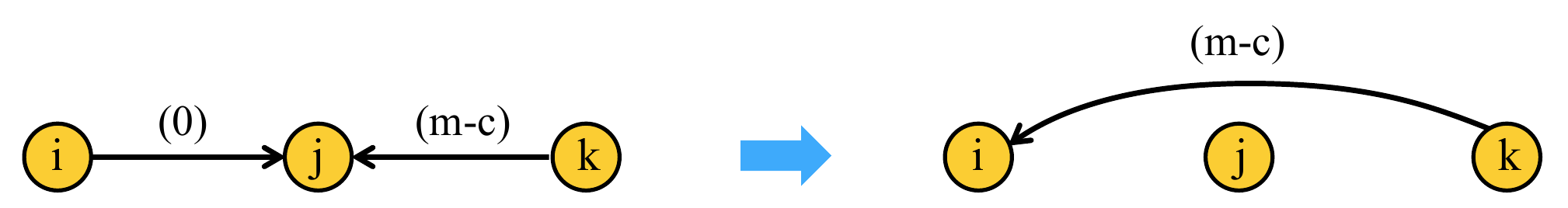}
\caption{Meson generated by anticomposition. Here $m-c\neq 0$. This rule is equivalent to the one in \fref{mutation_mesons}.}
	\label{mutation_ac_mesons}
\end{figure}

Equivalently, we can also understand the rule above as the composition of $\varphi_{ij}^{(0)}$ and $\varphi_{op,kj}^{(m-c)}$, with $m-c \neq 0$, even though their orientations seem to be incompatible. The result is a meson that we can call $\varphi_{op,ik}^{(m-c)}$, which is equivalent to a meson $\varphi_{ki}^{(c)}$. This phenomenon has been noted in both mathematics \cite{2015arXiv150402617O} and physics \cite{Franco:2016tcm} and we refer to it as {\it anticomposition}. Anticomposition becomes important for $m\geq 3$. This alternative formulation of the composition rule is illustrated in \fref{mutation_ac_mesons}.

Anticomposition becomes most shocking when focusing on the physical orientation of fields. In this case, a meson can correspond to the combination of a chiral field with another field of seemingly incompatible orientation, hence requiring the conjugation of the chiral field. This phenomenon was first noticed in \cite{Franco:2016tcm} in the context of $0d$ $\mathcal{N}=1$ quadrality. We will discuss it in further detail in \sref{section_mutations_QFT_dualities}.

It is worth noting we that we require $c\neq m$ in Rule $(2)$. In other words, the definition of composition forbids the anticomposition of two incoming chiral fields, i.e we cannot generate a meson by composing an incoming chiral with the conjugate of another incoming chiral.\footnote{Our mutation rule for mesons coincides with some works in the mathematics literature such as \cite[Sec. 6]{2015arXiv150402617O}, but slightly differs from others, particularly \cite{2008arXiv0809.0691B}. In \sref{sec:CC} we elaborate on the relation between our mutation prescription and \cite{2008arXiv0809.0691B}. Contrary to ours, those rules do not follow from a proper consideration of the potential, but have an equivalent effect when we restrict to higher cluster categories, which is the problem of interest in \cite{2008arXiv0809.0691B}.}

\subsection{Mutation of the Potential}

\label{section_mutation_potential}

We now explain how the potential transforms under mutation, which can be summarized by a short set of rules. In \cite{2015arXiv150402617O} Oppermann provided an alternative, but equivalent, prescription for mutating the potential. Our approach is more combinatorial than his differential geometric treatment. The connection between the two is discussed in Appendix \sref{section_mutation_diff}.

\begin{itemize}

\item[{\bf 2.a)}] {\bf Cubic dual flavors-meson couplings.} The first rule concerns new potential terms that are in one-to-one correspondence with mesons. For every $2$-path, \xymatrix{ i \ar@{->}[r]^{(0)} & j \ar@{->}[r]^{(c)} & k} in $\overline{Q}$, with $c\neq m$, add the new arrow \xymatrix{ i \ar@{->}[r]^{(c)} & k} in $\overline{Q}$ and the new cubic term $\varphi_{ik}^{(c)}\varphi_{jk}^{(c+1)}\varphi_{ij}^{(m)} =
\varphi_{ik}^{(c)}\varphi_{kj}^{(m-c-1)}\varphi_{ji}^{(0)}$ to $W$. \fref{mutation_cubic_couplings} shows the general form of these cycles.

\begin{figure}[H]
	\centering
	\includegraphics[width=11cm]{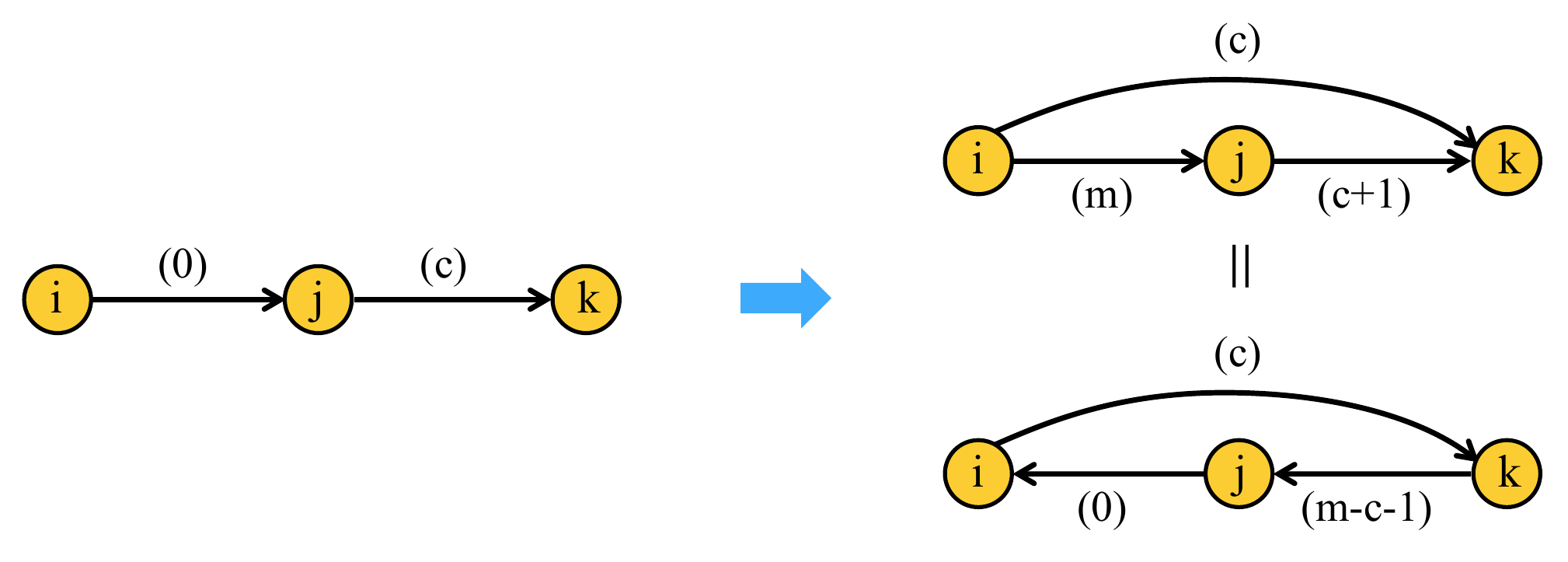}
\caption{New cubic terms coupling mesons to dual flavors.}
	\label{mutation_cubic_couplings}
\end{figure}

\end{itemize}

Besides adding these new terms to the potential, the original terms can also be altered.  Terms in the original potential that do not go through the mutated node remain unchanged.  However, let us consider what happens to terms in the potential that contain the mutated node.  There are two possibilities, depending on the degrees of the arrows that are connected to the mutated node in the corresponding cycle.  These are handled by Rule (2.b) in the first case, and Rules (2.c) and (2.d) in the second.

\begin{itemize}

\item[{\bf 2.b)}] Replace instances of $\varphi_{ij}^{(0)}\varphi_{jk}^{(c)}$ in $W$ with the meson $\varphi_{ik}^{(c)}$ obtained by composing the two arrows. Given the observation in \sref{section_potentials}, we know that $c \neq m.$\footnote{The case $c=m$ is ruled out since this arrow is part of a potential term which has degree $(m-1)$.}

\begin{figure}[H]
	\centering
	\includegraphics[height=3.8cm]{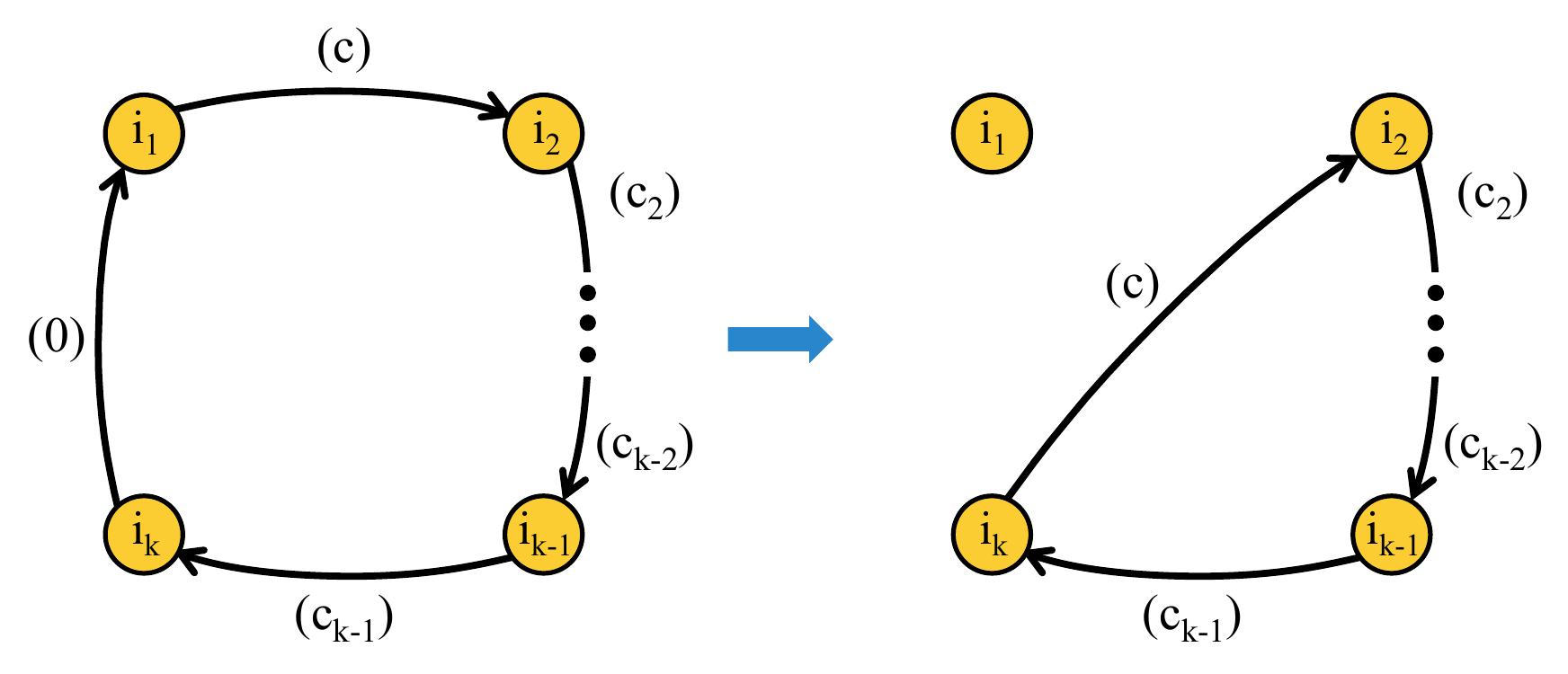}
\caption{Mutation of a potential term with a 2-path giving rise to a meson.}
	\label{W_1-mesons}
\end{figure}

\item[{\bf 2.c)}] Replace instances of $\varphi_{ij}^{(c)}\varphi_{jk}^{(d)}$ in $W$, where $c\neq 0$ and $d$ is arbitrary (again, the case $d=m$ is already ruled out) with the product of dual flavors $\varphi_{ij}^{(c-1)}\varphi_{jk}^{(d+1)}$.

\begin{figure}[H]
	\centering
	\includegraphics[height=3.8cm]{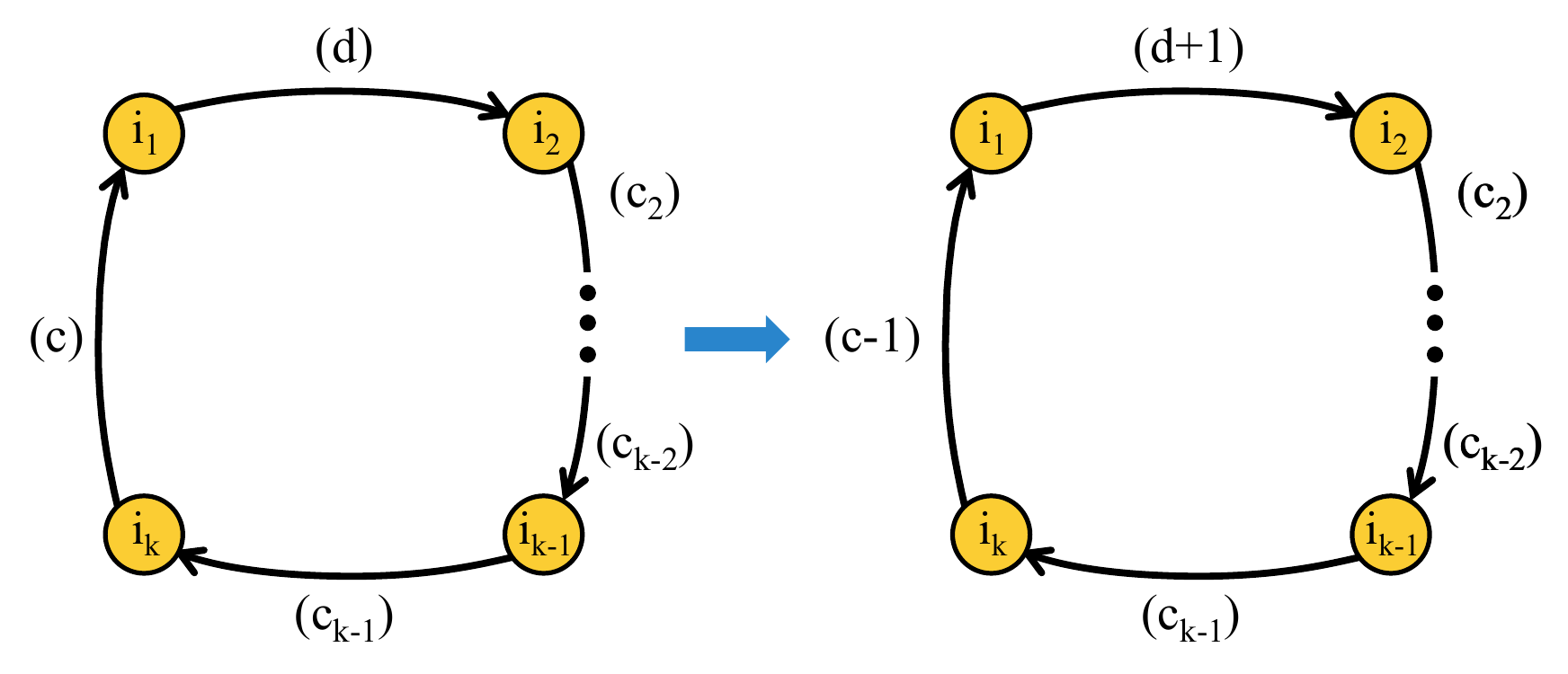}
\caption{Mutation of a potential term with a 2-path that goes through the mutated node but does not generate a meson.}
	\label{W_0-mesons}
\end{figure}

\item[{\bf 2.d)}] Additionally, if there is an incoming chiral arrow $\varphi_{i_0 j}^{(0)}$ at the mutated node an additional term in $W$ is generated by duplicating this cycle but replacing instances of $\varphi_{ij}^{(c)}\varphi_{jk}^{(d)}$ with the product of mesons $\varphi_{i i_0}^{(c)}\varphi_{i_0 k}^{(d)}$, which result from (anti)composing the original flavors  $\varphi_{ij}^{(c)}$ and $\varphi_{jk}^{(d)}$,with $\varphi_{i_0 j}^{(0)}$.\footnote{Technically, to preserve $\{W',W'\} = 0$ once the potential $W$ is mutated to $W'$, we in fact negate the coefficient in front of this additional term.  However, we will not worry about the signs of potential terms in this exposition.} It is clear that whenever we apply $(2.d)$, we also apply $(2.c)$. 

\begin{figure}[H]
	\centering
	\includegraphics[height=3.9cm]{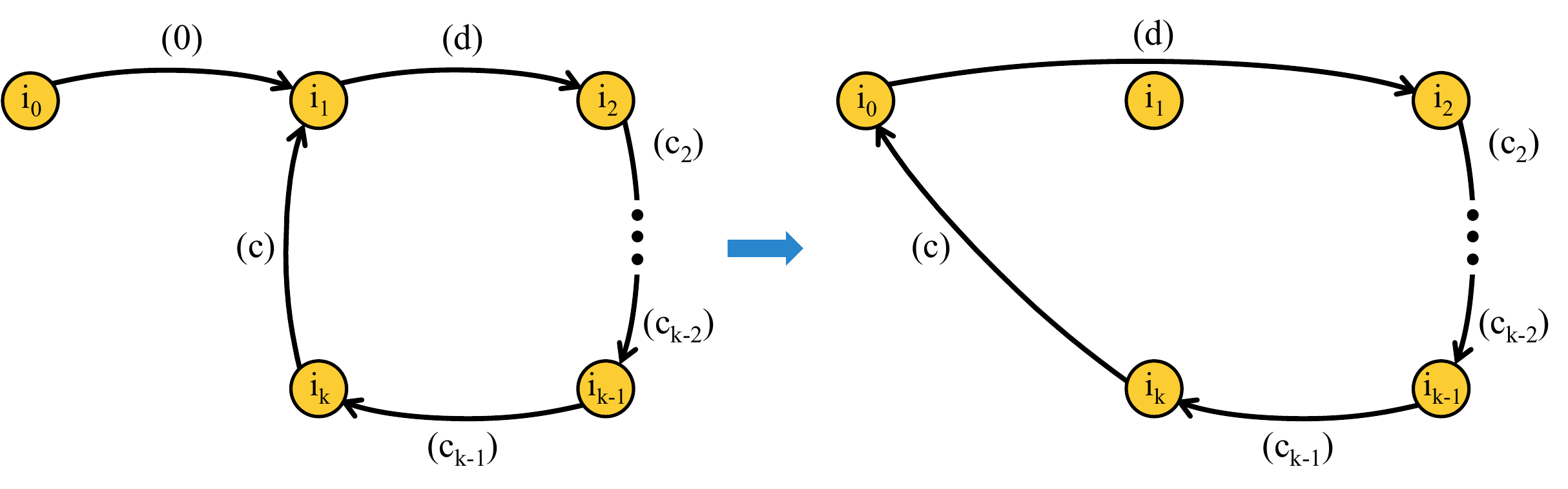}
\caption{Mutation of a potential term in the presence of an additional chiral field incoming into the mutated node.}
	\label{W_2-mesons}
\end{figure}

Rules $(2.c)$ and $(2.d)$ are new features of graded quivers and are only relevant for $m\geq 2$. Interestingly, it is possible to distinguish the previous three rules by the number of mesons in the new potential terms. $(2.b)$, $(2.c)$ and $(2.d)$ correspond to 1, 0 and $\geq 2$ mesons, respectively.

\item[{\bf 3)}] Finally, we can apply reductions of mass terms, see \sref{sec:mass-term}, to get an equivalent graded quiver with potential. Massive fields are eliminated using the relations coming from the corresponding cyclic derivatives of the potential.

\end{itemize}

\subsubsection*{Cycles that pass multiple times through the mutated node} 

Additionally, in the preceding discussion, we have deliberately kept the nodes in potential cycles arbitrary. In particular, our rules also apply when a cycle passes through the mutated node multiple times. When this occurs, we simply apply the appropriate rules to all appearances of the mutated node in the cycle.

\begin{figure}[ht]
	\centering
	\includegraphics[height=4cm]{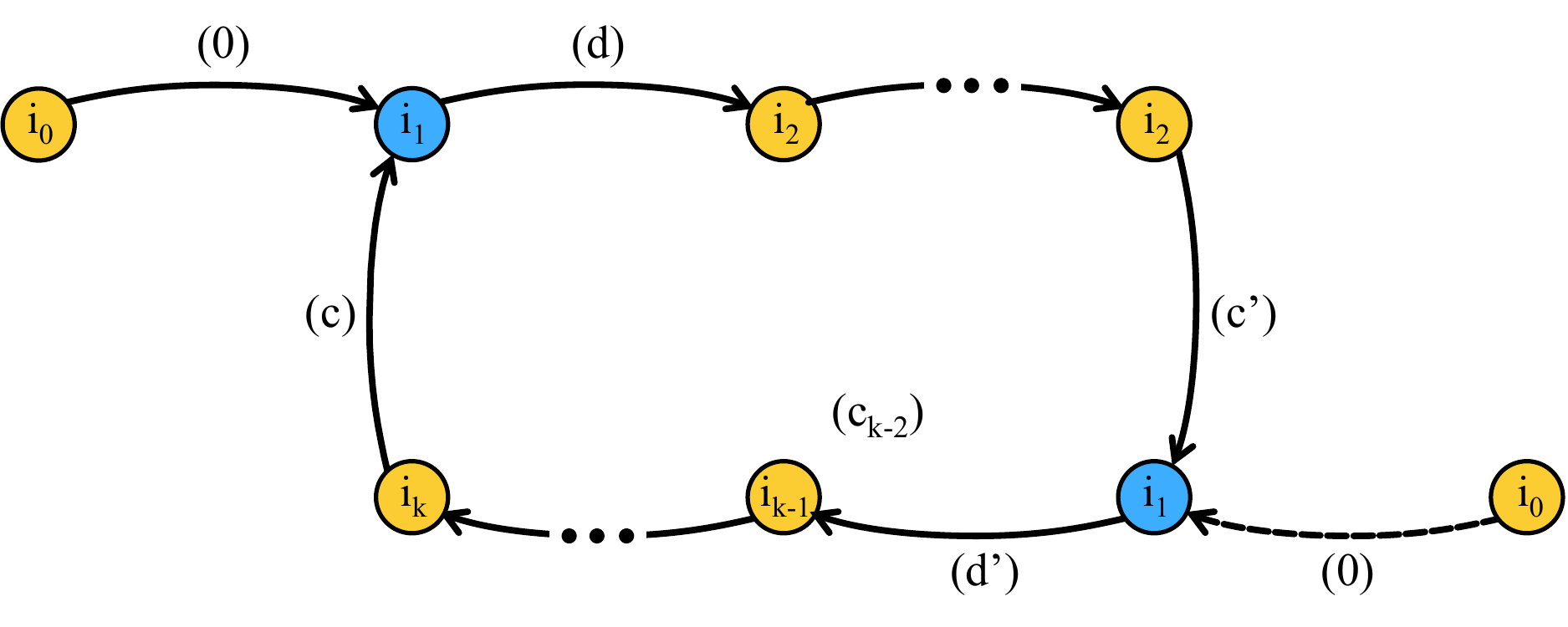}
\caption{A cycle going through the mutated node twice, such that the conditions for $(2.d)$ (and hence for $(2.c)$) hold for both passings. The mutated node is shown in blue.}
	\label{configuration_two_2d}
\end{figure}

\begin{figure}[ht]
	\centering
	\includegraphics[width=15cm]{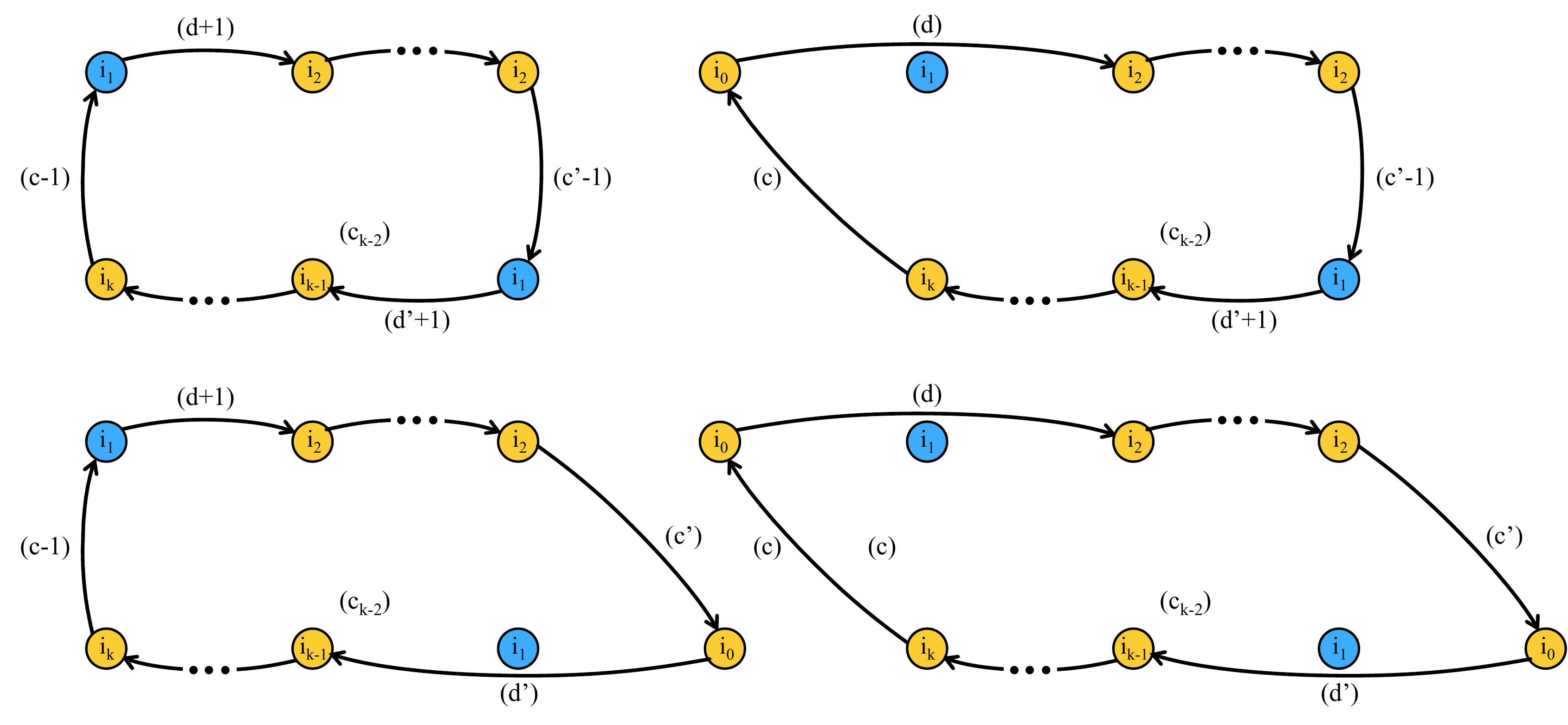}
\caption{The four terms generated by the cycle in \fref{configuration_two_2d} upon mutating on node $i_0$.}
	\label{Wterms_two_2d}
\end{figure}

Rather than going into a lengthy analysis, it is probably better to illustrate this discussion with an explicit example. \fref{configuration_two_2d} shows a potential cycle that passes twice through the mutated node $j$, which for clarity is shown in blue. Furthermore, $c,c'\neq 0$ and there is an incoming chiral $\varphi_{i_0 j}^{(0)}$. Both passings through node $j$ hence satisfy the conditions for rules $(2.c)$ and $(2.d)$. Applying $(2.c)$ and $(2.d)$ in all possible ways, we obtain the four terms shown in \fref{Wterms_two_2d}.\footnote{Technically, the terms corresponding to the second and third cycles illustrated in this figure would have negative signs in front of them.}

\subsubsection*{Allowable potential terms and mutations} 

As explained in \sref{section_potentials}, potential terms correspond to degree $(m-1)$ cycles. Having explained how the potential transforms under mutations, it is possible to show that any such cycle can be reached via a sequence of mutations from the basic configuration shown in \fref{basic_W_term}. We prove this claim in Appendix \ref{section_mathematics_potentials}.

\begin{figure}[ht]
	\centering
	\includegraphics[width=4.5cm]{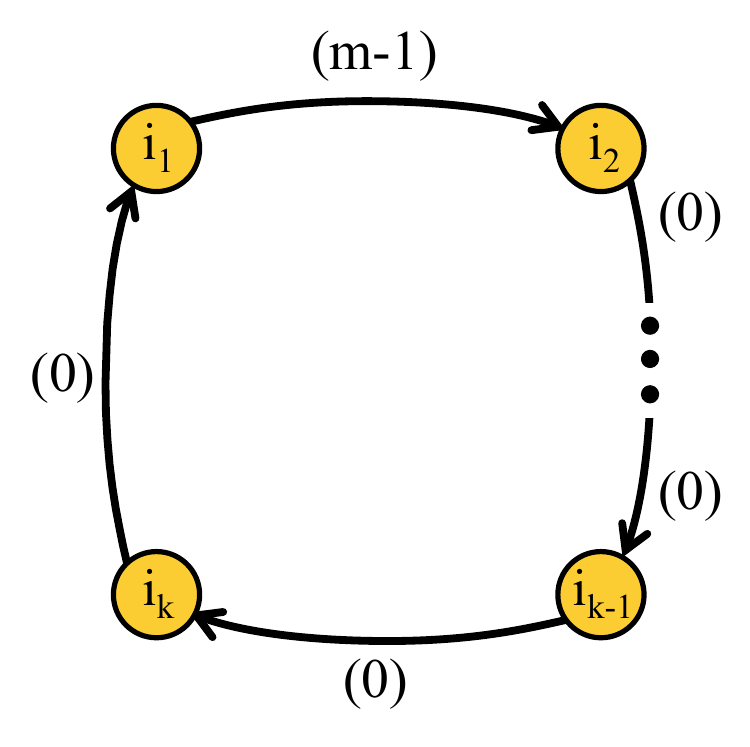}
\caption{Allowable potential terms can be connected by mutations to this basic configuration.}
	\label{basic_W_term}
\end{figure}

\subsubsection*{Kontsevich Bracket} 

The transformation rules for the potential that we introduced imply that the $\{W, W\} = 0$ condition is preserved by mutations. In Appendix \ref{section_mutation_diff} we show that this is the case and present an alternative proof based on Oppermann's ideas \cite{2015arXiv150402617O}. In the process, we will discuss connections between the two approaches.

\subsubsection*{Mutation of the Potential in Combinatorial Models} 

It is worth noting that there exist various combinatorial models that can be interpreted as certain classes of graded quivers with potentials where $m>1$. Going back to \cite{2008arXiv0809.0691B}, the case of type $A_n$ $m$-graded quivers (called colored quivers therein) corresponds to $(m+2)$-angulations of an $(mn+m+2)$-gon. In toric cases when $m=2$, graded quivers and triality corresponds to brane bricks and their transformations as studied in \cite{Franco:2015tna, Franco:2015tya, Franco:2016nwv, Franco:2016qxh}. Both $(m+2)$-angulations and brane bricks can be modeled by potentials as defined in \sref{section_potentials}. The $(m+2)$-gons give rise to ${m+2 \choose 3}$ potential terms, each of which corresponds to a choice of three not necessarily consecutive edges on the boundary of the $(m+2)$-gon. For the brane brick models, potential terms correspond to edges of these $3$-dimensional cell complexes. Using these combinatorial models, one can describe how mutation affects potentials by comparing the potentials associated to the $(m+2)$-angulation (respectively brane brick model) before or after mutation. The mutation of the potentials in these classes of combinatorial models coincide with our general rules, providing further motivation for their study.

\subsection{Mutation of the Ranks: Fractional Brane Charges and $c$-Vectors} 

\label{section_c-vectors_and_ranks}

We postulate that the rank $N_\star$ of a mutated node transforms as follows
\beq
N_\star'=N_0-N_\star ,
\label{mutation_ranks}
\eeq
where $N_0$ indicates the total number of incoming chiral fields. More generally, we will indicate with $N_c$ the total number of incoming arrows of degree $c$. While we opt for keeping our notation as simple as possible, allowing several nodes for each degree and multiple arrows between nodes is straightforward. 

Equation \eref{mutation_ranks} coincides with the transformation of ranks for  $m = 1$, $2$ and $3$, for which the mutations can be interpreted as quantum field theory dualities, as explained in \sref{section_mutations_QFT_dualities}. It is natural to assume, as we will do, that this mutation rule extends to arbitrary $m$. Below we motivate this proposal by discussing fractional brane charges and $c$-vectors, suggesting a connection between these two classes of objects along the way. Further motivation coming from higher cluster categories is provided in Appendix \ref{sec:CC}.

\subsubsection*{Fractional Brane Charges} 

In quivers with a brane realization, every node $i$ is associated to a {\it fractional brane charge} vector $Q_i$, whose dimension is equal to $n$, the total number of gauge groups in the quiver. The number of arrows between a pair of nodes $i$ and $j$ is controlled by the intersection number $\langle Q_i,Q_j \rangle$.\footnote{Whether the intersection pairing is symmetric or antisymmetric and the details regarding the degree and orientation of arrows depend on $m$.}  We assume that objects with these properties exist for quivers without a D-brane realization 

Without loss of generality, we can focus on a local configuration of the quiver as shown in \fref{initial_quiver_periodicity}.\footnote{This does not mean that there are no additional nodes in the quiver, which would determine the dimension $n$ of the $Q_i$ vectors.} In this case the multiplicity of fields is absorbed in the $N_c$'s. In particular, $\langle Q_0,Q_\star \rangle =1$. As it will be discussed in \sref{section_mirror_mutations}, mutation comes from a simple reorganization of the brane system, moving the branes associated to the mutated node over the ones that contribute incoming chirals to it. In this process, brane charges transform as follows
\be
\begin{array}{ccl}
Q_\star' & = & - Q_\star \\[.15cm]
Q_{0}'  & = & Q_0 +  \langle Q_0,Q_\star \rangle Q_\star = Q_{0} + Q_\star \\[.15cm]
Q_{1}' & = & Q_1 \\[.15cm]
 & \vdots &  \\[.15cm]
Q_m' & = & Q_m 
\end{array}
\label{mutation_charges}
\eeq
where we have naturally extended the known rule for $m\leq 3$ to arbitrary $m$. This rule leads to the appropriate transformation of the quiver. We refer the reader to \cite{Cachazo:2001sg,Franco:2016qxh,Franco:2016tcm} for discussions of the $m=1,2,3$ cases.

Focusing on the initial configuration in \fref{initial_quiver_periodicity}, the total initial brane charge is
\beq
Q_T = N_\star Q_\star + \sum_{i=0}^m N_i \, Q_i .
\eeq
After mutation, the total brane charge becomes
\beq
\begin{array}{ccl}
Q_T' & = & N_\star' Q_\star' + \sum_{i=0}^m N_i' \, Q_i' \\[.15cm]
& = & Q_T+\left[(N_0-N_\star)- N_\star' \right] Q_\star
\end{array}
\eeq
where we have used that only $N_\star$, $Q_\star$ and $Q_0$ are modified. Conservation of the total brane charge $Q_T'=Q_T$ requires that the second term vanishes, implying that the rank of the mutated node transforms as in \eref{mutation_ranks}.

\subsubsection*{$c$-Vectors: the $m=1$ Case} 

What are the mathematical counterparts of fractional branes charges? We now argue that they are (some generalization of) $c$-vectors.

The following discussion is restricted to the $m=1$ case. In the original formulation of cluster algebras from Fomin and Zelevinsky \cite{ClustI}, a seed for a cluster algebra is determined not only by an $n$-by-$n$ skew-symmetrizable matrix $B^0$ (equivalently a quiver on $n$ vertices) and by an initial cluster $\{x_1,x_2,\dots, x_n\}$, but also by the data of a coefficient $2n$-tuple $\{p_1^{\pm},p_2^{\pm},\dots,p_n^{\pm}\}$.  The coefficients play a role in the binomial exchange relations 
\beq
x_kx_k' = p_k^+\prod_{i=1}^n x_i^{[b^0_{ki}]_+} + p_k^-\prod_{i=1}^n x_i^{[-b^0_{ki}]_+}
\eeq 
where $[\alpha]_+ = \max(\alpha,0)$.  In their follow-up work \cite{ClustIV}, Fomin and Zelevinsky re-express such seeds\footnote{Here we focus on the case of cluster algebras of geometric type with principal coefficients.} using coefficient $n$-tuples $\{y_1,y_2,\dots, y_n\}$, extend the matrix $B^0$ to a $2n$-by-$n$ matrix starting with appending the $n$-by-$n$ identity matrix underneath, and use the binomial exchange relations
\beq
x_kx_k' = \prod_{i=1}^{n} x_i^{[b^0_{ki}]_+}\prod_{j=n+1}^{2n} y_{j-n}^{[b^0_{ki}]_+} + \prod_{i=1}^n x_i^{[-b^0_{ki}]_+}\prod_{j=n+1}^{2n} y_{j-n}^{[-b^0_{ki}]_+}
\eeq
instead.

As we mutate the seed of a cluster algebra the extended skew-symmetrizable matrix $\left[\begin{matrix} B^0 \\ I \end{matrix}\right]$ mutates according to the same rules as quiver mutation, and after a generic sequence of mutations, $\left[\begin{matrix} B^0 \\ I \end{matrix}\right]$ becomes  $\left[\begin{matrix} B \\ C \end{matrix}\right]$ where $C=[c_{ij}]_{i=1,j=1}^{n,n}$ is an invertible $n$-by-$n$ integer matrix.  We refer to the columns of this $C$-matrix as \emph{c-vectors}, and denote them as $\mathbf{c}_j$ as $j=1,\dots, n$.

Based on the quiver mutation rules, it follows that the c-vectors satisfy the following recurrence, e.g. see (2.9) of \cite{NakSt}:
\begin{equation} 
\label{eq-c-recurrence} c_{ij}' = \begin{cases} -c_{ij} &\mathrm{~if~}j = k \\ c_{ij} + c_{ik} [b_{kj}]_+ + [-c_{ik}]_+ b_{kj} &\mathrm{~if~} j \neq k \end{cases}
\end{equation}
This recurrence is also a tropicalization of the recurrence for coefficient tuples made up of $\mathbf{y}_j$'s, letting the c-vector $\mathbf{c}_j = \left[ \begin{matrix} c_{1j} & c_{2j} & \cdots & c_{nj}\end{matrix}\right]^{T}$ denote the exponent vector of $\mathbf{y}_j$ in terms of the initial coefficients $\{u_1,u_2,\dots, u_n\}$ (i.e. $\mathbf{y}_j = \prod_{i=1}^n u_i^{c_{ij}}$).
The coefficients, i.e. de-tropicalized c-vectors, also correspond to $\mathcal{X}$-coordinates of Fock and Goncharov \cite{FG1,FG2}.

We are now ready to investigate whether $c$-vectors are related to fractional brane charges. A quick comparison of \eref{eq-c-recurrence} and \eref{mutation_charges} reveals various similarities. In order to facilitate the comparison, it is convenient to translate \eref{eq-c-recurrence} to the language we used to discuss fractional branes. Given $c_{ij}$, the $j$ index indicates a node and $i$ runs over the components of the $c$-vector. In other words, we can identify $c_{ij}=Q_{j,i}$. Recall that we are working with $m=1$, so we only have $Q_\star$ for the mutated node, and $Q_0$ and $Q_1$ for the incoming and outgoing chirals, respectively. If we work in the convention in which $b_{\star 0} \geq 0$, it implies that $b_{\star 1} \leq 0$ and hence $(b_{\star 1})_+=0$. Furthermore, in supersymmetric configurations, $(-Q_{\star,i})_+=0$.\footnote{It is natural to conjecture that having $(-Q_{\star,i})_+\neq 0$ corresponds to the inclusion of anti-branes. It would be interesting to explore this idea in further detail.}  With all this, \eref{eq-c-recurrence} becomes
\be
\begin{array}{ccl}
Q_\star' & = & - Q_\star \\[.15cm]
Q_{0}'  & = & Q_0 +  b_{\star 0} \, Q_\star  \\[.15cm]
Q_{1}' & = & Q_1 
\end{array}
\eeq
in agreement with \eref{mutation_charges}.

We conclude that for $m=1$, $c$-vectors can be identified with fractional brane charges. Our discussion of fractional brane charges suggest that that the formulas for the ordinary quiver case, i.e. $m=1$, naturally lift to the case of $m\geq 1$. Thus, in future work, we wish to develop a theory of $c$-vectors for graded quivers with arbitrary $m$.

\section{Periodicity of the Mutations}

\label{section_order_mutation}

We now show that the mutation introduced in previous sections is indeed an order $(m+1)$ transformation, namely that after $(m+1)$ consecutive mutations acting on the same node of a quiver we obtain the original theory.

\begin{figure}[ht]
	\centering
	\includegraphics[width=6cm]{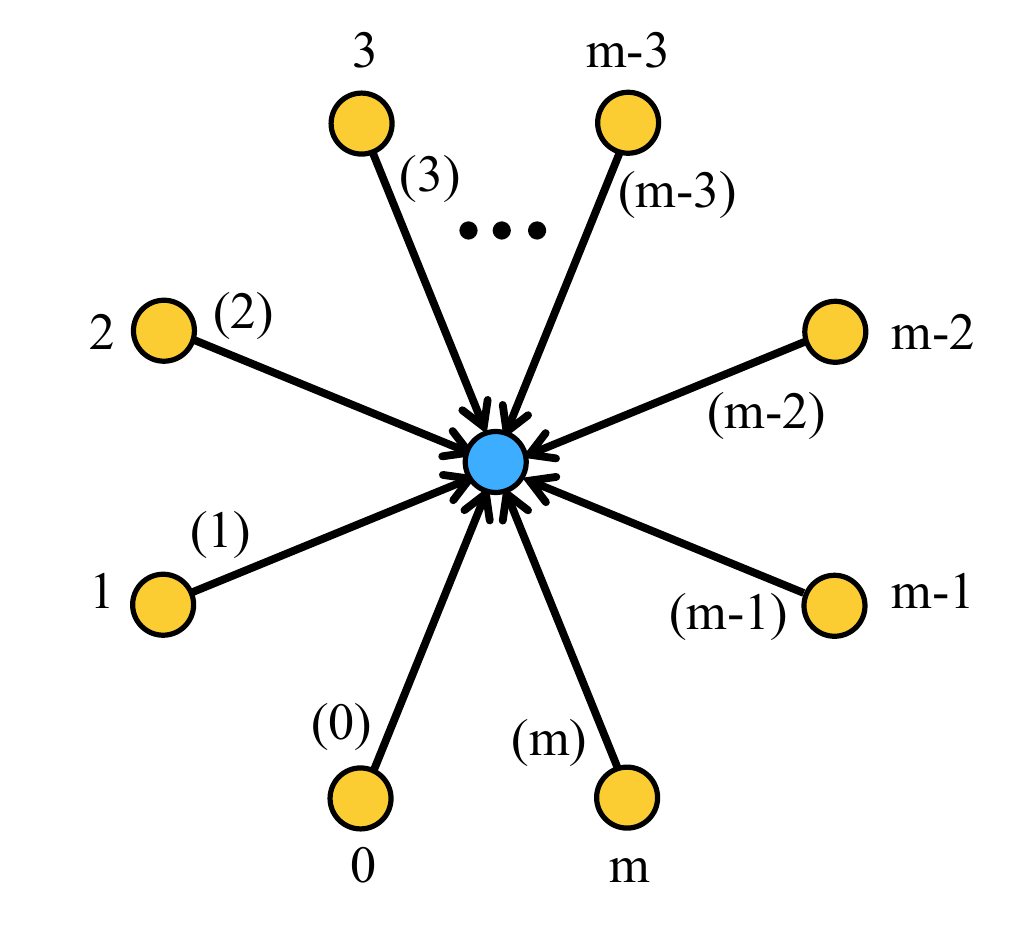}
\caption{Basic initial quiver. We consider consecutive mutations of the blue node.}
	\label{initial_quiver_periodicity}
\end{figure}

It is sufficient to focus on the basic configuration shown in \fref{initial_quiver_periodicity} as a starting point. In this figure, some of the ranks of the flavor nodes might be zero, i.e. flavors of the corresponding degrees might be absent. Our discussion extends to general initial configurations. We consider consecutive mutations of the blue node.

The transformation of flavors is simply given by a rotation, as shown in \fref{mutation_flavors}. It is thus clear that flavors return to the original configuration after $(m+1)$ mutations. 

Initially, there are no mesons stretching between external nodes in \fref{initial_quiver_periodicity}. Since mesons are created at every mutation, it is important to verify that all of them disappear by the end of the mutation sequence. Let us first focus on the pair of nodes $0$ and $c$, initially corresponding to flavors of degrees $(0)$ and $(c)$, both with non-vanishing ranks. If any of the ranks are zero, the corresponding flavors will be absent and no mesons between these two nodes will ever be generated. 

\begin{figure}[ht]
	\centering
	\includegraphics[width=15cm]{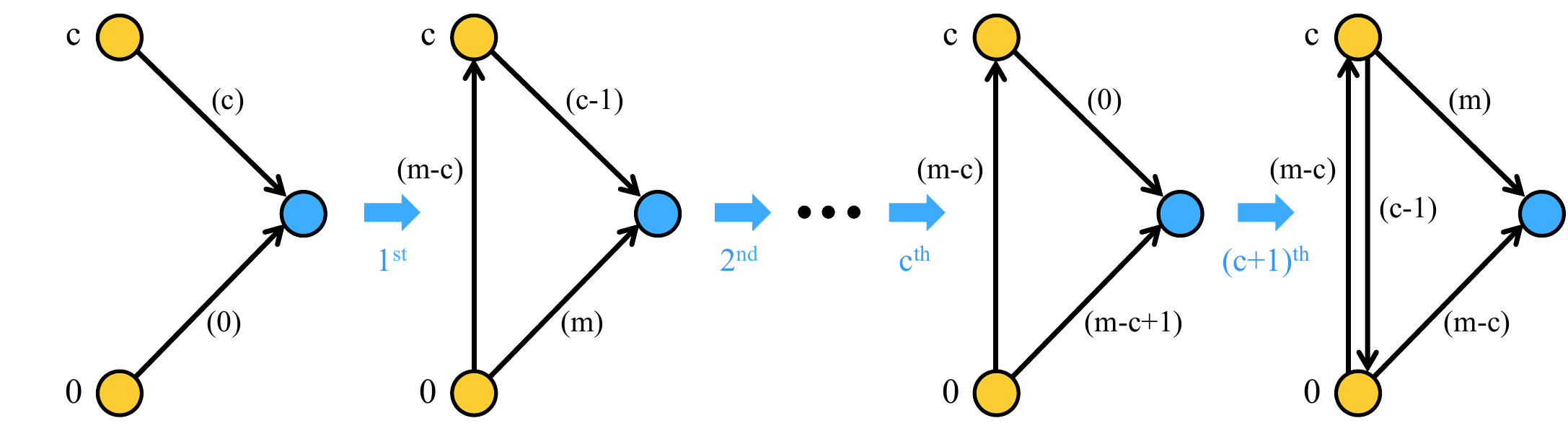}
\caption{Evolution of mesons connecting nodes $0$ and $c$ under $c+1$ consecutive mutations of the blue node.}
	\label{mutation_periodicity}
\end{figure}

\fref{mutation_periodicity} shows a sequence of $c+1$ mutations acting on the blue node. After the first mutation, a meson of degree $(m-c)$ is created between the two nodes. This remains the only meson between nodes $0$ and $c$ until the $(c+1)^{th}$ mutation, after which a meson of degree $(c-1)$ in the opposite direction is generated. The sum of the degrees of both mesons is $(m-1)$. In fact, they form a mass term in the potential and can be integrated out. We conclude that after $(c+1)$ mutations, the mesons connecting nodes $0$ and $c$ disappear. Given the transformation of flavors, only after $(m+2)$ mutations will we again generate a meson between nodes $0$ and $c$. 

\begin{figure}[ht]
	\centering
	\includegraphics[width=7cm]{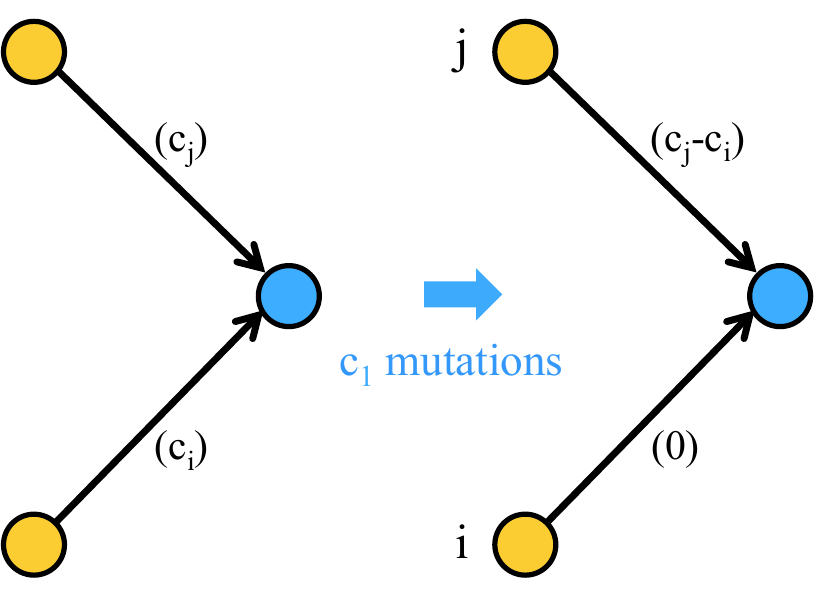}
\caption{Starting from a general pair of nodes, we reach the initial configuration in \fref{mutation_periodicity} after $c_i$ mutations.}
	\label{mutation_periodicity_general_initial}
\end{figure}

Let us now consider a pair of arbitrary nodes $i$ and $j$, initially connected to flavors of degrees $c_i<c_j$. As shown in \fref{mutation_periodicity_general_initial}, after $c_i$ mutations, we reach the starting configuration of \fref{mutation_periodicity}, with $c=c_j-c_i$. The analysis in the previous paragraph applies after this point. The meson generated at the $c_i^{th}$ mutation, is removed after the $(c_i+1)^{th}$ mutation by forming a massive pair with a meson going in the opposite direction. Since for any pair of nodes $c_j-c_i\leq m$, we conclude that after $(m+1)$ mutations we return to a configuration without mesons between external nodes. 

The mutation rules in \sref{section_mutations_quivers} preserve the global symmetries of the theory. This in particular requires/implies that if the final quiver is identical to the initial one, as we have just shown, then the final potential also coincides with the original one.

We complete the proof of periodicity in the next section, where we explain how, under certain conditions that generalize the cancellation of anomalies, the rank of the mutated node also returns to its original value after $(m+1)$ mutations.

\section{Generalized Anomaly Cancellation}

\label{section_gen_anom}

At present, physical interpretations of graded quivers as quantum field theories are only known for $m=1$, 2 and 3. This correspondence will be explained in \sref{section_gauge_theories}. Since anomalies play a central role in quantum field theories, it is reasonable to expect that they can be generalized to arbitrary $m$ and that they will remain important in this broader context.  For example, in the case of $m=1$, \emph{cancellation of gauge anomalies} requires at every node that the weighted number of incoming arrows equals the weighted number of outgoing arrows, with the weighting given by the ranks of the gauge groups. 

From the discussion in \sref{section_c-vectors_and_ranks}, in order for the rank of a node to return to the original value after $(m+1)$ consecutive mutations on it, we must have
\beq
N_\star = \sum_{c=0}^{m} (-1)^{m-c} N_c + (-1)^{m+1} N_\star .
\label{generalized_anomaly_cancellation}
\eeq

Here, $N_\star$ is the rank of the node $\star$, the node to be mutated at, and $N_c$ equals the number of arrows of degree $c$ incoming into that node as part of a double-arrow.  For $m=1$, 2 and 3, this agrees with the cancellation of the gauge anomaly, as will be discussed in \sref{section_gauge_theories}.\footnote{Here we focus on the anomalies associated to the $SU(N_i)$ factors in the $U(N_i)=SU(N_i)\times U(1)$ gauge groups. We will not consider anomalies involving the $U(1)$ factors, which in D-brane constructions can be cancelled by a stringy mechanism.} We will thus promote \eref{generalized_anomaly_cancellation} to a {\it generalized anomaly cancellation} condition for arbitrary $m$. It is in fact quite remarkable that anomaly cancellation emerges from the periodicity condition of theories under mutations!

Equation \eref{generalized_anomaly_cancellation} becomes (locally at every node)
\be
\begin{array}{rrl}
\mbox{{\bf odd m:  }} & 0 = N_m - N_{m-1}+ \ldots - N_1 + N_0 \\[.15 cm]
\mbox{{\bf even m:  }} & 2 N_* = N_m - N_{m-1}+ \ldots + N_1 - N_0
\end{array}
\label{anomalies_even_odd_m}
\eeq
The rank of the gauge group $N_*$ only enters anomaly cancellation for even $m$.

The preceding discussion of generalized anomalies is based on periodicity under $(m+1)$ consecutive mutations of the same node. Since we have not defined mutation in the presence of adjoint fields, our argument does not apply to nodes containing such fields.  However, for $m\leq 3$, anomalies can instead be computed by calculating the appropriate loop diagrams. It is then possible to include the contribution of matter in arbitrary representations of the gauge group, which are controlled by certain group theoretic factors. It is reasonable to expect that by directly generalizing such expressions we can incorporate the contributions of other representations, including adjoints, to generalized anomalies.

The convention of \eref{orientation_fields} ensures that the signs of the terms in \eref{anomalies_even_odd_m} coincide with the ones for the corresponding physical fields.

To conclude this section, let us mention that frozen nodes can be anomalous. In physics, they correspond to global symmetry groups. The invariance of their anomalies under mutations of other nodes is called {\it 't Hooft anomaly matching} and constitutes a powerful constraint on dualities. More generally, mathematically it is still interesting to consider theories with anomalous unfrozen nodes. 

\section{Map to Physics}

\label{section_map_to_physics}

We are now ready to explain the connection between graded quivers with potentials and physics. Quivers with a maximum degree $m$ correspond to $(6-2m)$-dimensional gauge theories with $2^{3-m}$ supercharges. More precisely, $m=1$, $2$ and $3$ correspond to $4d$ $\mathcal{N}=1$, $2d$ $\mathcal{N}=(0,2)$ and $0d$ $\mathcal{N}=1$ gauge theories, respectively.\footnote{In the coming section, we explain how many of the ideas presented here extend to the case of $m=0$.} Such theories are called minimally supersymmetric. This correspondence will be explained in detail in \sref{section_gauge_theories}. It is clear that, at least at present, only theories with $m\leq 3$ have a known physical interpretation, since $m>3$ would naively correspond to theories in a negative number of dimensions. It would be extremely interesting to determine whether there are physical systems described by graded quivers with $m>3$ or to detect a mathematical obstruction or qualitatively distinctive feature that first appears at $m=4$.

In physics, a quiver diagram summarizes the gauge symmetry and matter content of a quantum field theory. Nodes correspond to gauge groups, i.e. to vector superfields, and arrows indicate matter fields. As the dimension in which the field theory lives decreases, there are more types of matter superfields. This fact is nicely captured by the increasing number of possible degrees in the quiver as $m$ grows. The quiver diagram does not fully specify a minimally supersymmetric theory. In order to do so, additional information regarding interactions between matter fields needs to be provided. Such interactions are encoded in the potential. 

A large class of these $(6-2m)$-dimensional theories can be realized in Type IIB string theory on the worldvolume of  D$(5-2m)$-branes probing CY $(m+2)$-folds (see \cite{Morrison:1998cs,Beasley:1999uz,Feng:2000mi,Feng:2001xr,Franco:2015tna,Franco:2016tcm} and references therein). The probed CY manifolds emerge from the gauge theories as their classical moduli spaces. In this way, string theory provides a direct connection between these quivers and CY geometries, in nice parallel with the relationship based on (higher) Ginzburg algebras. Table \ref{sequence_theories} summarizes these setups and their mirror configurations. The use of mirror symmetry for understanding these theories is discussed in \sref{section_mirror_symmetry}.

\begin{table}[h]
\begin{center}
\begin{tabular}{|c|c|c|c|}
\hline
{\bf m} & {\bf QFT} & {\bf Original geometry} & {\bf Mirror} \\ \hline
1 & $4d$ $\mathcal{N}=1$ & IIB D3 probing CY$_3$ & IIA D6 on 3-cycles \\ \hline
2 & $2d$ $\mathcal{N}=(0,2)$ & IIB D1 probing CY$_4$ & IIB D5 on 4-cycles \\ \hline
3 & $0d$ $\mathcal{N}=1$ & IIB D(-1) probing CY$_5$ & IIA ED4 on 5-cycles \\ \hline
\end{tabular}
\end{center}
\vspace{-.3cm}\caption{D-brane configurations engineering quantum field theories in various dimensions.}
\label{sequence_theories}
\end{table}

When the CY $(m+2)$-folds are toric, a beautiful description of these theories in terms of objects that generalize dimer models exists. In this case, T-duality connects the D$(5-2m)$-branes probing CY $(m+2)$-folds to new configurations of branes living on tori $\mathbb{T}^{m+1}$. For $m=1$, 2 and 3, these configurations are brane tilings \cite{Franco:2005rj,Franco:2005sm}, brane brick models \cite{Franco:2015tya,Franco:2016nwv,Franco:2016qxh} and brane hyperbrick models \cite{Franco:2016tcm}. These constructions significantly streamline the connection between CY geometry and graded quivers in both directions. We envision profound connections between these combinatorial objects and the ideas presented in this paper. We postpone the exploration of this special toric case to future work.

\section{Gauge Theories for $m=0,1,2,3$}

\label{section_gauge_theories}

In this section we discuss how the general framework of graded quivers with potentials with $m=0,1,2,3$ captures and unifies the physics of supersymmetric gauge theories in $d=6,4,2,0$, respectively. We include references with in-depth presentations of such quantum field theories. We start from $m=1$ and comment on $m=0$ towards the end.

\subsection{$m=1$: $4d$ $\mathcal{N}=1$}

Here we explain how $m=1$ quivers correspond to $4d$ $\mathcal{N}=1$ gauge theories. There is a vast literature on these theories, see e.g. \cite{Intriligator:1995au,Signer:2009dx}.

\paragraph{Superfields.} 

\begin{figure}[ht]
	\centering
	\includegraphics[width=11cm]{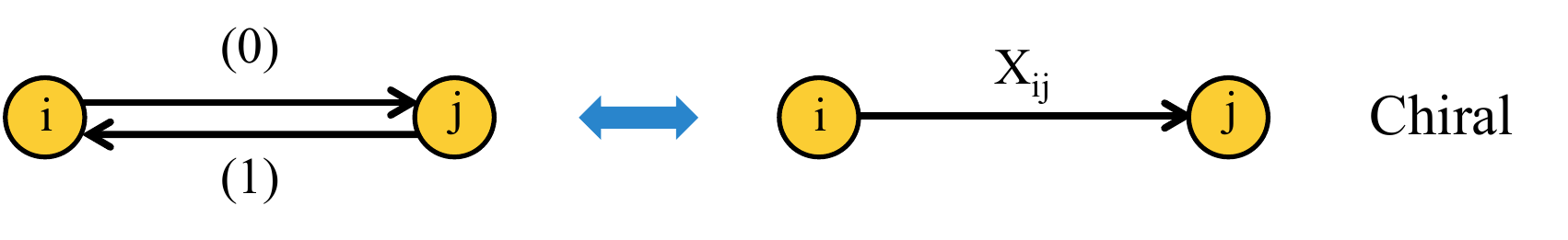}
\caption{(0,1) arrows corresponds to $4d$ $\mathcal{N}=1$ chiral fields.}
	\label{fields_m=1}
\end{figure}

Let us first discuss how the different elements in the quiver map to superfields. Nodes correspond to {\it vector multiplets}. As explained earlier, it is sufficient to focus on arrows with degrees $0\leq c \leq m/2$, which can also be completed into $(c,m-c)$ double arrows. This means that in this case there is a single type of arrow, i.e. of matter superfield, which corresponds to $c=0$, i.e. to a $(0,1)$ double arrow. We identify such arrows with {\it chiral superfields}, as shown in \fref{fields_m=1}. Here and in what follows, we determine the orientation of physical fields using the convention in \eref{orientation_fields}. In order to follow standard $4d$ $\mathcal{N}=1$ notation we call $X_{ij}\equiv \varphi_{ij}^{(0)}$. We conclude that $m=1$ quivers precisely match the most general field content of $4d$ $\mathcal{N}=1$ gauge theories.

\paragraph{Anomalies.} 
For every node, the generalized anomaly cancellation \eref{anomalies_even_odd_m} takes the form
\beq
0=N_{\chi_{in}}-N_{\chi_{out}} \, ,
\eeq
where $N_{\chi_{in}}$ and $N_{\chi_{out}}$ are the number of incoming and outgoing chiral fields, respectively. This is precisely the condition for the cancellation of the $SU(N_\star)^3$ gauge anomaly. The relative sign reflects the opposite contribution of fields transforming in the fundamental and antifundamental representations of $SU(N_\star)$. 

\paragraph{Potential.} 

\begin{figure}[H]
	\centering
	\includegraphics[width=11cm]{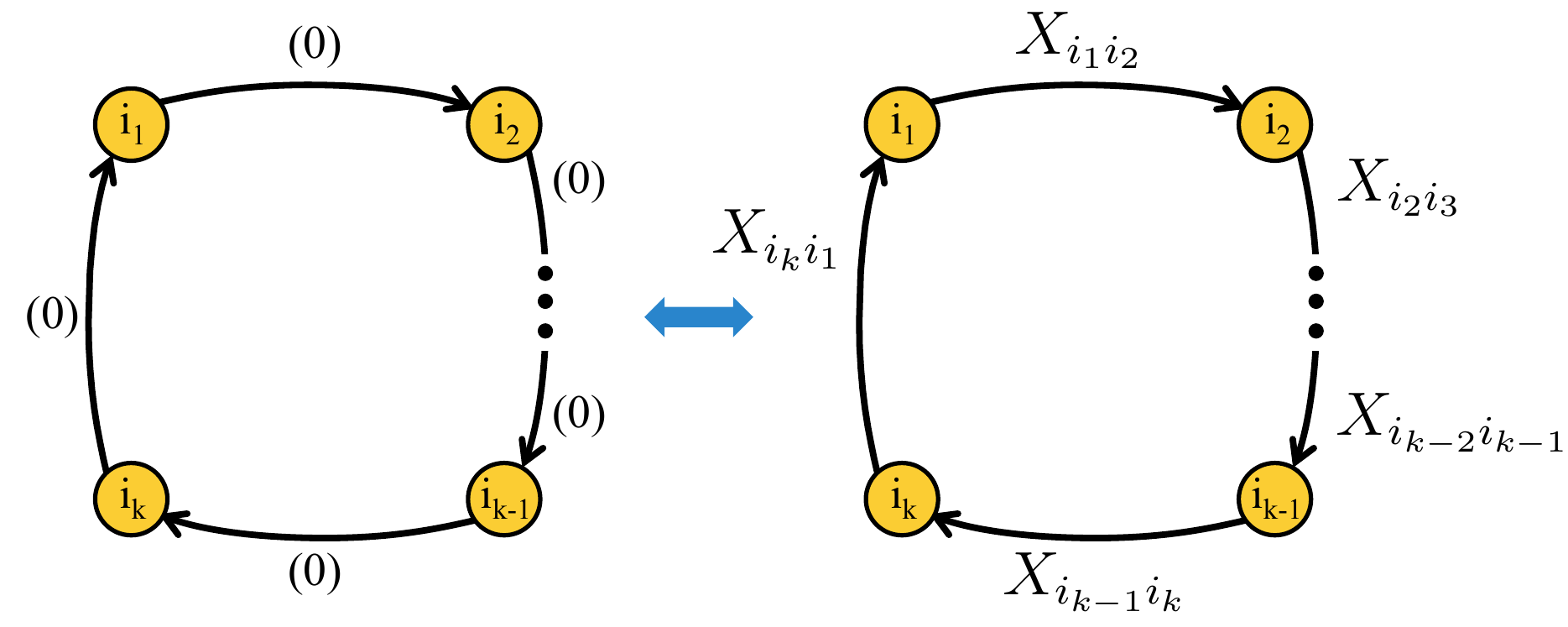}
\caption{The $m=1$ potential correspond to the $4d$ $\mathcal{N}=1$ superpotential.}
	\label{potential_m=1}
\end{figure}

Following the definition in \sref{section_potentials}, the degree of the potential for $m=1$ must be equal to 0. Then, terms in the potential correspond to oriented cycles of chiral fields, as shown in \fref{potential_m=1}. Different terms in the potential might involve different numbers of fields. In physics, we refer to the $m=1$ potential as the {\it superpotential}. An important property of the superpotential is that it is an holomorphic function of the chiral fields, i.e. it does not involve conjugate fields $\overline{X}_{ij}$.

The moduli space of these theories is determined by imposing vanishing $D$- and $F$-terms. The F-terms are the cyclic derivatives of the superpotential with respect to chiral fields. This agrees with \eref{eq:Jacobian_algebra}.

For clarity, the figures in this section show potential terms containing a large number of fields.

\paragraph{Kontsevich bracket.} 

Since the superpotential is holomorphic, the Kontsevich bracket vanishes automatically. This implies that, as it is well known from physics, there is no additional constraint on the superpotential.

\subsection{$m=2$: $2d$ $\mathcal{N}=(0,2)$}

We now consider $m=2$ quivers, which correspond to $2d$ $\mathcal{N}=(0,2)$ gauge theories. Thorough introductions to these theories can be found in \cite{Witten:1993yc,GarciaCompean:1998kh,Gadde:2013lxa,Franco:2015tna}.

\paragraph{Superfields.} 

Once again, every node corresponds to a vector superfield. There are two types of arrows, associated to $c=0,1$. The resulting $(0,2)$ and $(1,1)$ double arrows correspond to {\it chiral} and {\it Fermi} superfields, respectively. Following standard notation, we will refer to chiral fields as $X_{ij}$ and Fermi fields as $\Lambda_{ij}$. Fermi fields are the first examples of $(m/2,m/2)$ unoriented fields that we encounter in quantum field theories. Specifically, since $\varphi^{(1)}_{ij}$ and $\varphi^{(1)}_{op,ji}$ have the same degree, we can identify the pair with either $\Lambda_{ij}$ or $\Lambda_{ji}$. The symmetry under the exchange of $\Lambda_{ij} \leftrightarrow \bar{\Lambda}_{ij}$ for any Fermi field is an important property of $2d$ $\mathcal{N}=(0,2)$ gauge theories. Below we will discuss this symmetry on the potential.

\begin{figure}[ht]
	\centering
	\includegraphics[width=11cm]{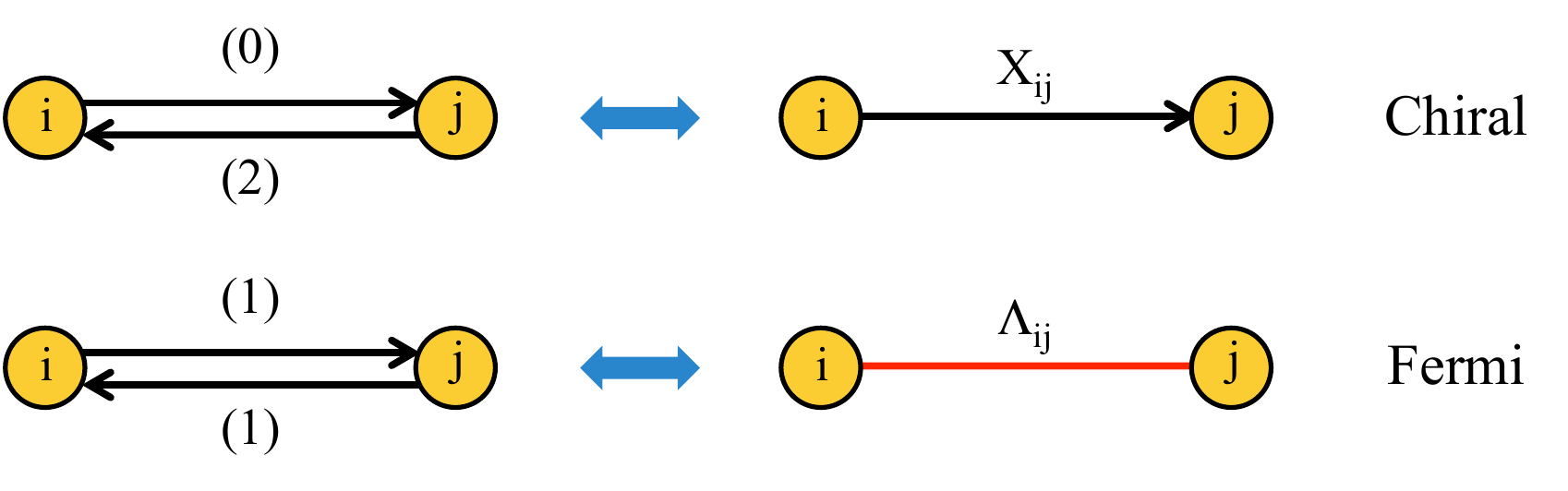}
\caption{(0,2) and (1,1) arrows correspond to $2d$ $\mathcal{N}=(0,2)$ chiral and Fermi fields, respectively.}
	\label{fields_m=2}
\end{figure}

\fref{fields_m=2} shows the map between arrows in an $m=2$ graded quiver and matter fields in a $2d$ $\mathcal{N}=(0,2)$ theory. Given the undirected nature of Fermi fields, it is standard to represent them by undirected lines.

\paragraph{Anomalies.} 

The generalized anomaly cancellation condition \eref{anomalies_even_odd_m} becomes
\beq
0=N_{\chi_{in}}+N_{\chi_{out}}-N_F-2N_\star \, ,
\eeq
with $N_{\chi_{in}}$, $N_{\chi_{out}}$, $N_F$ and $N_\star$ the numbers of incoming chirals, outgoing chirals, Fermis and the rank of the gauge group, respectively. A few words are in order for understanding this expression. First, we notice that the contributions of the incoming and outgoing chirals, i.e. of chirals transforming in the fundamental and antifundamental representations, have the same sign. This is because anomalies in $2d$ are quadratic. Second, unlike in the $4d$ case, there is a non-vanishing term proportional to $N_\star$. This is the contribution to the anomaly of gauginos in the vector multiplet. Finally, the contributions from chiral fields have an opposite sign to the ones of Fermis and vector multiplets. This is due to the opposite chirality of the fermions in these superfields. It is quite remarkable that all these details emerge from the simple requirement of periodicity under mutations.

\paragraph{Potential.} 

\begin{figure}[H]
	\centering
	\includegraphics[width=13cm]{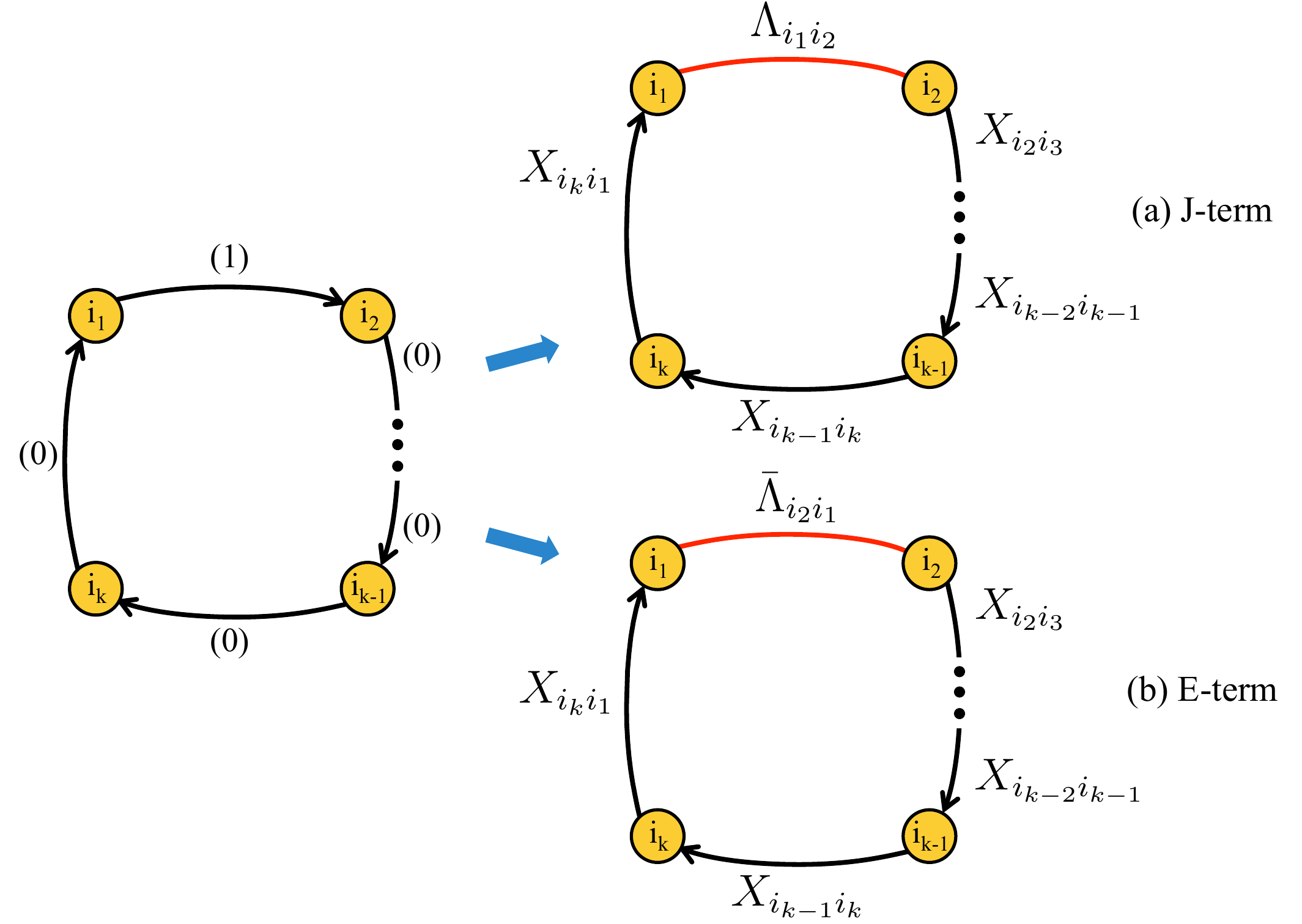}
\caption{There are two types of potential terms for $m=2$. They map to contributions to $J$- or $E$-terms in the corresponding $2d$ $\mathcal{N}=(0,2)$ gauge theories.}
	\label{potential_m=2}
\end{figure}

The degree of the potential for $m=2$ is 1. This means that all terms in the potential are of the general form shown on the left of \fref{potential_m=2}, namely they consist of a single Fermi field and an arbitrary number of chiral fields. The physical interpretation of such a potential term is interesting. In particular, following our previous discussion, a $c=1$ arrow connecting nodes $i_1$ and $i_2$ can be interpreted either as a Fermi field $\Lambda_{i_1 i_2}$ or as a conjugate Fermi field $\bar{\Lambda}_{i_2 i_1}$. The first possibility leads to a contribution to a so-called $J$-term while the second option gives a contribution to an $E$-term:
\beq
\varphi^{(1)}_{i_1 i_2} \varphi^{(0)}_{i_2 i_3} \ldots \varphi^{(0)}_{i_k i_1} \ \ \to \ \  
\left\{\begin{array}{lcc}
\mbox{$J$-term:} & &\Lambda_{i_1 i_2} X_{i_2 i_3} \ldots X_{i_k i_1} \\[.15 cm]
\mbox{$E$-term:} & & \bar{\Lambda}_{i_2 i_1} X_{i_2 i_3} \ldots X_{i_k i_1}
\end{array}\right.
\label{J_and_E_terms_from_colored_quivers}
\eeq

More precisely, we refer to the chiral field parts of the cycles in \eref{J_and_E_terms_from_colored_quivers} as $J$- and $E$-terms.  Every Fermi field in the theory is associated to a $J$- and an $E$-term, which are given by sums over contributions that, generically, can be of different orders. For the terms in \eref{J_and_E_terms_from_colored_quivers}, we have
\beq
\begin{array}{cl}
J_{\Lambda_{i_1 i_2}} & =X_{i_2 i_3} \ldots X_{i_k i_1} + \ldots \\[.15 cm]
E_{\Lambda_{i_2 i_1}} & = X_{i_2 i_3} \ldots X_{i_k i_1} + \ldots
\end{array}
\eeq
where the dots indicate possible additional terms.

The potential takes the form
\beq
W=\sum_{a} \left[ \Lambda_a J_a(X) + \bar{\Lambda}_a E_a(X) \right]\, ,
\label{m=2_potential}
\eeq
where $a$ is an index that runs over all the Fermi fields in the theory. 

The classical moduli space of the gauge theory requires vanishing $J$- and $E$-terms. This is in agreement with the discussion around \eref{eq:Jacobian_algebra}, which states that only the Jacobian algebra with respect to arrows of degree $(m -1)$ is important for the moduli space. In this case, $m-1=1$, implying that we must consider the Jacobian algebra with respect to both Fermis and conjugate Fermis. In addition, as always, we demand vanishing $D$-terms.

\paragraph{$\Lambda \leftrightarrow \bar{\Lambda}$ symmetry.} 

We have already mentioned that the unoriented nature of Fermi fields leads to a symmetry under the exchange of $\Lambda_a \leftrightarrow \bar{\Lambda}_a$ for any Fermi field. This symmetry corresponds to the exchange $J_a \leftrightarrow E_a$.

\paragraph{Kontsevich bracket.} 

The potential \eref{m=2_potential} contains both Fermi fields and their conjugates. This implies that the vanishing of the Kontsevich bracket gives rise to a non-trivial constraint, which takes the form
\beq
\sum_a J_a(X) E_a(X) = 0 . 
\eeq
This is precisely the {\it trace condition} of $2d$ $\mathcal{N}=(0,2)$ theories \cite{Witten:1993yc}.

\subsection{$m=3$: $0d$ $\mathcal{N}=1$}

Let us consider $m=3$ quivers. They correspond to $0d$ $\mathcal{N}=1$ gauge theories. These theories were recently studied in \cite{Franco:2016tcm}.

\paragraph{Superfields.} 

\begin{figure}[H]
	\centering
	\includegraphics[width=11cm]{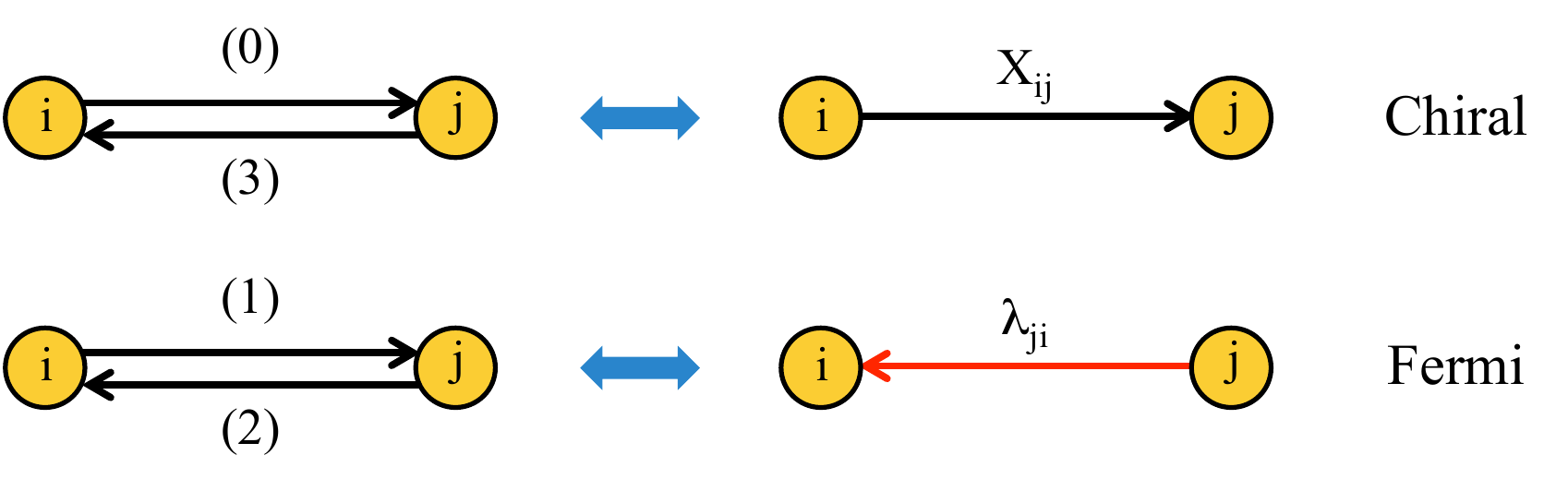}
\caption{(0,3) and (1,2) arrows correspond to $0d$ $\mathcal{N}=1$ chiral and Fermi fields, respectively.}
	\label{fields_m=3}
\end{figure}

Nodes correspond to gaugino superfields. There are two types of arrows, associated to $c=0,1$. The $(0,3)$ and $(1,2)$ double arrows map to {\it chiral} and {\it Fermi} superfields, respectively. We will refer to chiral fields as $X_{ij}$ and Fermi fields as $\lambda_{ij}$. Unlike what happens in the $m=2$ case, Fermi fields are oriented for $m=3$. The correspondence between double arrows and fields is illustrated in \fref{fields_m=3}.

\paragraph{Anomalies.} 

Anomaly cancellation becomes
\beq
0=N_{\chi_{in}}-N_{\chi_{out}}+N_{F_{in}}-N_{F_{out}} .
\eeq
Fields transforming in the fundamental and antifundamental representations of the gauge group contribute with opposite signs. The orientation prescription of \eref{orientation_fields} is crucial for obtaining this correlation.

\paragraph{Potential.} 

\begin{figure}[H]
	\centering
	\includegraphics[width=11cm]{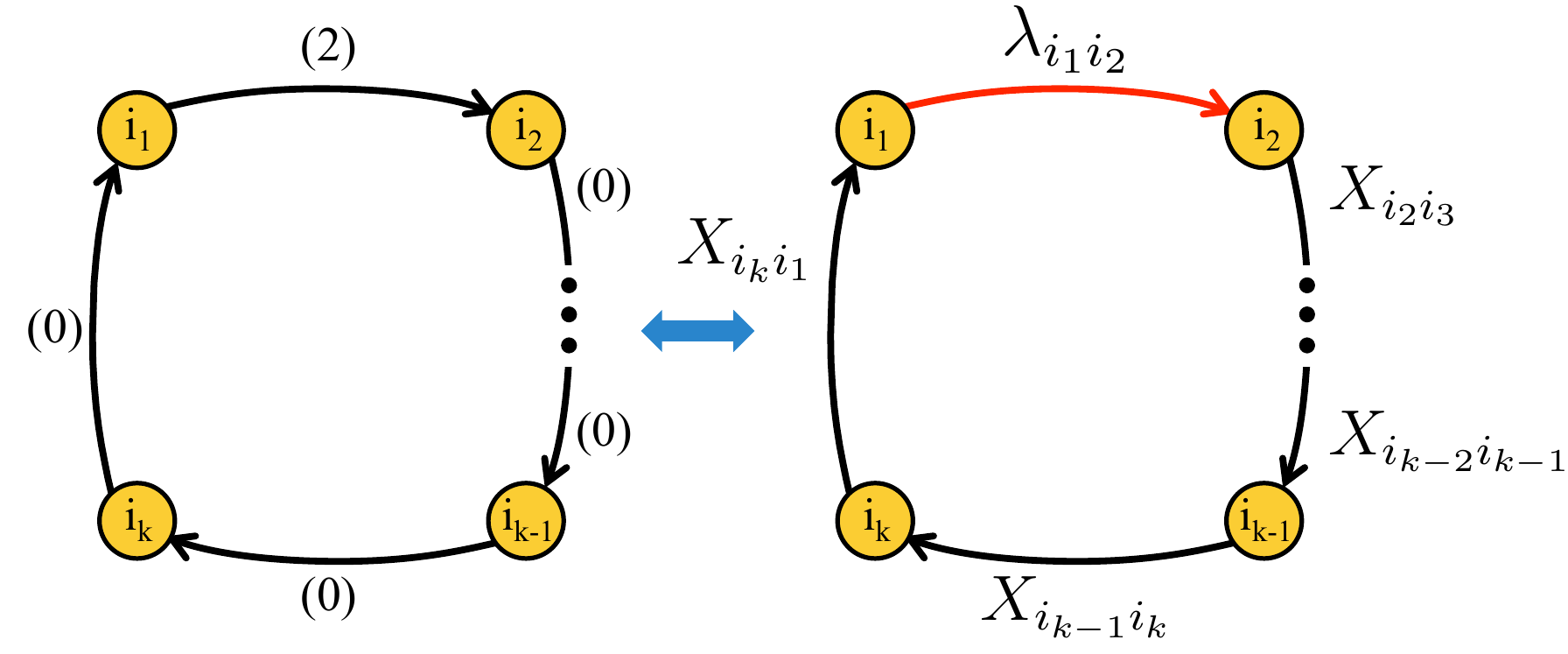}
\caption{Contribution to a $J$-term in a $0d$ $\mathcal{N}=1$ gauge theory.}
	\label{potential_m=3_J-term}
\end{figure}

The potential for $m=3$ has degree 2. This implies that there are two possible types of potential terms, which precisely reproduce the possible interaction terms of $0d$ $\mathcal{N}=1$ gauge theories \cite{Franco:2016tcm}. The first one, shown in \fref{potential_m=3_J-term}, has the form
\beq
\mbox{$J$-term:} \quad \varphi^{(2)}_{i_1 i_2} \varphi^{(0)}_{i_2 i_3} \ldots \varphi^{(0)}_{i_k i_1} \ \to \  \lambda_{i_1 i_2} X_{i_2 i_3} \ldots X_{i_k i_1} 
\label{J-terms_m=3}
\eeq
and corresponds to a contribution to a so-called $J$-term. $J$-terms are defined as the chiral field part of such loops. There is one $J$-term for every Fermi field, which becomes
\beq
J_{\lambda_{i_1 i_2}} = X_{i_2 i_3} \ldots X_{i_k i_1} + \ldots \, ,
\eeq
where the dots indicate the possibility of multiple contributions. 

\begin{figure}[H]
	\centering
	\includegraphics[width=15cm]{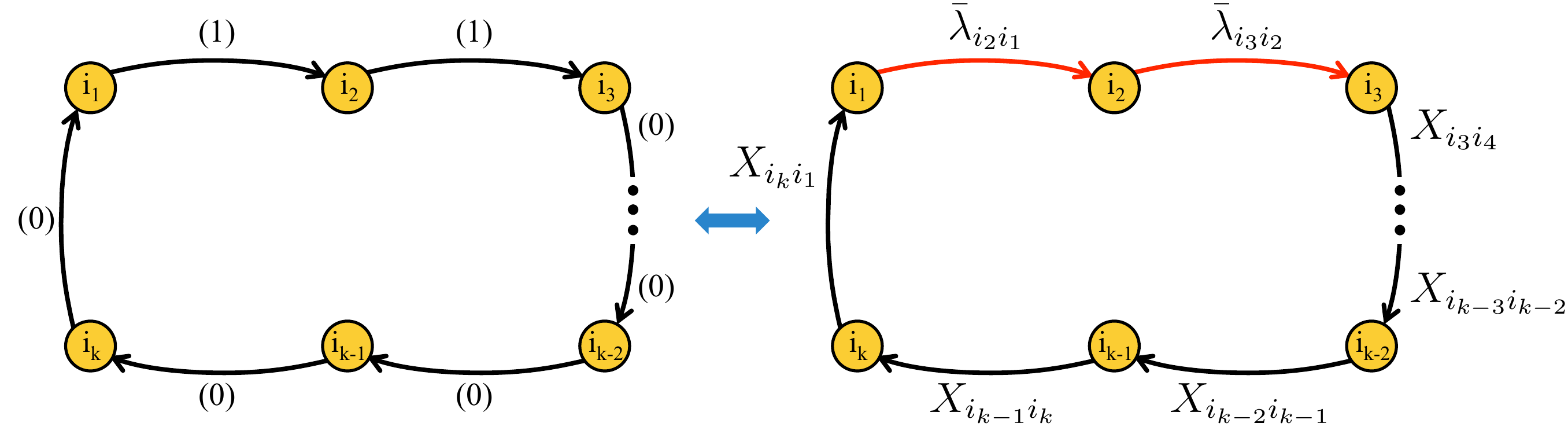}
\caption{Contribution to an $H$-term in a $0d$ $\mathcal{N}=1$ gauge theory.}
	\label{potential_m=3_H-term}
\end{figure}

The second type of potential term is a contribution to an $H$-term. It is shown in \fref{potential_m=3_H-term} and is given by
\beq
\mbox{$H$-term:}  \quad \varphi^{(1)}_{i_1 i_2}  \varphi^{(1)}_{i_2 i_3}  \varphi^{(0)}_{i_3 i_4} \ldots \varphi^{(0)}_{i_k i_1} \ \to \ \bar{\lambda}_{i_2 i_1} \bar{\lambda}_{i_3 i_2} X_{i_3 i_4}  \ldots X_{i_k i_1} \, ,.
\label{H-terms_m=3}
\eeq
There is an $H$-term for every pair of Fermi fields. Once again, $H$-terms are defined in terms of the chiral fields in the loops. For \eref{H-terms_m=3}, we have 
\beq
H_{\lambda_{i_2 i_1} \lambda_{i_3 i_2}} =  X_{i_3 i_4}  \ldots X_{i_k i_1} + \ldots
\eeq

The full potential can be written in terms of $J$- and $H$-terms as follows
\beq
W= \sum_a \lambda_a J_a + \sum_{a,b} \bar{\lambda}_a \bar{\lambda}_b H^{ab} \, ,
\eeq
where $a$ and $b$ run over Fermi fields.

In addition to vanishing $D$-terms, the moduli space of these theories is determined by only imposing vanishing $J$-terms. Once again, this is in full agreement with \eref{eq:Jacobian_algebra}, namely we only use the Jacobian algebra with respect to degree $(m-1)$ arrows, which in this case are the $\lambda_a$'s.

\paragraph{Kontsevich bracket.} 

The potential is holomorphic in chiral fields but contains both Fermi fields and their conjugates. Vanishing of the Kontsevich bracket requires
\beq
\sum_{a,b} \overline{H}^{ab} (\overline{X}) \bar{J}_a(\overline{X}) \lambda_b = 0 \,.
\eeq
Since every $\lambda_a$ is independent, this condition becomes
\beq
\sum_{b} \overline{H}^{ab} (\overline{X}) \bar{J}_a(\overline{X}) = 0 \quad 
\mbox{for every }a\,.
\label{H-constraint}
\eeq
This is the {\it $H$-constraint}, which in physics is necessary for preserving SUSY \cite{Franco:2016tcm}.

It is striking that the mathematical formulation of SUSY gauge theories in different dimensions in terms of graded quivers with potentials provides a unified explanation for seemingly unrelated constraints on the potential, such as the trace condition in $2d$ and the $H$-constraint in $0d$.

\subsection{Comments on $m=0$ and $6d$ $\mathcal{N}=(1,0)$}

Our discussion in previous sections started from $m=1$. This is a natural starting point since, mathematically, it corresponds to ordinary quivers and the first non-trivial example of mutations. However, our framework applies even for $m=0$, which becomes the natural initial case for the infinite tower of theories. 

Extending the dictionary in \sref{section_map_to_physics}, $m=0$ corresponds to $6d$ $\mathcal{N}=(1,0)$ gauge theories. Such theories can be realized on the worldvolume of D5-branes probing CY 2-folds. The case in which the CY$_2$ is toric is particularly tractable. Toric CY$_2$'s can only be $\mathbb{C}^2/\mathbb{Z}_n$ orbifolds, for which the toric diagrams are given by segments of $n+1$ points in $\mathbb{Z}$  \cite{Intriligator:1997kq,Blum:1997mm,Intriligator:1997dh,Brunner:1997gf}. These setups are T-dual to so-called {\it elliptic models} consisting of stacks of D6-branes suspended from $n$ parallel NS5-branes on $S^1$ \cite{Brunner:1997gf}. Elliptic models can be regarded as the simplest cousins of brane tilings. The corresponding {\it necklace quivers} realize the McKay correspondence for $\tilde{A}_{n-1}$ \cite{MR1886756}.

The interpretation of $m=0$ quivers as $6d$ $\mathcal{N}=(1,0)$ gauge theories works as follows. Nodes correspond to vector multiplets while $(0,0)$ unoriented arrows correspond to hypermultiplets. At present, we do not know whether tensor multiplets can be incorporated in this framework. This is certainly an interesting question that deserves further investigation. Since $m-1=-1$ in this case, these theories do not have a potential. Finally, there is no mutation, i.e. duality, in this case.

\section{Mutations as QFT Dualities}

\label{section_mutations_QFT_dualities}

For $m=1,2,3$ the mutations introduced in \sref{section_mutations_quivers} reproduce exactly the dualities of the corresponding quantum field theories. More precisely, for $m=1$ we obtain the Seiberg duality of $4d$ $\mathcal{N}=1$ theories \cite{Seiberg:1994pq}, for $m=2$ we recover the triality of $2d$ $(0,2)$ theories \cite{Gadde:2013lxa} and for $m=3$ we get the quadrality of $0d$ $\mathcal{N}=1$ theories \cite{Franco:2016tcm}. 

Seiberg duality is the prototypical and best understood example of a SUSY quantum field theory duality. It has passed numerous tests and found countless applications. The discovery of triality is far more recent \cite{Gadde:2013lxa} and it was initially motivated by the invariance of the elliptic genus.  By now, triality has been derived from Seiberg duality through compactification \cite{Honda:2015yha,Gadde:2015wta} and realized in terms of branes \cite{Franco:2016nwv,Franco:2016qxh}. Finally, quadrality was postulated based on mirror symmetry \cite{Franco:2016tcm}. These dualities are beautifully unified when realized in terms of geometric transitions using mirror symmetry \cite{Cachazo:2001sg,Franco:2016qxh,Franco:2016tcm}. Remarkably, the theory of graded quivers with potentials and their mutations achieves a similar algebraic unification of dualities in different dimensions. The existence of this subjacent mathematical structure adds further credence to the recently proposed dualities.\footnote{The term ``duality" might sound like a misnomer since we are referring to transformations that are not involutions, but this nomenclature has become standard in string theory so we adhere to it.}

As already noted, the mutation of graded quivers precisely reproduces the dualities of minimally supersymmetric gauge theories in $d=4,2,0$. Reformulating the mutation rules in physics language is straightforward. Rather than doing that, we now discuss how the mutations capture two distinctive features of these dualities. 

Anticomposition was first observed in physics in the context of $0d$ $\mathcal{N}=1$ quadrality \cite{Franco:2016tcm}. In this case, one type of mesons generated by a quadrality transformation corresponds to the composition between a Fermi field and a conjugate chiral field. In view of the results of this paper, this is not surprising, since $0d$ corresponds to $m=3$, which is the first nontrivial instance of anticomposition.\footnote{We can think that anticomposition is also present for $m=2$, but it acts rather trivially due to the Fermi-conjugate Fermi symmetry.} In \cite{Franco:2016tcm}, the evidence for anticomposition came from multiple fronts, including the transformation of brane charges under a geometric transition, the matching of anomalies and deformations between dual theories and the periodicity under four consecutive dualizations. The emergence of anticomposition from a simple mathematical structure is reassuring. 

The basic physics principle for determining the potentials of dual theories is that every term that is allowed by the symmetries of a theory must be present. The rules for mutating the potential introduced in \sref{section_mutation_potential} beautifully implement this principle. While the prescription based on symmetries is absolutely general, it can become hard to apply in complicated theories. The rules of \sref{section_mutation_potential} are {\it local}, namely they focus on the modification of the quiver in the neighborhood of the mutated node and are hence much more practical. In fact, they can be automatically implemented in a computer. Before this work, such local rules were only known for Seiberg duality ($m=1$). For triality, the most detailed understanding of the mutation of potentials was attained for toric theories \cite{Franco:2015tna,Franco:2015tya,Franco:2016nwv,Franco:2016qxh,Franco:2016fxm}. Even in this class of theories, the potential of dual theories must be read off from periodic quivers or brane brick models and doing so can become quite challenging. The rules in \sref{section_mutation_potential} hence represent a significant development in our understanding on how potentials mutate in $4d$, $2d$ and $0d$. The list of known explicit examples in $2d$ and $0d$ is still limited but rapidly growing (see e.g. \cite{Franco:2015tna,Franco:2015tya,Franco:2016nwv, Franco:2016qxh,Franco:2016fxm,Franco:2016tcm, Franco:2017cjj}). It is indeed possible to verify that our prescription reproduces all of them.

\section{Mirror Symmetry: Graded Quivers for Toric Calabi-Yaus}
 
 \label{section_mirror_symmetry}
 
 In this section we focus on toric CY $(m+2)$-folds, which give rise to a particularly nice family of $m$-graded quivers. We have already seen hints of this class of theories in \sref{section_map_to_physics} and \sref{section_gauge_theories}, where we discussed D-branes probing these geometries for $m=0,1,2,3$. A powerful way for connecting toric geometry to quivers involves mirror symmetry. Our primary goal is to emphasize that while this construction has a D-brane interpretation for $m\leq 3$, it actually applies to arbitrary $m$. Below we present a brief review of the key ideas and refer the reader to \cite{Feng:2005gw,Futaki:2014mpa,Franco:2016qxh,Franco:2016tcm} for further details. 

A toric CY$_{m+2}$ $\mathcal{M}$ is specified by a toric diagram $V$, which is a convex set of points in $\mathbb{Z}^{m+1}$. The corresponding mirror geometry \cite{Hori:2000kt,Hori:2000ck} is an $(m+2)$-fold $\mathcal{W}$ defined as a double fibration over the complex $W$-plane
\beq
\begin{array}{rl}
W = & P(x_1,\ldots, x_{m+1}) \\[.1cm]
W = & uv
\end{array}
\label{double_fibration}
\eeq 
where $u,v \in \mathbb {C}$ and $x_\mu\in \mathbb{C^*}$, $\mu=1,\ldots,m+1$. $P(x_1,\ldots, x_{m+1})$ is the Newton polynomial, which is defined as
\beq
P(x_1,\ldots, x_{m+1})=\sum_{\vec{v} \in V} c_{\vec{v}} \, x_1^{v_1} \ldots x_{n-1}^{v_{m+1}} , 
\label{Newton_polynomial}
\eeq
where the $c_{\vec{v}}$ are complex coefficients and the sum is over points $\vec{v}$ in the toric diagram. We can set $m+2$ of the coefficients to 1 by rescaling the $x_\mu$'s.

The critical points of $P$ are given by $(x_1^*,\ldots,x_{m+1}^*)$ satisfying
\beq
\left. {\partial \over \partial x_\mu} P(x_1,\ldots, x_{m+1})\right |_{(x_1^*,\ldots,x_{m+1}^*)}=0 \ \ \ \ \forall \, \mu .
\eeq
The corresponding critical values on the $W$-plane are $W^*=P(x_1^*,\ldots,x_{m+1}^*)$. For toric diagrams with at least one internal point, it can be shown that the number of critical points of $P$, which we call $G$, is equal to the normalized volume of the toric diagram \cite{Feng:2005gw}. In more detail, the normalization is determined with respect to the volume of a minimal ``tetrahedron" in  $(m+1)$ dimensions.

The double fibration consists of a holomorphic $m$ complex-dimensional surface $\Sigma_W$ coming from $P(x_1,\ldots, x_{m+1})$ and a $\mathbb{C}^*$ fibration from $uv$. The corresponding $S^{m}\times S^1$ is fibered over a {\it vanishing path}, which is a line segment connecting $W=0$ and $W=W^*$, and gives rise to an $S^{m+2}$.\footnote{More precisely, vanishing paths can be curved. See \cite{Franco:2016qxh} for a discussion.} We refer to these spheres as $\mathcal{C}_i$, $i=1,\ldots, G$.

The $\mathcal{C}_i$ are in one-to-one correspondence with {\it vanishing cycles} $C_i$ at $W=0$, where the $S^1$ fiber vanishes. Every $\mathcal{C}_i$ gives rise to a {\it vanishing cycle} $C_i$ with $S^{m+1}$ topology. The $C_i$ live on the Riemann surface $\Sigma_0$, defined by $P(x_1,\ldots, x_{m+1})=0$. As we explain below, the quiver theory is determined by how the $C_i$'s intersect.

\paragraph{Tomography.}

A convenient way of visualizing the geometry of the $S^{m}$'s is in terms of {\it tomography} \cite{Futaki:2014mpa,Franco:2016qxh}. The $x_\mu$-tomography corresponds to the projection of the $S^{n-2}$ spheres at $W=0$ onto the $x_\mu$-plane. An attractive feature of tomography is that it is easy to scale: every time $m$ is increased by one, we just include an additional $x_\mu$-plane. Various explicit examples of tomography for $m=1$ and $2$ can be found in \cite{Franco:2016qxh}.

\subsection{From Mirror Symmetry to Quivers}
 
The quiver diagram can be read from the mirror geometry. In fact, the mirror geometry specifies the full theory, namely not only its quiver but also its potential. The theories associated to toric CY $(m+2)$-folds are fully encoded by {\it periodic quivers} living on an $(m+1)$-dimensional torus $\mathbb{T}^{m+1}$. All the terms in the potential are mapped to {\it minimal plaquettes} in the periodic quiver. We refer the reader to \cite{Franco:2005rj,Franco:2015tna,Franco:2015tya,Franco:2016tcm} for implementations of this construction to $m\leq 3$. In this section we outline the basics of the map between mirror geometries and quiver theories.  A detailed study of the quiver theories associated to toric CY $(m+2)$-folds for arbitrary $m$ will be presented elsewhere \cite{toappear1}.

Every vanishing cycle $C_i$ corresponds to a node in the quiver. According to our previous discussion, the number of nodes is then equal to the normalized volume of the toric diagram. 
 
Every intersection between vanishing cycles gives rise to a field in the quiver. Depending on the coefficients in the Newton polynomial, intersections might not be fully resolved, i.e. they might have higher multiplicity. Explicit examples of this situation are studied in \cite{Franco:2016qxh}.

Finally, the periodic quiver is obtained by taking the {\it coamoeba projection}
\beq
(x_1,\ldots, x_{m+1}) \mapsto (\arg(x_1),\ldots, \arg(x_{m+1})) 
\eeq
which maps the intersections between vanishing cycles to the positions of the corresponding fields on the torus $\mathbb{T}^{m+1}$.

\subsection*{An Example: $\mathbb{C}^6/\mathbb{Z}_6$}
 
In order to illustrate these ideas, let us briefly consider the $\mathbb{C}^6/\mathbb{Z}_6$ orbifold with action $(1,1,1,1,1,1)$. The toric diagram for this geometry is given by the following collection of points in $\mathbb{Z}^5$:
\beq
\begin{array}{ccc}
(1,0,0,0,0) & \ \ \ \ \ \ &  \\
(0,1,0,0,0) & & (0,0,0,0,0) \\
\vdots & & (-1,-1,-1,-1,-1) \\
(0,0,0,0,1)  & & 
\end{array}
\eeq 

\begin{figure}[ht]
	\centering
	\includegraphics[width=14cm]{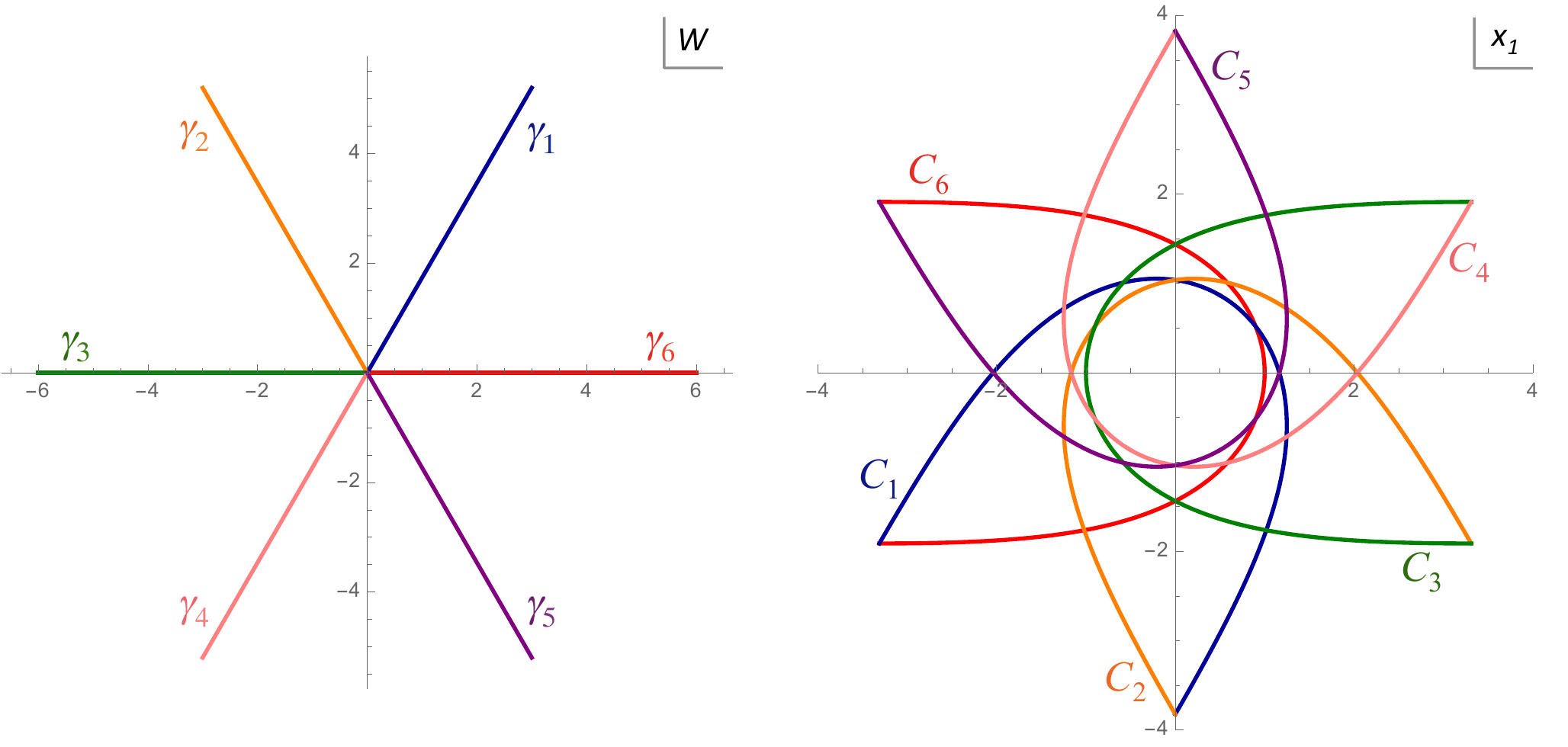}
\caption{Vanishing paths and $x_1$-tomography for the $\mathbb{C}^6/\mathbb{Z}_6$ orbifold with action $(1,1,1,1,1,1)$.}
	\label{mirror_C6_Z6}
\end{figure}

Six of the coefficients of the corresponding Newton polynomial can be scaled to 1, leaving a single free coefficient. In the notation of \eref{Newton_polynomial}, we pick this coefficient to be $c_{(0,0,0,0,0)}$. Setting it to zero, we obtain
\beq
P(x_1,x_2,x_3,x_4,x_5)=x_1 + x_2 + x_3 + x_4+ x_5 + {1\over x_1 \, x_2 \, x_3 \, x_4 \, x_5} \,.
\label{toric_diagram_C6_Z6}
\eeq
The normalized volume of the toric diagram defined by \eref{toric_diagram_C6_Z6} is six. This leads to six critical points of $P$, which in turn map to six nodes in the quiver, as expected for a $\mathbb{C}^6/\mathbb{Z}_6$ orbifold. The critical values are $W_j^*=6 \, \omega^j$, with $\omega=e^{i\pi/3}$ and $j=1,\ldots,6$. \fref{mirror_C6_Z6} shows the vanishing paths on the $W$-plane and the $x_1$ tomography.

The quiver theory associated to this geometry has $m=4$. The theory has an $SU(6)$ global symmetry corresponding to the isometry of the orbifold.\footnote{Notice that the $SU(6)$ global symmetry follows from the fact that we picked an orbifold that has the same action on each complex plane. Other $\mathbb{C}^6/\mathbb{Z}_6$ orbifolds have different global symmetries.} The quiver diagram consists of six nodes and arrows $\Phi^{(c)}_{i,i+c+1}$, $c=0,1,2$, transforming in the $(c+1)$-index antisymmetric representation of $SU(6)$. These properties generalize straightforwardly to $\mathbb{C}^{m+2}/\mathbb{Z}_{m+2}$ orbifolds with action $(1,\ldots,1)$, as it will be explained in \cite{toappear1}. The $\Phi^{(2)}_{i,i+3}$ fields are unoriented.

\begin{figure}[H]
	\centering
	\includegraphics[width=7cm]{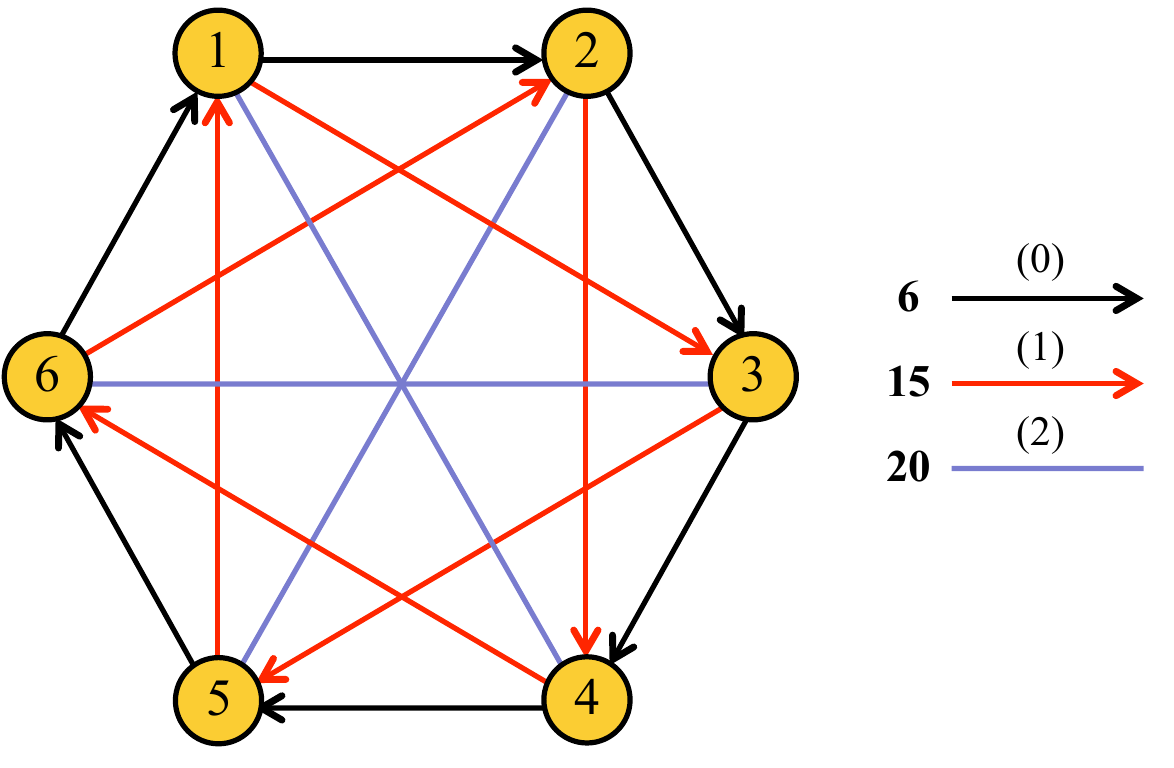}
\caption{Quiver diagram for the $\mathbb{C}^6/\mathbb{Z}_6$ orbifold with action $(1,1,1,1,1,1)$.}
	\label{quiver_C6_Z6}
\end{figure}

\fref{quiver_C6_Z6} shows the quiver diagram for this theory. Black, red and purple lines correspond to degree 0, 1 and 2, respectively. We also indicate the $SU(6)$ representation in which each type of fields transforms. 

The potential is purely cubic and takes the form
\beq
\begin{array}{rl}
W= \sum_{i=1}^6 \epsilon_{a_1 \ldots a_6}&  \left [\epsilon^{a_3 \ldots a_8} \, \Phi^{a_1(0)}_{i,i+1}\Phi^{a_2 (0)}_{i+1,i+2}\overline{\Phi}^{(1)}_{a_7 a_8;i,i+2}+\Phi^{a_1(0)}_{i,i+1}\Phi^{a_2 a_3(1)}_{i+1,i+3}\Phi^{a_4 a_5 a_6(2)}_{i+3,i} \right. \\ \\[-.3cm]
+ & \left. \epsilon^{a_4 \ldots a_9} \, \Phi^{a_1(0)}_{i,i+1}\Phi^{a_2 a_3(1)}_{i+1,i+3}\overline{\Phi}^{(2)}_{a_7 a_8 a_9;i,i+3}+\Phi^{a_1 a_2(1)}_{i,i+2}\Phi^{a_3 a_4(1)}_{i+2,i+4}\Phi^{a_5 a_6(1)}_{i+4,i} \right] .
\end{array}
\eeq

The bifundamental indices are taken mod$(6)$. The $a_\mu$ superscripts and subscripts are $SU(6)$ fundamental and antifundamental indices, respectively. They are contracted such that the potential is $SU(6)$ invariant. This is the first example of a graded quiver theory associated to a toric CY$_6$ to appear in the literature.

\subsection*{Mirror Symmetry and Degree}
 
So far, we have discussed how every intersection between vanishing cycles gives rise to a field in the quiver. We have not, however, explained how to determine the corresponding degree. Understanding this is crucial ingredient for completing the map between the mirror geometry and graded quivers. 

This question has been already addressed for the cases of $m=1$ and $2$. For $m=1$, the two possible degrees correspond to the two orientations of chiral fields and follow directly from the orientation of the intersecting cycles. This prescription is equivalent to the one based on the directions of intersecting zig-zag paths on brane tilings \cite{Hanany:2005ss,Feng:2005gw}. For $m=2$, the degree can also be determined by a detailed analysis of the intersection \cite{Franco:2016qxh}. An alternative systematic approach consists of connecting the geometry of interest to an orbifold by partial resolution. It is straightforward to determine field degrees for orbifolds and to follow them through the process of partial resolution. This method has been successfully exploited for $m=1$ and $2$ (see e.g. \cite{Morrison:1998cs,Beasley:1999uz,Feng:2000mi,Feng:2001xr,Franco:2015tna} and references therein) and will be studied for general $m$ in \cite{toappear1}.
 
\begin{figure}[H]
	\centering
	\includegraphics[width=7cm]{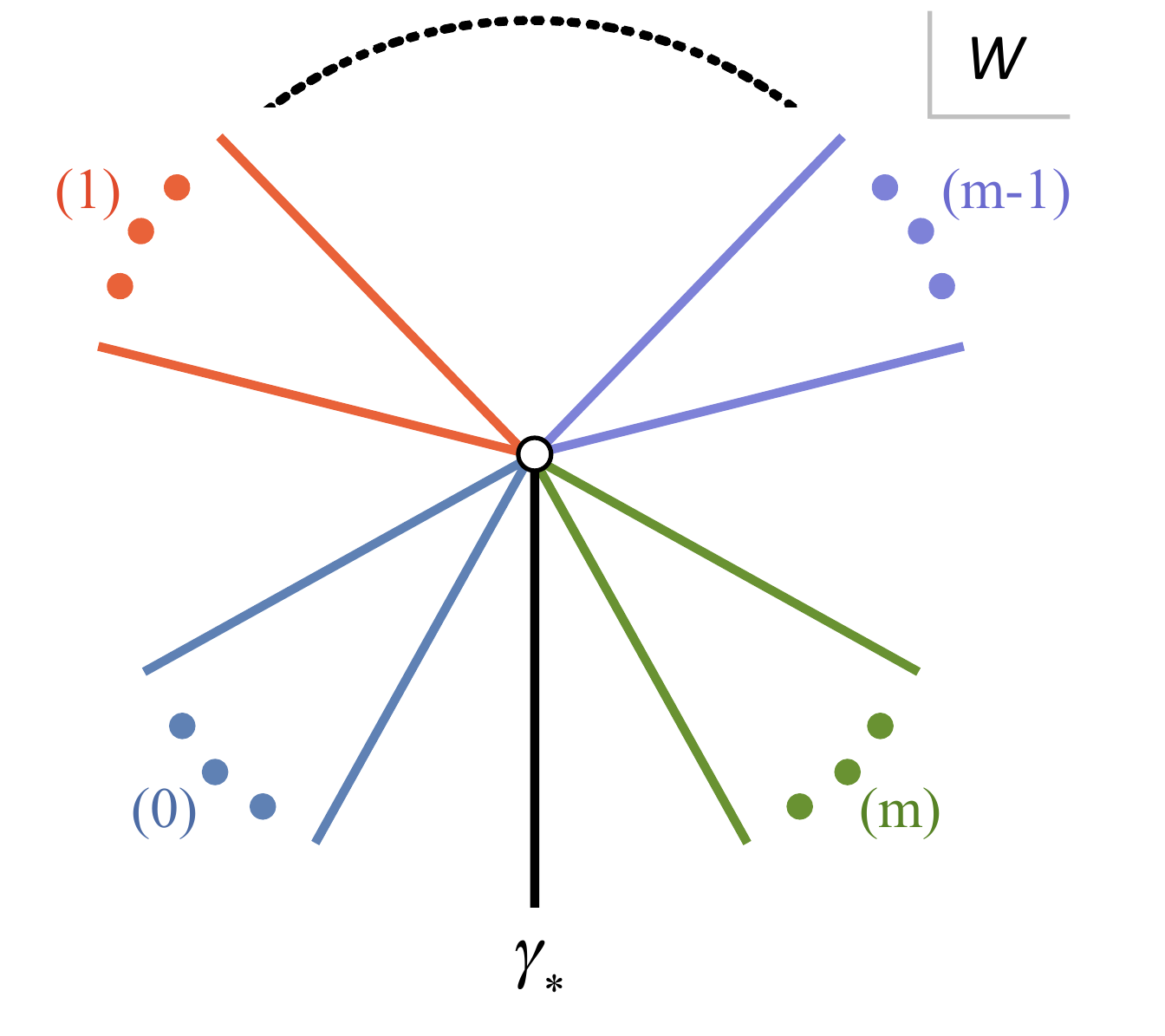}
\caption{For any reference cycle, the other vanishing paths are degree ordered on the $W$-plane.}
	\label{colors_mirror_W_plane}
\end{figure}

Let us mention a general connection between degree and the mirror. Consider an arbitrary vanishing cycle $C_\star$. Other vanishing cycles intersecting with $C_\star$ provide flavors to the corresponding node. For $m=1$, $2$ and $3$, it has been observed that the corresponding vanishing paths are arranged on the $W$-plane according to the cyclic order of increasing degree discussed in \sref{gradquivs}. The cyclic order emerges from geometry. This is more than an empirical observation: as we explain in the coming section, it is at the heart of the geometric realization of Seiberg duality, triality and quadrality in terms of geometric transitions in the mirror \cite{Cachazo:2001sg,Franco:2016qxh,Franco:2016tcm}. We thus expect this property, which is schematically shown in \fref{colors_mirror_W_plane}, to hold for arbitrary $m$.

It is reasonable to expect that a prescription for reading the degree of an intersection directly from the mirror might exist. We leave this interesting question for future work.
Note that the tomography of the mirror of a CY$_{m+2}$ involves $(m+1)$ $x_i$-planes. This is precisely the same number of possible degrees in the corresponding quiver.

\subsection{Mirror Symmetry and Mutations}
 
 \label{section_mirror_mutations}
 
For completeness, let us briefly discuss how the mutations of \sref{section_mutations_quivers} are realized as geometric transitions in the mirror geometry. This understanding was developed for $m=1$ in \cite{Cachazo:2001sg}, for $m=2$ in \cite{Franco:2016qxh} and for $m=3$ in \cite{Franco:2016tcm}. Indeed, in \cite{Franco:2016tcm} it was emphasized that mirror symmetry provides a unification of the mutations for different values of $m$. The previous works focused on the physically understood instances of $m\leq 3$. Our main new statement is that this geometric implementation of mutations applies to arbitrary $m$.

\begin{figure}[H]
	\centering
	\includegraphics[width=7cm]{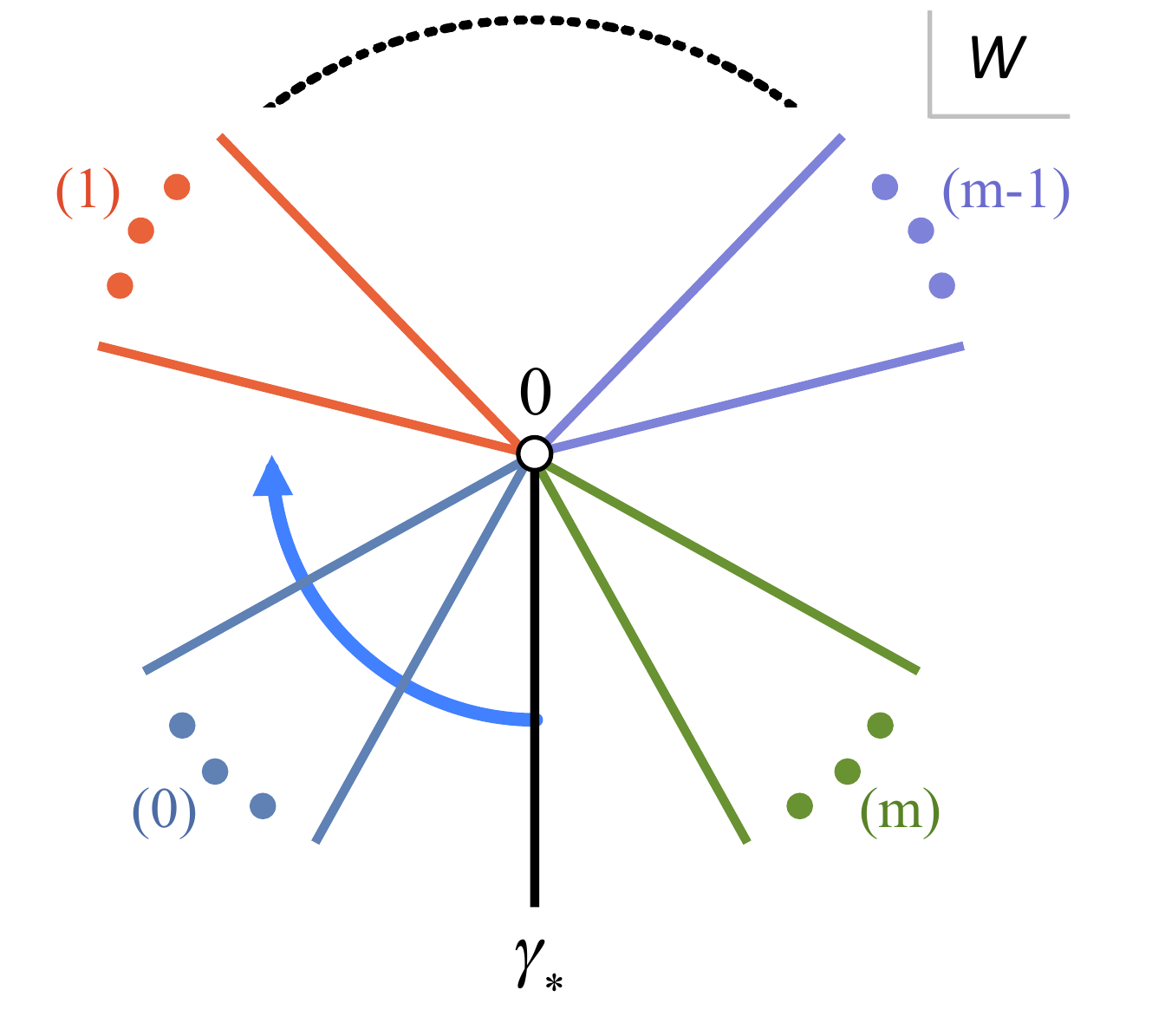}
\caption{A mutation is realized as a geometric transition in the mirror.}
	\label{mutation_mirror}
\end{figure}

The mutation of a node in the quiver associated to the vanishing cycle $C_{*}$ corresponds to the geometric transition shown in \fref{mutation_mirror}. The moduli of the underlying CY $(m+2)$-fold are changed until the vanishing path $\gamma_*$ moves past all the degree zero vanishing paths, i.e. those contributing incoming chirals to the mutated node, on the $W$-plane.
 
 \fref{mutation_mirror} makes it clear that the mutation is an order $(m+1)$ operation, since the vanishing paths associated to different degrees divide the $W$-plane into $(m+1)$ wedges.

\section{Algebraic Dimensional Reduction}

\label{section_dimensional_reduction}

In physics, {\it dimensional reduction} is the process that starts from a gauge theory in $d$ dimensions, assumes that fields are independent of $\Delta d$ of them and results in a new gauge theory in $d'=d-\Delta d$ dimensions. The relevant cases for this paper are the $6d \to 4d$, $4d \to 2d$ and $2d \to 0d$ reductions of minimally supersymmetric gauge theories. Each of them decreases the dimension $d\to d'=d-2$ while increasing $m\to m'=m+1$. From the perspective of the number of dimensions in which the gauge theory lives, it is clear that dimensional reduction cannot proceed beyond $m=3$, since it would require us to go below $0d$. We instead focus on the corresponding CY. Dimensional reduction increases the dimension of the CY by 1 in a very special way, simply adding a $\mathbb{C}$ factor to the original geometry. The underlying geometry, i.e. the moduli space, thus changes as follows
\beq
\mbox{CY}_{m+2}\to \mbox{CY}_{m+3}=\mbox{CY}_{m+2} \times \mathbb{C} .
\label{dimensional_reduction}
\eeq
It is natural to adopt \eref{dimensional_reduction} as the definition of dimensional reduction acting on graded quivers. This procedure coincides with dimensional reduction for the physical cases and generalizes it to arbitrary $m$.\footnote{In the Type IIB string theory constructions discussed in \sref{section_map_to_physics}, the upper bound on the dimension of the CY is 5. This follows from the fact that the ambient space is 10-dimensional.} A more appropriate name for this operation would perhaps be {\it CY dimensional increase}.

We now explain how this generalized notion of dimensional reduction admits an elegant implementation within the framework of graded quivers.

\paragraph{The Quiver.}

Let us first discuss how the quiver transforms under dimensional reduction. We use the notation of \eref{orientation_fields} for arrows. In order to facilitate their distinction, we refer to the fields in the original theory as $\Phi$ and those in the dimensionally reduced theory as $\Psi$. Dimensional reduction is given by
\beq 
\begin{array}{ccc}
m & & m+1 \\ \hline \\[-.4 cm]
\mbox{node}_i & \ \ \ \ \to \ \ \ \ &  \mbox{node}_i + \mbox{adjoint chiral } \Psi^{(0)}_{ii} \\[.12 cm]
\Phi^{(c)}_{ij} & \ \ \ \ \to \ \ \ \ & \Psi^{(c)}_{ij} + \tilde{\Psi}^{(c+1)}_{ij} 
\end{array}
\label{dim_red_quiver}
\eeq
where $0\leq c\leq m/2$. The previous table also applies when $i=j$, namely when the starting theory contains adjoint fields. Notice that, modulo the new $\Psi^{(0)}_{ii}$ fields, this procedure preserves the chiral field content of the theory. 

Equation \eref{dim_red_quiver} fully determines the dimensional reduction of a quiver. It is interesting to consider the undirected fields of degree $m/2$ that can be present in theories with even $m$ in further detail. According to \eref{dim_red_quiver}, we get
\beq
\begin{array}{ccc}
\mbox{even } m & & m+1 \\ \hline \\[-.3 cm]
\Phi^{(m/2)}_{ij} & \ \ \ \ \to \ \ \ \ & \Psi^{(m/2)}_{ij} +\tilde{\Psi}^{(m/2+1)}_{ij} = \Psi^{(m/2)}_{ij}+\tilde{\Psi}^{(m/2)}_{ji} 
\end{array}
\eeq
where we have used the fact that $\tilde{\Psi}^{(m/2+1)}_{ij}=\tilde{\Psi}^{(m/2)}_{ji}$. We went from a degree greater than $m'/2$ to a degree smaller than $m'/2$ by reversing the orientation of the arrow. In summary, the undirected fields $\Psi^{(m/2)}_{ij}$ are dimensionally reduced to a pair of fields of degree $m/2$ with opposite orientations.
 
\medskip
 
\paragraph{The Potential.}
 
 Let $W_m$ denote the original potential and let $W_{m+1}$ be the one for the dimensionally reduced theory. There are two types of contributions to $W_{m+1}$:
 
 \begin{itemize}
 \item[{\bf 1)}] {\bf Dimensional reduction of terms in $W_m$.} 
The degree of the potential increases by 1, going from $m-1$ in the parent theory to $m'-1=m$ in the dimensionally reduced one. According to \eref{dim_red_quiver}, every arrow $\Phi^{(c)}_{ij}$ in the initial theory gives rise to a pair of arrows $\Psi^{(c)}_{ij} + \tilde{\Psi}^{(c+1)}_{ij} $. The dimensional reduction of $W_m$ is then straightforward. For every term in $W_m$, we replace every field with the corresponding $\Psi^{(c)}_{ij}$ except for one, which we instead replace by $\tilde{\Psi}^{(c+1)}_{ij}$. We repeat this process for all fields in the term. This procedure generates a series of terms in $W_{m+1}$ for every term in $W_m$. 
 
Schematically, for any term in $W_m$ we have
\beq 
\begin{array}{ccc}
m & & m+1 \\ \hline \\[-.3 cm]
\Phi^{(c_1)}_{i_1 i_2} \Phi^{(c_2)}_{i_2 i_3} \ldots \Phi^{(c_k)}_{i_k i_1} & \ \ \ \ \to \ \ \ \ & 
\begin{array}{cccl}
& \tilde{\Psi}^{(c_1+1)}_{i_1 i_2} \Psi^{(c_2)}_{i_2 i_3} \ldots \Psi^{(c_k)}_{i_k i_1}& + & \Psi^{(c_1)}_{i_1 i_2} \tilde{\Psi}^{(c_2+1)}_{i_2 i_3} \ldots \Psi^{(c_k)}_{i_k i_1} \\[.12 cm] 
+ & \ldots & + & \Psi^{(c_1)}_{i_1 i_2} \Psi^{(c_2)}_{i_2 i_3} \ldots \tilde{\Psi}^{(c_k+1)}_{i_k i_1}.
\end{array}
\end{array}
\label{dim_red_potential}
\eeq
 
 \item[{\bf 2)}]  {\bf New terms involving adjoints.} 
In additon, $W_{m+1}$ contains a new class of terms. For every arrow $\Phi^{(c)}_{ij}$ in the original quiver, we introduce the following pair of potential terms in the dimensionally reduced one
 \be
\Psi^{(0)}_{ii} \Psi^{(c)}_{ij} \tilde{\Psi}^{(m-c-1)}_{ji} - \tilde{\Psi}^{(m-c-1)}_{ji} \Psi^{(c)}_{ij} \Psi^{(0)}_{jj} \, ,
\label{dim_red_potential_adjoints}
 \eeq
which use the adjoint chiral fields arising from the dimensional reduction of every node.\footnote{These fields should not be confused with other adjoint fields that descend from pre-existing adjoints in the initial theory.} The choice of relative sign is a convention.
 
\end{itemize}

These two steps generate all possible terms in $W_{m+1}$ that are consistent with the symmetries of the theory.
 
It is straightforward to check that the prescription introduced in this section reproduces the $6d\to 4d$, $4d\to 2d$ and $2d \to 0d$ dimensional reduction of minimally supersymmetric gauge theories.

\subsection*{Dimensional Reduction of the Moduli Space}

We now explain how, under dimensional reduction, the moduli space transforms simply as in \eref{dimensional_reduction}. For simplicity, let us restrict to theories in which the ranks of all nodes are equal to 1. Following the discussion in \sref{section_moduli_spaces}, for computing the moduli space we should focus exclusively on the chiral fields. Let us first consider the adjoint chiral fields $\Psi_{ii}^{(0)}$ descending from nodes in the parent theory. They only appear in the potential $W_{m+1}$ through the terms \eref{dim_red_potential_adjoints}. The relations arising from the cyclic derivatives of those terms with respect to either $\Psi^{(c)}_{ij}$ or  $\tilde{\Psi}^{(m-c-1)}_{ji}$ imply that $\Psi_{ii}^{(0)}= \Psi_{jj}^{(0)}$ for any $i$ and $j$. All of the $\Psi_{ii}^{(0)}$'s are thus equal and give rise to the decoupled $\mathbb{C}$ factor in \eref{dimensional_reduction}. Next, we know that the chiral fields in the original theory $\Phi^{(0)}_{ij}$ subject to vanishing $D$-terms and the relations coming from the initial potential $W_m$ correspond to the geometry that we call $\mbox{CY}_{m+2}$ in \eref{dimensional_reduction}. According to \eref{dim_red_quiver}, every such chiral gives rise to a chiral $\Psi^{(0)}_{ij}$ in the dimensionally reduced theory. The $D$-terms remain the same. Furthermore, the potential terms in \eref{dim_red_potential} guarantee that the relations for the $\Psi^{(0)}_{ij}$'s that follow from $W_{m+1}$ are precisely the same as the ones for the $\Phi^{(0)}_{ij}$'s due to $W_m$. We conclude that the $\Psi^{(0)}_{ij}$'s give rise to the $\mbox{CY}_{m+2}$ factor in \eref{dimensional_reduction}.

\section{Conclusions and Outlook}

\label{section_conclusion}

We have shown that the mathematical frameworks of graded quivers with potentials and higher Ginzburg algebras provide a unified description of minimally supersymmetric quantum field theories in even dimension. Moreover, the mutations of these quivers precisely correspond to dualities of the associated quantum field theories, some of which were discovered only recently.

We also explained how to exploit mirror symmetry to connect toric CY $(m+2)$-folds to the corresponding graded quivers. Finally, we discussed the simple implementation of dimensional reduction within this mathematical framework.

Our work suggests several interesting directions for future investigation. A few of them are:

\begin{itemize}
\item A striking revelation of our work is the fact that SUSY gauge theories in $6d$ to $0d$ are actually part of an infinite family of theories and that the same is true for the corresponding dualities. This naturally raises the following questions. Is there a physical realization of $m>3$ theories? If so, what is the meaning of the corresponding order $(m+1)$ dualities? We plan to address these issues in the near future \cite{toappear2}.

\item Graded quivers with potentials nicely describe SUSY gauge theories in even dimensions. Does a similar unified description exist for gauge theories and their dualities in odd dimensions?

\item In \sref{section_mirror_symmetry}, we outlined how higher dimensional generalizations of dimer models living on $\mathbb{T}^{m+1}$ can be constructed for toric CY $(m+2)$-folds for arbitrary $m$. For $m=2$, such generalization are called brane brick models and have been introduced in \cite{Franco:2015tya}. Similarly, brane hyperbrick models correspond to $m=3$ and were first postulated in \cite{Franco:2016tcm}. In future work, we plan to develop these constructions for general values of $m$ and investigate how they bridge geometry to the corresponding quivers. In particular, this will require the generalization of combinatorial notions such as perfect matchings, zig-zag paths, etc.

\item It would be interesting to extend the mathematical understanding of mutations of graded quivers with potentials to the case in which the mutated node contains adjoints fields. This is a promising direction for uncovering new order $(m+1)$ dualities for $m>1$. For $m=2,3$, i.e. $d=2,0$, these would be generalizations of triality and quadrality analogous to the generalization of Seiberg duality to $4d$ SQCD with an adjoint chiral field \cite{Kutasov:1995ve,Kutasov:1995np,Kutasov:1995ss}.

\end{itemize}

\acknowledgments

We would like to thank A. Garver, S. Gurvets, A. Hasan,  R.-K. Seong and especially S. Lee and C. Vafa for enjoyable discussions. We are also indebted to S. Oppermann for useful correspondence. S. F. gratefully acknowledges support from the Simons Center for Geometry and Physics, Stony Brook University, where some of the research for this paper was performed during the 2017 Simons Summer Workshop. The work of S. F. is supported by the U.S. National Science Foundation grant PHY-1518967 and by a PSC-CUNY award. The work of G. M. is supported by NSF grant DMS-1362980.

\newpage

\appendix

\section{Allowable Potential Terms and Mutations}

\label{section_mathematics_potentials}

\paragraph{Definition.} 
A configuration of $k$ arrows, with orientations of double arrows chosen so that it is an oriented cycle, graded as $(c_1,c_2,\dots, c_k)$ is an {\it allowable potential term} if and only if 
\begin{equation} \label{csum} c_1+c_2 + \dots + c_k = m-1. 
\end{equation}

\paragraph{Definition.} 
Given a graded quiver $\overline{Q}$, we call a node $i \in Q_0$ is {\it mesonic} if there is an incoming chiral, i.e. an arrow of degree zero, incident to $v$.  We call that $i$ {\it non-mesonic} otherwise.

\paragraph{Claim.} 
Any allowable potential term is mutation-equivalent to a configuration of the form $(m-1, 0, 0, \dots, 0)$ with exactly one non-chiral with degree $(m-1)$ as an oriented cycle via a sequence of non-mesonic mutations.\footnote{This statement applies in the obvious way in the case $m=1$.}

\paragraph{Proof.} 
A configuration $(m-1,0,0,\dots, 0)$ clearly satisfies \eref{csum}.  Furthermore, a non-mesonic mutation replaces 
a $2$-path of arrows having degrees $(d,e)$ with one of degrees $(d-1,e+1)$ and hence the sum on the left-hand-side of \eref{csum} is unchanged by such mutations.  Thus it is clear that all allowable potential terms will satisfy identity \eref{csum}.

We now show that any $(c_1,c_2,\dots, c_k)$ satisfying this equation is indeed reachable from $(m-1,0,0,\dots, 0)$ via non-mesonic mutations.\footnote{In the proof we assume that no node containing adjoint fields needs to be mutated along the sequence.}
First, we note that 
{\footnotesize \beq (c_1,c_2,\dots, c_d, 0,0,\dots, 0,c_k) \sim (c_1,c_2,\dots, c_d-1, 1,0,\dots, 0,c_k) \sim (c_1,c_2,\dots, c_d-1, 0,1,\dots, 0,c_k) \nonumber \eeq
\beq {\footnotesize \sim \dots \sim (c_1,c_2,\dots, c_d-1, 0,0,\dots, 1,c_k) \sim (c_1,c_2,\dots, c_d-1, 0,0,\dots, 0,c_k+1)} \eeq} by mutating 
at the $(d+1)^{st}$, $(d+2)^{nd}$, $\dots$, $k^{th}$ node in order.  Using this identity repeatedly, 
we convert between configurations
\beq
(c_1,c_2,\dots, c_{d-1}, c_d, 0, 0, \dots, 0, c_k) \sim (c_1,0,\dots, 0, 0, 0, 0, \dots, 0, c_2 + c_3 + \dots + c_d + c_k) .
\eeq
Since we assumed that $c_1+c_2 + \dots + c_k = m-1$ up front, this last entry is still from $\{0,1,2,\dots, m-1\}$.  And one final sequence of applications of this identity yields 
\beq
(0,0,\dots, 0, 0, 0, 0, \dots, 0, c_1+c_2 + c_3 + \dots + c_k) = (0,0,\dots, 0, 0, 0, 0, \dots, 0, m-1).
\eeq 
Up to cyclic rotation, we have a configuration of the desired form.

\section{Mutation of Differentials and Relation to Oppermann's Work}

\label{section_mutation_diff}

We now discuss how the differential structure transforms under mutation. In particular, we will show that $\{W,W \}=0$ is preserved by mutations. While doing so, we will discuss the connections between our approach and Oppermann's work in \cite{2015arXiv150402617O}. 

\noindent\paragraph{Claim.} 
If the potential $W$ vanishes under the Kontsevich backet, i.e. $\{W,W\}=0$, then after mutation, the resulting potential $W'$ still vanishes under the Kontsevich bracket, i.e. $\{W',W'\} = 0$.

\paragraph{Proof.} 
This has been proven as Theorem 8.1 of \cite{2015arXiv150402617O}.  We now provide an alternative proof using the description of differentials given in \sref{section_Ginzburg_algebras} and the mutation rules for the potential in \sref{section_mutation_potential}.

\begin{figure}[ht]
	\centering
	\includegraphics[width=16cm]{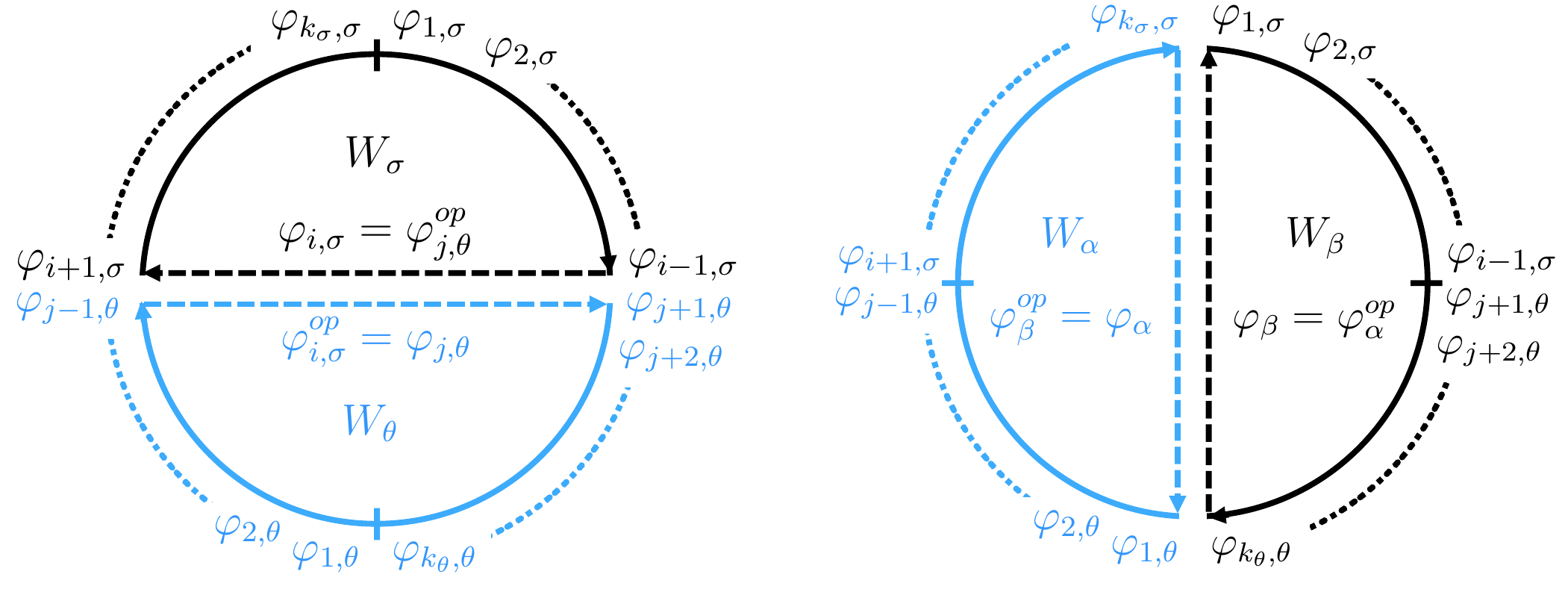}
\caption{Graphical representation of the two terms in \eref{eq:altsum}, which follow from four potential terms $W_\sigma + W_\theta + W_\alpha + W_\beta$.}
	\label{fig-altsum}
\end{figure}

Let us write the potential as
\beq
W=\sum_{\sigma \in W} c_\sigma W_\sigma ,
\eeq
where $\sigma$ indicates a cycle in the potential, $W_\sigma = \varphi_{1,\sigma}\varphi_{2,\sigma}\cdots \varphi_{k_\sigma,\sigma}$ is the corresponding term involving $k_\sigma$ arrows and we explicitly indicate the possibility of numerical coefficients $c_\sigma$. For the proof, we set all $c_\sigma$'s to be $1$ and introduce a new notation for arrows, in which $\varphi_{i,\sigma}$ indicates the $i^{th}$ arrow in the cycle $\sigma$. Then 
$$\{W,W\} = dW = \sum_{\sigma} d(W_\sigma)
 = \sum_{\sigma} 
 d(\varphi_{1,\sigma})\varphi_{2,\sigma}\cdots \varphi_{k_\sigma,\sigma}  + 
 (-1)^{|\varphi_{1,\sigma}|} \varphi_{1,\sigma} d(\varphi_{2,\sigma}\cdots \varphi_{k_\sigma,\sigma})$$
 {\footnotesize $$ =\sum_{\sigma}  
(\partial_{\varphi_{1,\sigma}^{op}}W) \varphi_{2,\sigma}\cdots \varphi_{k_\sigma,\sigma}  + 
 (-1)^{|\varphi_{1,\sigma}|} \varphi_{1,\sigma} d(\varphi_{2,\sigma})\varphi_{3,\sigma}\cdots \varphi_{k_\sigma,\sigma}
 + (-1)^{|\varphi_{1,\sigma}|+|\varphi_{2,\sigma}|}\varphi_{1,\sigma}\varphi_{2,\sigma} d(\varphi_{3,\sigma}\cdots \varphi_{k_\sigma,\sigma})$$}
\beq
 =\sum_{\sigma} \sum_{i=1}^{k_\sigma} 
 (-1)^{\sum_{j=1}^i |\varphi_{j,\sigma}|} \varphi_{1,\sigma}\varphi_{2,\sigma} \cdots \varphi_{i-1,\sigma}(\partial_{\varphi_{i,\sigma}^{op}}W)
 \varphi_{i+1,\sigma}\cdots \varphi_{k_\sigma,\sigma}.
 \label{eq_WW_1}
 \eeq 
In this previous expression, the factors involving derivatives are given by 
\beq
\partial_{\varphi_{i,\sigma}^{op}}W = 
\sum_{\theta \in W, \, \, W_\theta \supset \varphi_{i,\sigma}^{op}} \varphi_{1,\theta}\varphi_{2,\theta}\cdots \varphi_{j-1,\theta}\varphi_{j+1,\theta}
\cdots \varphi_{k_\theta,\theta} ,
\label{eq_WW_2}
\eeq
where $\varphi_{i,\sigma}^{op}$ coincides with $\varphi_{j,\theta}$ and is hence removed from the corresponding summand in \eref{eq_WW_2}. Consequently, the condition $\{W,W\} = 0$ in the initial theory implies that the double sum in \eref{eq_WW_1} decomposes into a sum of alternating sums of the form 
$$ \varphi_{1,\sigma}\varphi_{2,\sigma} \cdots \varphi_{i-1,\sigma}\varphi_{j+1,\theta}\cdots \varphi_{k_\theta,\theta} 
 \varphi_{1,\theta}\varphi_{2,\theta}\cdots\varphi_{j-1,\theta} \varphi_{i+1,\sigma}\cdots \varphi_{k_\sigma,\sigma}~~
 - $$
\begin{equation} \label{eq:altsum}
(-1)^{\sum_{j=1}^i |\varphi_{j,\sigma}|} \varphi_{j+1,\theta}\cdots \varphi_{k_\theta,\theta} 
 \varphi_{1,\theta}\varphi_{2,\theta}\cdots\varphi_{j-1,\theta} \varphi_{i+1,\sigma}\cdots \varphi_{k_\sigma,\sigma}
 \varphi_{1,\sigma}\varphi_{2,\sigma} \cdots \varphi_{i-1,\sigma}.
 \end{equation}
The first term comes from the potential terms $W_\sigma$ and $W_\theta$, while the second one comes from the potential terms 
\beq
\begin{array}{ccl}
W_\beta & = & \varphi_{j+1,\theta} 
\cdots \varphi_{k_\theta,\theta}\varphi_{\beta}\varphi_{1,\sigma}\varphi_{2,\sigma}\cdots \varphi_{i-1,\sigma} \\ [.15cm]
W_\alpha & = & \varphi_{i+1,\sigma} 
\cdots \varphi_{k_\sigma,\sigma} \varphi_{\alpha}\varphi_{1,\theta}\varphi_{2,\theta}\cdots \varphi_{j-1,\theta}
\end{array}
\eeq
where $\varphi_\alpha^{op} = \varphi_\beta$. Equation \eref{eq:altsum} can then be written as 
\beq
\partial_{\varphi_{i,\sigma}}W_\sigma \partial_{\varphi_{j,\theta}}W_\theta 
- \partial_{\varphi_{\alpha}}W_\alpha \partial_{\varphi_{\beta}}W_\beta.
 \label{eq_WW_6}
\eeq
We thus conclude that in $\{W,W\}$ every pair of terms of the form \eref{eq_WW_6} independently cancels due to signed cyclic equivalence \eqref{eq:sgneq}. Furthermore, for each cancellation, it is only necessary to focus on a set of four potential terms of the type $W_\sigma + W_\theta + W_\alpha + W_\beta$. \fref{fig-altsum} shows a graphical representation of this process. This provides a clear strategy for proving that $\{W',W'\}=0$ after mutation: it is sufficient to follow the evolution of such combinations of four potential terms. There are two possibilities, depending on whether some of the arrows in \eref{eq_WW_6} pass through the mutated node $k$ or not. Below we analyze each of them independently.

Let us first consider the case in which no arrow in \eref{eq_WW_6} runs through node $k$. Then, mutating at node $k$ will leave $W_\sigma + W_\theta + W_\alpha + W_\beta$ and the alternating sum of \eqref{eq:altsum} invariant and it will thus still vanish after mutation.

Next, let us consider the case in which some arrows in \eref{eq_WW_6} go through node $k$. This case can be separated into two possibilities. First, let us assume that node $k$ is not incident to $\varphi_{i,\sigma}^{op} = \varphi_{j,\theta}$ nor $\varphi_\alpha^{op} = \varphi_\beta$. Instead, node $k$ is at some intermediate point of the $2$-path $\xymatrix{ \bullet \ar@{->}[r]^{\varphi_{r,\sigma}^{(c_r)}} & k \ar@{->}[r]^{\varphi_{r+1,\sigma}^{(c_{r+1})}} & \bullet}$, where we have expanded our notation to indicate the degree of the arrows in the exponents. If the degree $c_r=0$, i.e. if $\varphi_{r,\sigma}^{(0)}$ is a chiral going into $k$, then a mesonic arrow corresponding to the composition $\varphi_{r,\sigma}\varphi_{r+1,\sigma}$ is created according to Rule $(2)$. Following Rule $(2.b)$, the product $\varphi_{r,\sigma}\varphi_{r+1,\sigma}$ is replaced by the meson in both $W_\sigma$ and $W_\alpha$ (resp. $W_\beta$). The resulting alternating sum in $\{W',W'\}$ still vanishes.  Additionally a new cubic term is added to the potential according to Rule $(2.a)$ but since the opposite of this mesonic arrow is not also adjoined, this contributes zero to $\{W',W'\}$ as well. If instead $c_r \neq 0$, then we use Rule $(2.c)$ instead and replace $\varphi_{r,\sigma}^{(c_r)}\varphi_{r+1,\sigma}^{(c_{r+1})}$ with  
$\varphi_{r,\sigma}^{(c_r-1)}\varphi_{r+1,\sigma}^{(c_{r+1}+1)}$. Again, after this replacement $\{W',W'\}=0$ still holds.

\begin{figure}[ht]
	\centering
	\includegraphics[width=14cm]{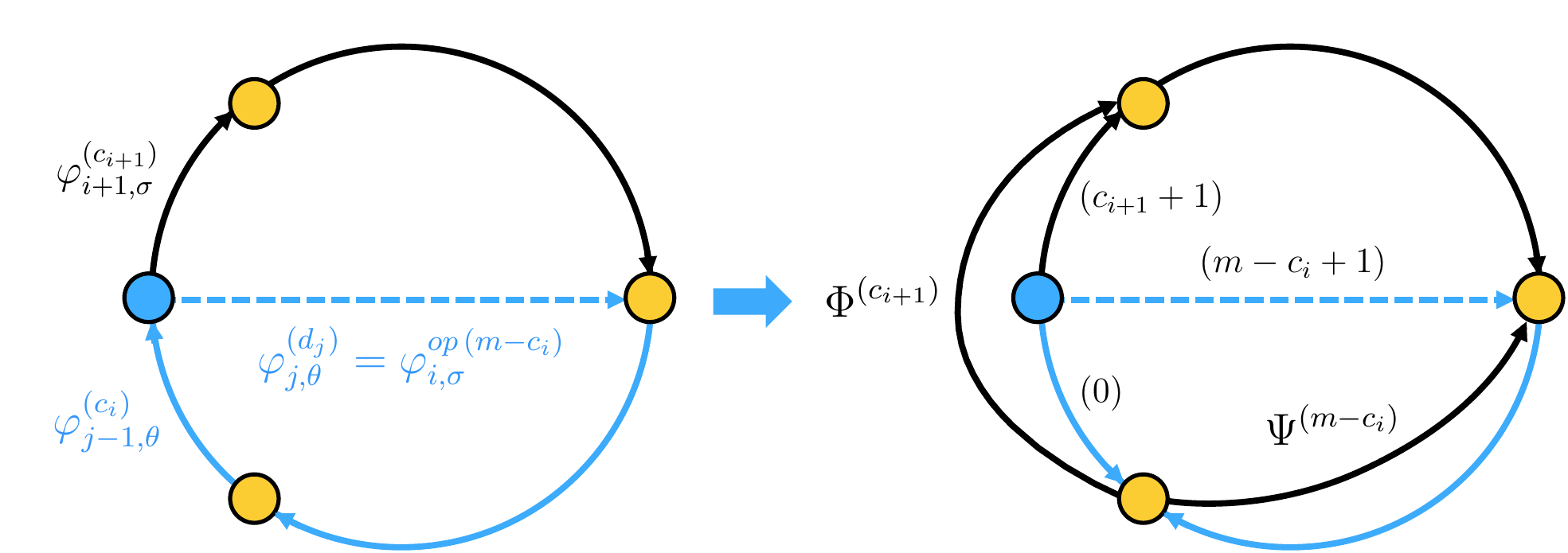}
\caption{Mutation at a node sitting at a triple intersection of arrows in \fref{fig-altsum}. In this figure, we assume $c_i=0$ and $d_j=m-c_i$.}
	\label{fig-3node}
\end{figure}

Lastly, let us consider the other possibility, in which we mutate at $k$ which is incident to $\varphi_{j,\theta}$ or $\varphi_\beta$. This means that the mutated node is located at the endpoint of one of the dashed arrows in \fref{fig-altsum}, at the triple intersection between a dashed arrow and the two solid lines of a different color.\footnote{We assume such cycles go through node $k$ only once. The proof for cycles that pass multiple times through the mutated node is analogous but requires a more lengthy analysis.} We illustrate this situation in \fref{fig-3node}, where we explicitly indicate the mutated node (blue) and the three other relevant nodes connected to it (yellow).

Without loss of generality, we consider the node at the intersection of the three arrows $\varphi_{j-1,\theta}$ (incoming), $\varphi_{j,\theta} = \varphi_{i,\sigma}^{op}$ (outgoing) and $\varphi_{i+1,\sigma}$ (outgoing). We then apply rules $(2.a)$-$(2.d)$ as appropriate, depending on whether  $\varphi_{j-1,\theta}$ is chiral or not.  We replace $W_\sigma + W_\theta + W_\alpha + W_\beta$ with $W_\sigma' + W_\theta' + W_\alpha' + W_\beta' + W_D'$ where $W_D'$ is the new term arising from Rule $(2.d)$. We obtain: 
\beq
\begin{array}{ccl}
W_\sigma' & = & \varphi_{i+1,\sigma}^{(c_{i+1}+1)}\varphi_{i+2,\sigma}^{(c_{i+2})}\cdots \varphi_{i-1,\sigma}^{(c_{i-1})}\overline{\varphi_{j,\theta}}^{(c_i-1)}\\[.2cm]
W_\theta' & = & \varphi_{j+1,\theta}^{(d_{j+1})}\varphi_{j+2,\theta}^{(d_{j+2})}\cdots \varphi_{j-2,\theta}^{(d_{j-2})}\Psi^{(m-c_i)}
+ \Psi^{(m-c_i)}\overline{\varphi_{j, \theta}}^{(c_i-1)}\overline{\varphi_{j-1,\theta}}^{(0)}\\[.2cm]
W_\alpha' & = & \varphi_{1,\theta}^{(d_1)}\varphi_{2,\theta}^{(d_2)}\cdots \varphi_{j-2,\theta}^{(d_{j-2})}\Phi^{(c_{i+1})}
\varphi_{i+2,\sigma}^{(c_{i+2})}\cdots \varphi_{k_\sigma,\sigma}^{(c_{k_\sigma})}\varphi_{\alpha}^{(a)}
+ \Phi^{(c_{i+1})} \overline{\varphi_{i+1,\sigma}}^{(m-c_{i+1}-1)} \overline{\varphi_{j-1,\theta}}^{(0)}\\[.2cm]
W_\beta' & = & W_\beta = \varphi_{j+1,\theta}^{(d_{j+1})} 
\cdots \varphi_{k_\theta,\theta}^{(d_{k_\theta})}\overline{\varphi_{\alpha}}^{(m-a)}\varphi_{1,\sigma}^{(c_1)}\varphi_{2,\sigma}^{(c_2)}\cdots \varphi_{i-1,\sigma}^{(c_{i-1})} \\[.2cm]
W_D' & = & - \overline{\Psi}^{(c_i)}\Phi^{(c_{i+1})}\varphi_{i+2,\sigma}^{(c_{i+2})}\cdots \varphi_{i-1,\sigma}^{(c_{i-1})}
\end{array}
\eeq
where, in order not to clutter the notation, we used $\overline{\varphi}$ for $\varphi^{op}$. Keeping track of the extra negative term resulting from Rule $(2.d)$, we indeed obtain $\{W',W'\}=0$ in this case as well. In particular, $\{W_\sigma' + W_\theta' + W_\alpha' + W_\beta' + W_D', W_\sigma' + W_\theta' + W_\alpha' + W_\beta' + W_D'\}$ equals 
{\footnotesize
\beq
\begin{array}{c}
\partial_{\overline{\Psi^{(c_i)}}}W_{D'}~\partial_{\Psi^{(m-c_i)}}W_{\theta'}
+ \partial_{\varphi_\alpha^{(a)}}W_{\alpha'}~\partial_{\overline{\varphi_\alpha}^{(m-a)}}W_{\beta'}
+ \partial_{\varphi_{i+1,\sigma}^{(c_{i+1}+1)}}W_{\sigma'}~\partial_{\overline{\varphi_{i+1,\sigma}}^{(m-c_{i+1}-1)}}W_{\alpha'}
\\[.25cm]
= - ~\Phi^{(c_{i+1})}\varphi_{i+2,\sigma}^{(c_{i+2})}\cdots \varphi_{i-1,\sigma}^{(c_{i-1})}\varphi_{j+1,\theta}^{(d_{j+1})}\varphi_{j+2,\theta}^{(d_{j+2})}\cdots \varphi_{j-2,\theta}^{(d_{j-2})}
- ~\Phi^{(c_{i+1})}\varphi_{i+2,\sigma}^{(c_{i+2})}\cdots \varphi_{i-1,\sigma}^{(c_{i-1})}\overline{\varphi_{j, \theta}}^{(c_i-1)}\overline{\varphi_{j-1,\theta}}^{(0)}
\\[.25cm]
+  ~\varphi_{1,\theta}^{(d_1)}\varphi_{2,\theta}^{(d_2)}\cdots \varphi_{j-2,\theta}^{(d_{j-2})}\Phi^{(c_{i+1})}
\varphi_{i+2,\sigma}^{(c_{i+2})}\cdots \varphi_{k_\sigma,\sigma}^{(c_{k_\sigma})}
\varphi_{1,\sigma}^{(c_1)}\varphi_{2,\sigma}^{(c_2)}\cdots \varphi_{i-1,\sigma}^{(c_{i-1})}\varphi_{j+1,\theta}^{(d_{j+1})}
\cdots \varphi_{k_\theta,\theta}^{(d_{k_\theta})}
\\[.25cm]
+ ~\varphi_{i+2,\sigma}^{(c_{i+2})}\cdots \varphi_{i-1,\sigma}^{(c_{i-1})}\overline{\varphi_{j,\theta}}^{(c_i-1)}\overline{\varphi_{j-1,\theta}}^{(0)}\Phi^{(c_{i+1})} = 0 
\end{array}
\eeq
}
up to signed cyclic equivalence.  

The prescription for mutating the potential and reducing it via massive terms that we introduced in \sref{section_mutation_potential} was partially motivated by Section 6 of Oppermann's work \cite{2015arXiv150402617O} which is formulated in terms of the higher Ginzburg algebra. The latter has the advantage of showing that the derived endomorphism ring of the graded quiver with potential is invariant under mutation \cite[Thm. 1.1]{2015arXiv150402617O}. In the remainder of this section, we discuss how the terminology of \cite{2015arXiv150402617O} compares with ours.

In order to help the reader interested in a more detailed comparison with \cite{2015arXiv150402617O} we will clarify the notation in that paper and use it in the discussion that follows. There are two new operations acting on arrows $\alpha$ of the quiver:

\begin{itemize}
\item $\alpha^{-1}$: roughly speaking, this is the same as $\alpha_{op}$. The distinction between the two is subtle and depends on whether we regard the quiver as consisting of single arrows, in which case we use $\alpha^{-1}$ to label a piece of a mesonic arrow, or double arrows, for which we use the notation $\alpha_{op}$.

\item $\alpha^*$: this is a compact way of indicating a mutated flavor, namely the mutation of an arrow connected to the mutated node. The rule for mutating flavors was given in \sref{section_mutation_quiver}. Oppermann's convention is to also flip the orientation of the flavors. Hence, in his notation, our mutation takes the form $\alpha \to \alpha^*_{op}$.
\end{itemize}

In \cite{2015arXiv150402617O}, the transformation of the potential in a mutation on node $k$ is described in an extremely compact form as
\beq
W \to W' = \mathrm{dec}_\mathrm{cyc} W + \mathop{ \sum_{\alpha: \xymatrix{~ \ar@[->][r]^{(0)} & k } } }_ 
{\varphi: \xymatrix{k\ar@[->][r]^{(c)} & } } \alpha ~\mathrm{dec}(\varphi\overline{\varphi}) \alpha^* .
\eeq

Such a compact expression becomes possible thanks to the introduction of the functions $\mathrm{dec}$ and $\mathrm{dec}_\mathrm{cyc}$, defined as an action on a cycle $\gamma$ and then extended linearly to act on a potential. In particular, if $\gamma$ is a cycle that is never incident to node $k$, the node where mutation is occurring, then $\mathrm{dec}(\gamma)$ and $\mathrm{dec}_\mathrm{cyc}(\gamma)$ are defined to be equal to $\gamma$.  Otherwise, for every $2$-path
$\varphi_{i,k} \varphi_{k,j}$ contained\footnote{For convenience, we now think of cycle $\gamma = \varphi_{i_1,i_2}\varphi_{i_2,i_3}\cdots \varphi_{i_\ell,i_1}$ where the subscripts denote the head and tail of the constituent arrows.  We also let $i,k,j$ label three of these vertices in a row.} inside of $\gamma$, we replace such a $2$-path with the element 
$\left(\varphi_{i,k} \varphi_{k,j}  - \sum_{\alpha:   \xymatrix{~\ar@[->][r]^{(0)} & k } } \varphi_{i,k}\alpha^{-1}\alpha \varphi_{k,j} \right)$.  The result is $\mathrm{dec}(\gamma)$.  For the case of  $\mathrm{dec}_\mathrm{cyc}(\gamma)$, this operation is taken cyclically, meaning that if $\gamma$ starts and ends at the node $k$, then $\left(1-\sum_{\alpha:   \xymatrix{~\ar@[->][r]^{(0)} & k } } \alpha^{-1}\alpha \right)$ is also multiplied at the beginning of the cycle.  
In particular, to compare our combinatorial rule for mutation of potentials to that in \cite{2015arXiv150402617O}, we split his rule into two pieces: 
$(a)$ $W \to \mathrm{dec}_\mathrm{cyc} W$; and $(b)$ $W' \to W' + \mathop{ \sum_{\alpha: \xymatrix{~\ar@[->][r]^{(0)} & k } } }_ 
{\varphi: \xymatrix{k\ar@[->][r]^{(c)} & ~} } \alpha ~\mathrm{dec}(\varphi \overline{\varphi}) \alpha^*$.

The first piece agrees with our rules $(2.b)$ and $(2.c)$ implicitly, since terms appearing in the potential remain in it after mutation.  The only differences are that the degrees of the arrows are updated accordingly and, in the case of a composition or anticomposition involving a degree zero incoming arrow $\alpha$, the $2$-path $\alpha\varphi$ (resp. $\varphi \alpha^{-1}$) becomes a single arrow $[\alpha\varphi]$ (respectively $[\varphi \alpha^{-1}]$) of degree matching that of $\varphi$ prior to mutation. The square bracket notation $[\alpha \beta]$ indicates this single mesonic arrow corresponding to the composition of $\alpha$ and $\beta$.

On the other hand, Rule $(2.d)$ is applied explicitly since the insertion of the factor $\left(1-\sum_{\alpha:   \xymatrix{~\ar@[->][r]^{(0)} & k } } \alpha^{-1}\alpha \right)$ yields\footnote{Originally, our Rule $(2.d)$ did not specify the sign of the new potential term.  However, as motivated by \cite{2015arXiv150402617O} and to ensure $\{W',W'\}=0$, it is mathematically natural to give such new potential terms from the case of two mesons a coefficient of $-1$.} one new potential term where the $\alpha$ and $\alpha^{-1}$ are paired with two different flavors to yield new mesonic arrows $[\varphi_{ik}\alpha^{-1}]$ and $[\alpha \varphi_{kj}]$.  Focusing on the prescence of a single chiral from a vertex denoted as $i_0$, i.e. denoted as $\alpha : i_0 \to^{(0)} k$, we have
$[\varphi_{ik}\alpha^{-1}] : i \to^{(c)} i_0$ and 
$[\alpha \varphi_{kj}]: i_0 \to^{(d)} j$.  This is consistent since this insertion of the above factor turns the term $W_\gamma = \varphi_{i_1,i_2}\cdots \varphi_{i_r,i}\varphi_{i,k}\varphi_{k,j}\varphi_{j,i_{r+4}}\cdots \varphi_{i_\ell,i_1}$ into 
$W_\gamma - \varphi_{i_1,i_2}\cdots \varphi_{i_r,i}[\varphi_{i,k}\alpha^{-1}][\alpha\varphi_{k,j}]\varphi_{j,i_{r+4}}\cdots \varphi_{i_\ell,i_1}$.

Lastly, Rule $(2.a)$, which adjoins a new cubic term involving a product of three arrows for every composition 
$[\alpha \varphi]$, involving an incoming arrow $\alpha$ of degree zero and any outgoing arrow $\varphi$, is exactly of the form\footnote{In \cite{2015arXiv150402617O}, potential terms are read right-to-left rather than left-to-right and outgoing chirals are used instead of incoming chirals.  These two reversals from our convention cancel each other out.}
\beq
\alpha \mathrm{dec}(\varphi \overline{\varphi})\alpha^* = [\alpha \varphi] \overline{\varphi} \alpha^*.
\eeq
Notice that since we are assuming that there are not adjoints at the node under mutation, we indeed have $\mathrm{dec}(\varphi \overline{\varphi}) = \varphi \overline{\varphi}$. In particular, the source of $\overline{\varphi}$ is a node other than $k$.

Note that Sections 7 and 8 of \cite{2015arXiv150402617O} explicitly describe the cyclic derivatives with respect to certain arrows. Putting this together, the vanishing of the square of the differential in the mutated theory is proved. This is equivalent to showing that the Kontsevich bracket vanishes for the mutated potential, hence giving an alternative to our proof based on the mutation rules given in \sref{section_mutation_potential}.

\section{Background on Cluster Categories}

\label{sec:CC}

We include a brief background on cluster categories and higher cluster categories for the interested reader. This provides some of the motivation behind graded (i.e. colored) quivers from the mathematical perspective.

We first focus on the $m=1$ case of ordinary quivers.  The path algebra $kQ$ of a finite acyclic quiver $Q$, i.e. a quiver that contains no cycles, satisfies a lot of important mathematical properties.  Such a path algebra is a {\it hereditary} {\it finite dimensional} {\it basic} algebra and its modules form an {\it abelian} $k$-{\it category} that is also {\it Krull-Schmidt}.  In other words, given two representations (equivalently, modules) of the path algebra $M$ and $N$, the set of homomorphisms between $M$ and $N$ is a $k$-vector space denoted as $Hom(M,N)$.  Further, any representation can be written as a direct sum of indecomposable representations in a unique way up to re-ordering.  

From this, we form $\mathcal{D}(kQ)$ the {\it bounded derived category} of $kQ$ with shift functor $[1]$, whose indecomposable objects are all of the form $M[i]$ where $M$ is an indecomposable of $kQ$ and $i \in \mathbb{Z}$ signifies an application the shift functor (or its inverse) a certain number of times.  The bounded derived category is a {\it triangulated category} which means that we can write down short certain exact sequences $0 \to A \to B \to C \to 0$ known as {\it almost split sequences}.  Such a sequence is not split, i.e. $B \not \cong A\oplus C$, but are irreducible in a certain sense meaning they are as close to being split as possible without being split.  In particular in an almost split sequence, $A$ and $C$ are both indecomposable and if we replace $B$ with any subrepresentation of it, we would induce a split exact sequence.  Moreover, the maps $A \to B$ and $B\to C$ are irreducible meaning that (i) they do not have left- or right-inverses and (ii) if these maps decompose as a composition, i.e. $f = gh$, then $g$ has a right-inverse or $h$ has a left-inverse.

Given an indecomposable representation $C$, there is a unique almost split sequence of the form $0 \to \_ \to \_ \to C \to 0$.  We define the {\it Auslander-Reiten translation} of indecomposable $C$ to be $\tau C = A$, unique indecomposable that fills in to the left-most object of the almost split sequence with $C$ as the right-most object.  The Auslander-Reiten (or AR) translation has the property that it sends projective indecomposable objects to zero and otherwise sends non-projective indecomposables to indecomposables.  

We work with a certain quotient of the bounded derived category known as the {Cluster Category} $\mathcal{C}_1(H)$ defined as$\mathcal{D}(H) / (\tau^{-1}\circ [1])$ where $\tau$ is {Auslander-Reiten translation} and $[1]$ is the shift functor.  Because of this identification, if $P_i$ is the projective indecomposable associated to node $i \in Q_0$, then $\tau P_i = P_i[1]$ rather than zero in $\mathcal{C}_1(H)$. Furthermore, $\tau P_i[1] = I_i$, the injective indecomposable associated to node $i \in Q_0$.  

\paragraph{Remark.} For a finite acyclic quiver $Q$, and a node $i \in Q_0$, the projective indecomposable $P_i$ is the module with basis given by all paths beginning at $i$ and ending at any other node.  In particular, $P_i = kQ e_i$.  In contrast, the injective indecomposable $I_i$ is the module with basis given by all paths ending at $i$, i.e. $I_i = e_i \, kQ$.  Lastly, $kQ = P_1 \oplus P_2 \oplus \cdots \oplus P_n$ as a module because of the orthogonality of idempotents.  

\vspace{0.5em}

The cluster category is again triangulated and Krull-Schmidt, and also has the property that it is a $2$-Calabi-Yau category, meaning that the Serre functor $\nu = [1]\tau$ is equivalent to $[2]$ (since $\tau^{-1}\circ [1] \sim id$).  Cluster categories provide a categorification for cluster algebras and by the Caldero-Chapoton map, the Laurent expansions of cluster variables even correspond to rigid indecomposables of $\mathcal{C}_1(H)$.  

This motivated the {\it higher $m$-cluster category}, which is the triangulated $(m+1)$-Calabi-Yau category obtained by the {quotient $\mathcal{D}(H) / (\tau^{-1}\circ [m])$}. Here $(m+1)$-Calabi-Yau signifies that the Serre function $\nu = [1]\tau \sim [m+1]$.

\vspace{1em}

The {Cluster-Tiling Objects} are maximally dimensional direct sums of indecomposables which have no self-extensions.  They can also be organized into what are called {\it exchange triangles}, each of which are short almost split sequences:
\beq
\begin{array}{ccccc}
X_i & \to & B_0 & \to & X_i' \\[.12 cm]
X_i' & \to & B_1 & \to & X_i'' \\[.12 cm]
X_i'' & \to & B_2 & \to & X_i''' \\[.12 cm]
& & \vdots & & \\[.12 cm]
X_i^{(m)} & \to & B_m & \to & X_i
\end{array}
\eeq
where
\beq
B_c = \bigoplus_{\varphi : \xymatrix{i\ar@{->}[r]^{(c)}&j}} X_j.
\eeq
{\it Higher tilting} objects are $X_1 \oplus X_2 \oplus \dots \oplus X_n$ where we take one element out, i.e. $X_i$, and then $B_0$ is some direct sum of the other $X_j$'s. These category theory definitions generalizing tilting theory (from 1970's) lead to the work of \cite{2008arXiv0809.0691B} to define graded quivers combinatorially from this algebra. See \cite{MR3406522} for a related but different treatment.

From a physical point of view, these exchange triangles also give rise to a relationship between ranks 
\beq
N_{B_0} = N_{X_i} + N_{X_i'},
\eeq 
which nicely agrees with the transformation rule for ranks under mutation \eref{mutation_ranks}.

Note that in the $m=1$ case, this sequence reduces to 
\beq
X_i \to B_0 \to X_i' \mathrm{~~and~~} X_i' \to B_1 \to X_i
\eeq
because of the $(m+1)=2$-periodicity.  In fact, this leads to a single exchange relation
\beq
CC(X_i)CC(X_i') = CC(B_0) + CC(B_1),
\eeq where $CC$ denotes the Caldero-Chapton map \cite{CalChap, clus-cat}. This relates the cluster category to cluster variables and cluster algebras in the $m=1$ case.  An analogue of such an algebraic structure for higher $m$ is an open question.

\subsubsection*{Constraints in the Potential and Higher Cluster Categories}

\begin{figure}[ht]
	\centering
	\includegraphics[width=10.5cm]{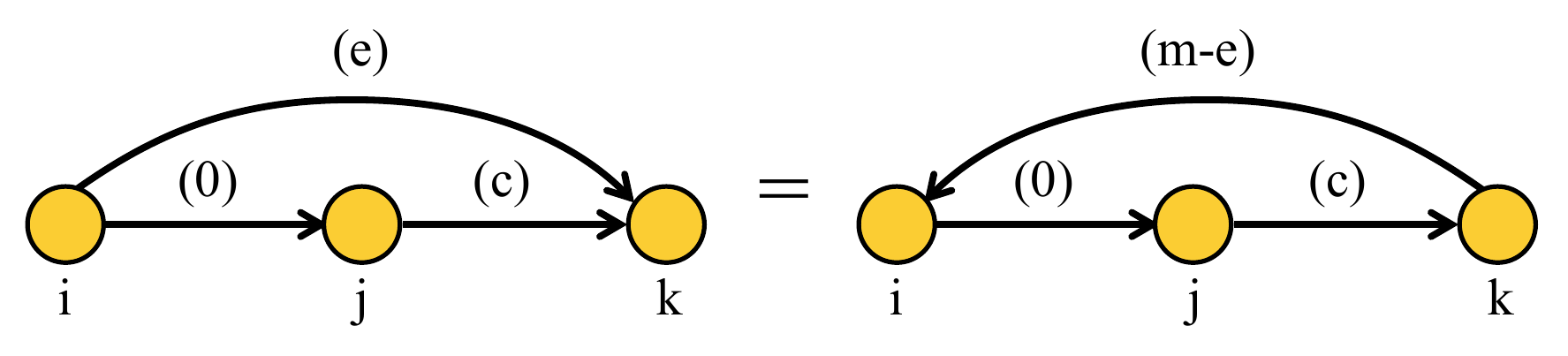}
\caption{A special local configuration considered in \cite{2008arXiv0809.0691B}.}
	\label{proposition_Buan_Thomas}
\end{figure}

The treatment of graded quivers and higher cluster categories in \cite{2008arXiv0809.0691B} does not include potentials.\footnote{While that work does not consider potentials explicitly, it is possible to argue that their manipulations are mostly consistent with assuming a totally generic potential. This statement is true, modulo the observation we made in \sref{section_mutation_potential} regarding the inconsistent removal of chiral-chiral pairs for $m>1$, for which they cannot form mass terms in the potential, in \cite{2008arXiv0809.0691B}.} However, that work proposes a condition that is closely related to our general discussion of potentials. Proposition 5.1 of \cite{2008arXiv0809.0691B} states that if a graded quiver associated to a higher-cluster tilting object has a local configuration of the form shown in \fref{proposition_Buan_Thomas}, then the degree $(e)$ must be $(c)$ or $(c+1)$. This local configuration is an {allowable potential term} if and only if $(e)=(c+1)$.

Forbidding other cyclic configurations of three arrows is important because otherwise mutation at the middle node $j$ would lead to a configuration where monochromaticity breaks.  In particular, as we saw in \sref{sec:mass-term}, the only massive potential terms are of the form $\varphi_{ik}^{(c)}\varphi_{ki}^{(m-c-1)} \sim \varphi_{ik}^{(c)}\varphi_{ik}^{(c+1)}$. As a result, only $2$-cycles arising from mutation at the middle term where Proposition 5.1 of \cite{2008arXiv0809.0691B} holds, i.e. $e=c+1$, can be deleted. The other case of $(e)=(c)$ yields multiple arrows of degree $(c)$.

\subsubsection*{Comments on Mesons at Nodes with Multiple Incoming Chirals}

Notice that Rule (2) of \sref{section_mutation_quiver} regarding the generation of mesons under mutation forbids new mesons coming from $2$-paths of the form 
$\xymatrix{ i \ar@{->}[r]^{(0)} & j \ar@{->}[r]^{(m)} & k}$, or equivalently from the anticomposition 
$\xymatrix{ i \ar@{->}[r]^{(0)} & j & k \ar@{->}[l]_{(0)}}$. This differs from Buan and Thomas \cite[Sec. 10]{2008arXiv0809.0691B}, which creates new arrows from such compositions. In fact, they not only generate a new arrow $\xymatrix{ i  \ar@/^0.5pc/[rr]^{(m)} & j & k}$ from the composition $\xymatrix{ i \ar@{->}[r]^{(0)} & j \ar@{->}[r]^{(m)} & k}$ but also a new arrow $\xymatrix{ i   & j & k \ar@/^0.5pc/[ll]^{(m)}}$ (which is equivalent to $\xymatrix{ i  \ar@/^0.5pc/[rr]^{(0)} & j & k}$).  This second arrow arises since every arrow in sight is in reality a double arrow and thus the configuration $\xymatrix{ i \ar@{->}[r]^{(0)} & j \ar@{->}[r]^{(m)} & k}$ is not only equivalent to $\xymatrix{ i \ar@{->}[r]^{(0)} & j & k \ar@{->}[l]_{(0)}}$ but is also equivalent to 
$\xymatrix{ i & j\ar@{->}[l]_{(m)}  & k \ar@{->}[l]_{(0)}}$, and so we have compositions in both directions.  The authors of \cite{2008arXiv0809.0691B} include a rule that removes any chiral-chiral 2-cycle, which results in the cancellation of these two arrows between nodes $i$ and $k$. Since they do not work with potentials for graded quivers, the result of constructing two new mesons from such compositions and then canceling them as a pair is identical to our rule of this paper forbidding compositions from $\xymatrix{ i \ar@{->}[r]^{(0)} & j \ar@{->}[r]^{(c)} & k}$ with $c=m$ in the first place.

Not only does Rule $(2.a)$ as we stated it avoid extra bookkeeping that would arise from the creation and deletion of such $2$-cycles but, more importantly, it is in fact necessary since as explained in \sref{sec:mass-term} chiral-chiral pairs cannot correspond to mass terms in the potential for $m>1$. Hence this $2$-cycle could not correspond to a mass term and Rule $(3)$ would not apply for reducing it. Nonetheless, up to this nuance,\footnote{As well as a reversal of the role of incoming and outgoing that we include to better match physics.} the definition of colored quiver mutation in  \cite{2008arXiv0809.0691B} indeed agrees with our definition of graded quiver mutation. Furthermore, while they do not implement the full treatment that would follow from a potential, the results they focus on are insensitive to this discrepancy.

\section{Silting}

\label{sec:silting}

There is a variant of tilting, known as \emph{silting}. Roughly speaking, it can be regarded as the $m\to \infty$ or CY$_\infty$ limit of graded quivers and their dualities. In particular, in this case there is no upper limit on the degree of arrows, which can grow arbitrarily under mutations. Similarly, sequences of repeated mutations on the same node are not periodic. Oppermann's work on potentials for graded quivers \cite{2015arXiv150402617O} was in fact motivated by considering this particular setting.  In this appendix, we sketch the relationship between silting and higher cluster categories, thereby providing the mathematical bridge between Buan-Thomas \cite{2008arXiv0809.0691B} and Oppermann \cite{2015arXiv150402617O}.

Following Buan-Reiten-Thomas \cite{2010arXiv1005.0276B}, we now compare silting objects and $m$-cluster tilting objects.  
Let $H$ be a finite dimensional hereditary algebra.  For $m \geq 1$, the $m$-cluster category $\mathcal{C}_m$ is 
defined as the quotient category $\mathcal{D}/\tau^{-1}[m]$.  Here $\mathcal{D}$ is the bounded derived category of $H$, 
$\tau$ is Auslander-Reiten translation in $\mathcal{D}$ and $[m]$ signifies applying the shift functor $[1]$ $m$ times.  

The $m$-cluster category is a Krull-Schmidt category, meaning that we can decompose objects into finite direct sums of indecomposables.  Let $mod ~H$ denote the indecomposables objects of $H$, and $mod ~H[i]$ denote the set after applying $i$ copies of the shift functor to each indecomposable.  
Because $\mathcal{C}_m$ is a quotient of $\mathcal{D}$, a fundamental domain for the set of indecomposables is given by 
\beq
\mathcal{S}_m = mod~ H[0] \cup mod~ H[1] \cup \dots \cup mod~ H[m-1] \cup \{P_1[m],P_2[m],\dots, P_n[m]\}
\eeq
where $P_1,P_2,\dots, P_n$ are projective objects in $H$.  

\paragraph{Proposition [Proposition 2.4 of \cite{2010arXiv1005.0276B}].} With the above setup, let $T$ be an object of $\mathcal{D}$ given as a direct sum of indecomposables all of which lie in $\mathcal{S}_m$.  Then $T$ is a {\it silting object} if and only if $T$ is an $m$-cluster tilting object in $\mathcal{C}_m$.

\paragraph{Theorem [Theorem 3.5, Corollary 3.6 of \cite{2010arXiv1005.0276B}].} Let $T_1\oplus T_2 \oplus \cdots T_n$ be a {\it basic} silting ojbect in $\mathcal{D}$, where $T_i$ are indecomposable and $n$ is the number of isomorphism-classes of simple $H$-modules (here basic means that each of the $T_i$'s are distinct). Assume that all $T_i = M[j]$ where $M$ is an indecomposable of $H$ and $j \geq 0$.  Choose $m$ large enough so that each $T_i \in \mathcal{S}_m$.

Then for each $i \in \{1,2,\dots, n\}$, there are $(m+1)$ non-isomorphic complements $M_0, M_1, \dots, M_m$ lying in $\mathcal{S}_m$ for the almost complete silting object 
\beq
T/T_i = T_1\oplus T_2 \oplus \cdots \oplus T_{i-1} \oplus T_{i+1} \oplus \cdots \oplus T_n.
\eeq

Furthermore, $T/T_i$ has a countably infinite number of non-isomorphic complements $M_i$ for $i \in \mathbb{Z}$ where there exists $M_{-1}$ and $M_{m+1}$ such that $M_{j+m+1} = M_{m+1}[j]$ and $M_{-j-1} \cong M_{-1}[-j]$ for $j\geq 0$.

\bibliographystyle{JHEP}
\bibliography{mybib}

\end{document}